\documentclass{aa}

\usepackage{graphicx}
\usepackage{txfonts}
\usepackage{color}
\usepackage{bm}
\usepackage{morefloats}
\usepackage{booktabs,siunitx}

\newcommand{\brac}[1]{\left\langle #1 \right\rangle}

\newcommand{\lH}{\ell_\mathrm{H}}
\newcommand{\bnab}{\bm{\nabla}}
\newcommand{\dd}{\,\mathrm{d}}
\newcommand{\au}{\mathrm{au}}
\newcommand{\K}{\mathrm{K}}
\newcommand{\id}[2]{#1_{\mathrm{#2}}}

\begin{document} 
 
   \title{Global simulations of protoplanetary disks with net magnetic flux}
   \subtitle{I. Non-ideal MHD case}

   \author{William B\'ethune
          \and
          Geoffroy Lesur 
          \and
          Jonathan Ferreira}

   \institute{Univ. Grenoble Alpes, CNRS, IPAG, F-38000 Grenoble, France \\
   			\email{william.bethune@univ-grenoble-alpes.fr}
             }

   \date{Received 14 November 2016 / Accepted 25 January 2017}

 
  \abstract
   {The planet-forming region of protoplanetary disks is cold, dense, and therefore weakly ionized. For this reason, magnetohydrodynamic (MHD) turbulence is thought to be mostly absent, and another mechanism has to be found to explain gas accretion. It has been proposed that magnetized winds, launched from the ionized disk surface, could drive accretion in the presence of a large-scale magnetic field. }
   {The efficiency and the impact of these surface winds on the disk structure is still highly uncertain. We present the first global simulations of a weakly ionized disk that exhibits large-scale magnetized winds. We also study the impact of self-organization, which was previously demonstrated only in non-stratified models. }
   {We perform numerical simulations of stratified disks with the PLUTO code. We compute the ionization fraction dynamically, and account for all three non-ideal MHD effects: ohmic and ambipolar diffusions, and the Hall drift. Simplified heating and cooling due to non-thermal radiation is also taken into account in the disk atmosphere.}
   {We find that disks can be accreting or not, depending on the configuration of the large-scale magnetic field. Magnetothermal winds, driven both by magnetic acceleration and heating of the atmosphere, are obtained in the accreting case. In some cases, these winds are asymmetric, ejecting predominantly on one side of the disk. The wind mass loss rate depends primarily on the average ratio of magnetic to thermal pressure in the disk midplane. The non-accreting case is characterized by a meridional circulation, with accretion layers at the disk surface and decretion in the midplane. Finally, we observe self-organization, resulting in axisymmetric rings of density and associated pressure ``bumps''. The underlying mechanism and its impact on observable structures are discussed. }
   {}


   \maketitle
%
\section{Introduction}

Protoplanetary disks are commonly observed around young stellar objects, linking the young star to its protostellar envelope. Optical and near-infrared images show disks with strong extinctions near the midplane \citep{PBS99}. They also reveal bipolar outflows, normal to the disk plane, with inner jets collimated on large scales \citep{BSW96}. The structure of these disks can be probed by infrared and radio wavelengths. These images have unveiled a number of features, such as spiral arms \citep{MGH12,BJ15}, asymmetric dust concentrations \citep{MD13,FTM13}, or axisymmetric rings \citep{ALMA15,TWHYA16}. 

The diversity of processes occurring in protoplanetary disks is ultimately constrained by the disk lifetime \citep{HLL01,CPS07}. On the one hand, any viable planetary formation scenario must be able to build planetesimals out of submicron-sized dust grains in a few million years. On the other hand, the rapid disk dispersal calls for efficient transport mechanisms. Part of the gas is accreted onto the central star, as deduced from their excess continuum emission and redshifted absorption lines \citep{EHGA14,HEG95}. The rest of it must be evacuated in the form of an outflow. What drives accretion and ejection remains a central research topic. 

Hydrodynamic models fail to account for the observed jets' collimation and ejection efficiency \citep{Cabrit07}. Including magnetic fields opens a number of promising perspectives. Magnetized Keplerian disks are prone to the magnetorotational instability \citep[MRI,][]{BH91}, which could drive accretion by turbulent angular momentum transport \citep{HGB95}. Large-scale magnetic fields could also extract angular momentum by the launching of a magnetohydrodynamic (MHD) wind \citep{BP82,FP95}. In this ideal MHD picture, accretion and ejection are co-dependent aspects of one global process \citep{Pudritz85}. Thermodynamical effects, and in particular heating, are also important in order to characterize outflow properties. Heating at the disk surface has been shown to increase the outflow mass \citep{CF00B}, a result recently recovered by \cite{BGY16}, albeit with a prescribed magnetic topology and without solving the disk dynamics.

This ideal MHD picture is compromised when realizing that, due to their optical thickness and low temperatures, protoplanetary disks should be weakly ionized beyond $0.1$---$1$ $\au$ \citep{G96}. The concept of a magnetic dead zone emerged, wherein the gas would be weakly coupled to the magnetic field. In principle, such a plasma should be described using a multi-fluid approach (e.g., \citealt{SD06}). However, the recombination time in protoplanetary disks is in general much shorter than the orbital time (\citealt{Bai11,RRD16}, but see \citealt{IN08}), making the single-fluid approximation perfectly valid and actually preferable to reduce computational costs.
In the single-fluid approximation, the reduced ionization fraction is taken into account by incorporating three non-ideal MHD effects: ohmic diffusion, ambipolar diffusion, and the Hall drift \citep{NU86,WN99}. Both diffusive effects can damp magnetic structures, and potentially quench the MRI \citep{J96,KB04}, as supported by numerous non-linear simulations \citep[e.g.,][]{HS98,FSH00,BS11}. 

The Hall drift leads to qualitatively new behavior, due to its dispersive nature. In particular, it induces new branches of instability, and a polarity dependence on the large-scale magnetic field \citep{W99,BT01,SS02,KD14}. Local simulations, using the shearing-sheet approximation \citep{GLB65}, found the emergence of organized structures from Hall-MHD turbulence. In non-stratified simulations, produces zonal flows \citep{KL13} and magnetized vortices \citep{BLF16}. These structures were not observed in the vertically stratified simulations of \cite{LKF14} and \cite{Bai15}. Instead, the magnetic field was laminar and filled the disk midplane. As these authors point out, the results were influenced by their vertical boundary conditions. 

The only way to ascertain whether these features are caused by boundary effects is to perform global models. Most of the properties of global, stratified and magnetized disks has been studied in the framework of ideal MHD \citep{FN06,FDKTH11}, albeit without any global poloidal field, preventing magnetized outflows. The inclusion of an global poloidal field in numerical simulations is notoriously difficult. Inner boundary conditions easily produce numerical artifacts, and high Alfv\'en speeds in the atmosphere can lead to dramatically small numerical time steps. 
For these reasons, relativistic simulations of disks around black holes (e.g., \citealt{BH08}) are technically easier to achieve. In the context of protoplanetary disks, initial attempts were either severely limited in computational time \citep{SH01} or did not resolve the small-scale dynamics of the disk \citep{MFZ10}. The more recent models of \cite{SI14} improved the situation, but they could not access the outflow acceleration and collimation regimes due to their limited vertical extent. Global simulations including both ohmic and ambipolar diffusion were carried out by \cite{GTNN15}, using realistic estimates for the ionization state of the disk. However, the vertical extent of these simulations was again somewhat limited, casting doubts on the asymptotic properties of the wind. Moreover, they left open the question of Hall-driven self-organization, as this effect was not included in their simulations.

This paper presents a series of global simulations of protoplanetary disks, including all three non-ideal MHD effects in a domain suitable to study disk winds and integrated over many orbital periods. Our primary objective is the characterization of accretion and ejection in a weakly magnetized disk, freed from the limitations of local simulations, and resolving the internal dynamics of the disk. A second goal is to ascertain the role of the Hall drift in producing large-scale, organized structures. The physical framework, including our assumptions, conventions and definitions, are given in Sect. \ref{sec:framework}. The specific method of resolution is described in Sect. \ref{sec:method}, and our results are presented in Sect. \ref{sec:results}. A discussion of our results is made in Sect. \ref{sec:discussion} before summarizing of our main findings in Sect. \ref{sec:summary}.

\section{Framework} \label{sec:framework}

\subsection{Physical model} \label{sec:physmod}

We wish to model the dynamics of a protoplanetary disk orbiting a young stellar object. We do not include the star in our model, but rather look at the outer regions of the disk, down to one astronomical unit ($\au$). We will consider the portion of a disk contained in a spherical shell, with an inner radius $r_0$, and covering the largest possible polar extent. We enumerate the main features of our disk model: 
\begin{enumerate}
\item the disk is geometrically thin;
\item the disk is made of weakly ionized gas;
\item the disk is threaded by a weak magnetic field;
\item the disk is embedded in a warm corona.
\end{enumerate}

We use the non-ideal MHD framework, and thus describe neutral and charged particles as one fluid with internal currents (cf. Sect. \ref{sec:dyneq}). The disk chemistry is evolved in a very simplified way, always assuming chemical equilibrium (cf. Sect. \ref{sec:nonideal}). This is made possible by the short time scale of chemical processes in comparison with the local orbital period \citep{Bai11}. The surrounding radiation field is not solved for; instead, we prescribe the temperature distribution via a cooling / heating function (see Appx. \ref{app:inbound}). We use two coordinate systems: 
\begin{itemize}
\item a cylindrical system $(r,\varphi,z)$, with the vertical coordinate $z$ along the cylindrical axis of the disk;
\item a spherical system $(R,\theta,\varphi)$, with the polar angle $\theta \in [0,\pi]$ increasing from the northern to the southern hemisphere. 
\end{itemize}

The disk will be denoted by the letter $\mathcal{D}$, while the symbols $\mathcal{C}^{+}$ and $\mathcal{C}^{-}$ will stand for the northern and southern coronal regions respectively. 

\subsubsection{Disk} \label{sec:framework_disk}

Let $c_s$ be the midplane sound speed and $\Omega$ the local orbital frequency. Under the influence of gravity and thermal pressure alone, a disk in hydrostatic equilibrium must be vertically stratified, its density varying on a characteristic vertical scale $h \equiv c_s / \Omega$. Protoplanetary disks are geometrically thin in the sense that the ratio $h/r$ is typically within $0.03$ to $0.2$, increasing with cylindrical radius $r$ \citep{BCM13}. We set a constant ratio $h/r = 5\%$ at all radii by imposing a constant ratio of sound over Keplerian velocity $c_s / \id{v}{K} = 5\%$ within the disk. 

We only consider disks with a radial profile of surface density $\Sigma(r) \propto 1/r$. This is flatter than the $r^{-3/2}$ often found in the literature since \cite{H81}, but seems to better fit recent observations \citep{ACH01,WC11}. 

The ionization fraction controls the coupling of the magnetic field with the gas. We are not interested in the detailed chemical composition of the medium: the computation of non-ideal MHD effects already comes with computational overhead, and the self-consistent resolution of a full chemical network is prohibitively demanding in CPU hours\footnote{Reduced chemical models can nonetheless be computationally affordable, see for example \cite{TSD07,IN08}.}. We reduce the disk chemistry to a local ionization fraction, computed dynamically with the evolution of the density distribution, and following the prescriptions of previous dedicated studies (cf. Sect. \ref{sec:nonideal}). 

Large-scale magnetic fields are observed in the vicinity of protoplanetary disks, hosted by the parent molecular cloud and presumably concentrated near the young stellar object \citep{RGLM14,SLS15,YLLS16}. Their global topology is still not well constrained, but their intensity is always weak in the sense that the midplane ratio of thermal over magnetic pressure $\beta \equiv 2 P/B^2 \gtrsim 1$ \citep{DPBF05}. Assuming that the global magnetic field has a non-zero vertical component, we will consider disks with initial average (midplane) $\beta \gg 1$, constant with radius. Given our radial density profile, a constant $\beta$ means that the average (midplane) Alfv\'en velocity $\id{\bm{v}}{A} \equiv \bm{B} \,/ \! \sqrt{\rho}$ is a constant fraction of the local Keplerian velocity. 

\subsubsection{Corona} \label{sec:framework_corona}

To study the global dynamics of protoplanetary disks via direct numerical simulations, previous studies used locally isothermal disk models, where the temperature decreases as a function of the cylindrical radius \citep{FN06, SI14, GTNN15}. These solutions have their density decreasing exponentially fast with height, $\rho/\rho_0 \sim \exp\left(-z^2 / 2 h^2\right)$. The Alfv\'en velocity increases exponentially with height; this is particularly appreciable in our case for we want a midplane $\beta \lesssim 10^{6}$ and over ten scale heights of vertical extent. Resolving the dynamics of these Alfv\'en waves over Keplerian time scales is numerically impractical, so we cannot rely on such cold disk equilibria alone\footnote{Unless limiting the Alfv\'en velocity by some artificial procedure; see for example the appendix of \cite{MS00}.}. 

It is known that protoplanetary disks are surrounded by a warm environment of optically thin and well ionized gas \citep{AKM11, BCM13, Wetal2016}. Assuming hydrostatic equilibrium, the predicted coronal temperatures can go as high as ten thousand degrees at $1\au$, above a disk at about $300\K$. The corresponding ratios of corona to midplane sound speed range typically from $3$ to $6$. 

To mimic this structure, we make the gas warmer above a certain height $H(r)$. We take a constant corona to midplane temperature ratio at all radii for the sake of self-similarity \citep{CL94,CF00B}. This ratio is set to $6$ in most simulations in order to avoid excessively low densities, so that the maximal Alfv\'en velocity remains comparable to the maximal Keplerian velocity (see Appx. \ref{app:inbound}).

We set the transition height $H$ based on the expected chemistry at the disk surface. Our ionization model, essentially the same as in \cite{LKF14}, predicts a high ionization degree for $z \gtrsim 3.6h$ (cf. their figure 1). More complete radiative transfer models also find a disk-corona transition at $z \approx 3h$ for $r\in \left[1,10\right]\au$ \citep{AKM11}. For these reasons, we heat the gas smoothly from $H \equiv 3.7h$ to $4.7h$ (cf. Fig. \ref{fig:profz_xecs} and Appx. \ref{app:initcond}). The ratio $H/h$ is held constant for simplicity. A self-consistent treatment of the thermodynamics and radiative transfer in the corona is beyond the scope of the present paper.

\subsection{Units and conventions} \label{sec:units}

Unless otherwise stated, all quantities will be evaluated in our code unit system. We use the inner radius $r_0$ and the Keplerian velocity $v_0$ at the inner radius as our distance and velocity units. We use both the orbital frequency $\Omega_0 \equiv v_0 / r_0$ and period $T_0 \equiv 2\pi/\Omega_0$ to measure time. Densities are measured in comparison to the initial density $\rho_0$ in the midplane and at the inner radius. The intensity of the magnetic field is measured by the corresponding Alfv\'en velocity $B/ \!\sqrt{\rho_0}$. 

The estimation of the ionization fraction introduces dimensional constants in the problem. We give in Table \ref{table:units} the conversion from code to physical units in the fiducial case of a disk around a  $1 \mathrm{M}_{\odot}$ star, with a surface density of $\Sigma(r) = 500 \,(r/1\au)^{-1}\mathrm{g} \mathrm{cm}^{-2}$, and for the two considered simulation inner radii $r_0 = 1 \au$ and $r_0 = 10 \au$. 

\begin{table*}[th]
Convertion from code to physical units
\centering                          
\begin{tabular}{l l l l}        
\hline\hline                 
Quantity & code unit & physical unit for $r_0=1 \au$ & physical unit for $r_0=10 \au$  \\
\hline                        
   distance & $r_0$ & $1 \;\au$ & $10 \;\au$\\
   time & $\Omega_0^{-1} \equiv T_0/2\pi$ & $0.159 \;\mathrm{yr}$ & $5.03\;\mathrm{yr}$ \\
   velocity & $v_0 \equiv \Omega_0  r_0$ & $29.8 \;\mathrm{km}.\mathrm{s}^{-1}$ & $9.42 \;\mathrm{km}.\mathrm{s}^{-1}$\\
   surface density & $\Sigma_0 $ & $500 \;\mathrm{g}.\mathrm{cm}^{-2}$ & $50 \;\mathrm{g}.\mathrm{cm}^{-2}$ \\
   volume density & $\rho_0 \equiv \Sigma_0 / \!\sqrt{2\pi}\,(h/r) r_0$ & $2.67 \times 10^{-10} \mathrm{g}.\mathrm{cm}^{-3}$ & $2.67 \times 10^{-12} \mathrm{g}.\mathrm{cm}^{-3}$\\
   mass & $m_0 \equiv \rho_0 r_0^3$ & $4.49 \times 10^{-4} \;\mathrm{M}_{\odot}$ & $4.49 \times 10^{-3} \;\mathrm{M}_{\odot}$ \\
   magnetic field & $B_0 \equiv v_0 \sqrt{\rho_0}$ & $48.6 \;\mathrm{G}$ & $1.54 \;\mathrm{G}$\\
\hline                                   
\end{tabular}
\vspace{2mm}
\caption{Correspondence between code and physical units for a disk around a $1 \mathrm{M}_{\odot}$ star, with a surface density of $\Sigma = 500 \,(r/1\au)^{-1}\mathrm{g}.\mathrm{cm}^{-2}$ for a simulation inner radius at $r_0 = 1 \au$ (third column) and at $r_0 = 10 \au$ (fourth column). } 
\label{table:units}      
\end{table*}

We use the pressure scale height $h(r) = 0.05 r$ to normalize curvilinear abscissa in most figures; if the cylindrical radius $r$ varies along the profile (for example along a streamline), then $h(r)$ varies as well. 

\subsection{Dynamical equations} \label{sec:dyneq}

We consider the equations of inviscid, non-ideal MHD. We denote by $\rho$ the bulk mass density, $P$ the gas pressure, $\bm{v}$ the bulk velocity, $\bm{B} = \vert B \vert \,\bm{e}_{b}$ the magnetic field, $\bm{J} \equiv \nabla \times \bm{B}$ the electric current, $\Phi$ the gravitational potential. The three non-ideal MHD effects, namely ohmic diffusion, ambipolar diffusion and the Hall effet, can be characterized by their effective diffusivities $\id{\eta}{O}$, $\id{\eta}{A}$ and $\id{\eta}{H}$ respectively \citep{W07}. It will be useful to rewrite the last two as
\begin{align}
&\id{\eta}{H} = \lH \,|\id{v}{\mathrm{A}}|, \label{eqn:deflH}\\
&\id{\eta}{A} = \id{\tau}{in} \,\id{v}{\mathrm{A}}^2, \label{eqn:defetaA}
\end{align}
where the Hall length $\lH$ and the ion-neutral coupling time $\id{\tau}{in}$ primarily depend on the ionization fraction and the nature of the charge carriers, but not on the magnetic field amplitude. In a singly charged ion-electron plasma, these coefficients read:
\begin{align}
\nonumber \eta_O&=\frac{c^2m_e}{4\pi e^2}\frac{n}{n_e}\langle \sigma v\rangle_e\\
\nonumber \lH&=\Bigg(\frac{c^2\rho}{4\pi e^2n_e^2}\Bigg)^{1/2}\\
\nonumber \id{\tau}{in}&=\frac{m_n+m_i}{\langle \sigma v\rangle_i\rho_i}
\end{align}
where $\langle \sigma v\rangle_{e,i}$ are respectively the electron-neutral and ion-neutral collision rates, $\rho_i$ is the ion mass density, $n$ is the neutral number density, $n_e$ is the electron neutral density, $m_{n,i,e}$ are the neutral, ion and electron masses and $e$ is the elementary charge.

The evolution of mass density, momentum and magnetic field are governed by the following equations: 
\begin{align}
&\partial_t\, \rho = - \bm{\nabla \cdot} \left[ \rho \bm{v} \right], \label{eqn:dyn-rho}\\
&\partial_t \!\left[ \rho \bm{v} \right] = - \bm{\nabla \cdot} \left[ \rho \bm{v
\otimes v} \right] - \bnab P + \bm{J \times B} - \rho\nabla\Phi, \label{eqn:dyn-v}\\
&\partial_t \bm{B} = \bm{\nabla \times} \left[ \bm{v \times B} -
\eta_{\mathrm{O}} \bm{J} - \eta_{\mathrm{H}} \bm{J \times e_b} +
\eta_{\mathrm{A}} \bm{J \times e_b \times e_b} \right]. \label{eqn:dyn-b}
\end{align}
The pressure distribution is locally isothermal, i.e., the isothermal sound speed $c_s(r,z) \equiv \sqrt{P/\rho}$ is prescribed beforehand. The detailed treatment of the gas energetics is given in Appx. \ref{app:inbound}.

\subsection{Ionization fraction} \label{sec:nonideal}

The intensity of the non-ideal MHD effects involves the local ionization fraction. We compute it via the same model as \cite{LKF14}. This model includes stellar X-rays and far ultraviolet (FUV) photons, cosmic rays and radio-active decay. The ionization rates and penetration depths are taken from previous studies \citep{UN81,BG09,UN09,PBC11}. The ionization rates from cosmic rays and radioactive decay are respectively $\zeta_{\mathrm{CR}} = 10^{-17} \mathrm{s}^{-1}$ and $\zeta_{\mathrm{RAD}} = 10^{-19} \mathrm{s}^{-1}$. The X-ray ionization rate is given by Eq. (21) of \cite{BG09}, with a luminosity $L_X = 10^{30} \mathrm{erg}.\mathrm{s}^{-1}$. The total ionization rate is balanced by dissociative recombination, without dust grains nor metals \citep{FTB02}. 

By soaking up electrons, dust grains can become one of the main charge carriers of the gas. Numerical simulations have not shown a drastic qualitative change when including ``large'' (micron-sized) grains \citep{LKF14, GTNN15}. However, submicron-sized grains can increase the diffusivities by several orders of magnitude when they are sufficiently abundant \citep{SW08,XB16}. The impact of small grains on the dynamics is therefore controlled by the dust size distribution, which is largely unconstrained. For this reason, our reaction network does not include grains.

Unlike in the local simulations of \cite{LKF14}, we compute the ionization fraction every $1/8$ inner orbit, consistently with the global evolution of the density distribution. Because we do not track the propagation of ionizing radiation, we simply use the gas column density integrated along the density gradient. The integration is thus performed in the polar direction instead of the vertical one. Both yield similar ionization profiles, because the regions of interest are located near the midplane. 

Regarding the X and FUV radiations, they should originate from the central star before being scattered and absorbed in the disk's upper layers. However, with $h/r$ constant and $\rho \sim R^{-2}$, the integration of density along spherical radius strongly depends on the lower integration bound. In addition, shadowing effects of the disk by the inner regions, which are excluded from the computational domain, could have an important dynamical impact \citep{BK12}. We however ignore this additional degree of complexity in our model. We therefore use the polar column density to compute the X and FUV ionizing rates, affecting only the surface layers of the disk and mimicking a global flaring \citep{Wetal2016,AKM11}. 

It is important to note that our ionization model is unable to model thermal ionization in the hot corona. For this reason, we assume the gas is fully ionized in the hot corona region by setting the diffusion coefficients to zero (see Sect. \ref{sec:fullcond}).

\subsection{Main diagnostics and definitions} \label{sec:diagno}

The averaging of a quantity $X$ over the azimuthal angle is:
\begin{equation}
\brac{X}_{\varphi}(r,z,t) \equiv \frac{1}{\Delta \varphi} \int_{\varphi=0}^{\Delta \varphi} X(r,\varphi,z,t)  \,\dd \varphi, 
\end{equation}
where $\Delta \varphi$ is the total azimuthal extent of the considered disk. Multiple indices denote consecutive averagings. Radial profiles are produced by integration in the disk region only:
\begin{equation}
\brac{X}_{\varphi,z}(r,t) \equiv \frac{1}{2H(r)} \int_{z=-H(r)}^{H(r)} \brac{X}_{\varphi}(r,z,t) \,\dd z. 
\end{equation}
The brackets and indices will be omitted when obvious from the context. Let $\brac{X}_{\rho} \equiv \brac{\rho X}_{\varphi,z} / \brac{\rho}_{\varphi,z}$ be the density weighted average of $X$. We define the fluctuating Reynolds stress tensor by $\mathcal{R}\equiv \rho \,\bm{\tilde{v}} \otimes \bm{\tilde{v}}$, where $\bm{\tilde{v}} = \bm{v} - \brac{\bm{v}}_{\rho}$. The Maxwell stress tensor is defined as $\mathcal{M} \equiv -\bm{B} \otimes \bm{B}$, and we define the sum $\mathcal{T} \equiv \mathcal{R} + \mathcal{M}$. We normalize these quantities by the local, vertically averaged pressure; although the flow might not be turbulent, we use the same notation as \cite{SS73}:
\begin{equation} \label{eqn:alphadef}
\alpha^{\mathcal{R}}(r,z,t) \equiv \brac{\mathcal{R}}_{\varphi} / \brac{P}_{\varphi,z},
\end{equation}
and similarly for $\alpha^{\mathcal{M}}$ and $\alpha^{\mathcal{T}}$ separately. The average radial mass flux can be deduced from the conservation of angular momentum \citep{BP99}:
\begin{align}
\begin{split}
\Sigma \brac{v_r}_{\rho} &\simeq -\frac{1}{\partial_r \left[ r \brac{v_{\varphi}}_{\varphi,z}\right]} \left( \frac{1}{r} \partial_r \left[2 r^2 H(r) \brac{\mathcal{T}_{r\varphi}}_{\varphi,z}\right] - r \left[ \brac{\mathcal{T}_{z\varphi}}_{\varphi}\right]_{-H}^{+H} \right) \\
&\equiv \tau_r + \tau_{z}, \label{eqn:accretor}
\end{split}
\end{align}
where $\Sigma(r) \equiv 2 H(r) \brac{\rho}_{\varphi,z}$ is the mass surface density, $\tau_r$ is the mass accretion rate due to to the radial transport of angular momentum, and $\tau_{z}$ the mass accretion rate due to angular momentum extracted by a wind through the surfaces $z = \pm H(r)$. We compute the wind mass loss rates through spherical shells at the outer simulation radius $R_{\mathrm{out}}$ in both hemispheres separately:
\begin{equation} \label{eqn:mdotwind}
\dot{m}^{\pm}_w \equiv \iint\rho v_R  \,R_{\mathrm{out}}^2 \sin(\theta) \dd \theta \dd \varphi, \quad (\theta,\varphi) \in \,\mathcal{C}^{\pm}
\end{equation}
and the total mass lost in the wind per unit time is $\dot{m}_W = \dot{m}^{-}_w + \dot{m}^{+}_w$. We quantify the net toroidal magnetic flux through the disk with
\begin{align} \label{eqn:symphi}
\iota_{\varphi} \equiv \frac{\int_{\mathcal{D}} \mathrm{sign}\left(\brac{B_{\varphi}}_{\varphi}\right) \dd r \dd \theta}{\int_{\mathcal{D}} \dd r \dd \theta}, 
\end{align}
so that $\iota_{\varphi} = \pm 1$ when a poloidal section of the disk is threaded by positive/negative toroidal field only, and zero when both polarities occupy the same area over this plane. We define
\begin{align} \label{eqn:modphi}
\sigma_{\varphi} \equiv \frac{\int_{\mathcal{D}} \mathrm{sign}\left(\brac{\mathcal{M}_{z\varphi}}_{\varphi}\right) \times \mathrm{sign}(z) \dd r \dd \theta}{\int_{\mathcal{D}} \dd r \dd \theta}, 
\end{align}
such that $\sigma_{\varphi} > 0$ when the angular momentum flux is directed from the disk to the corona in both hemispheres. The symmetry coefficients $\iota_{\varphi}$ and $\sigma_{\varphi}$ are analogous to the zeroth and first Fourier coefficients of $B_{\varphi}(z)$, respectively related to the net flux $\int_{\mathcal{D}} B_{\varphi}$, and to the phase of the large-scale fluctuations of $B_{\varphi}(z)$.

\section{Method} \label{sec:method}

\subsection{Numerical scheme} \label{sec:numschem}

We use the finite-volume code PLUTO \citep{M07} to integrate equations \eqref{eqn:dyn-rho} to \eqref{eqn:dyn-b} in time. The simulations are either three-dimensional or axisymmetric two-dimensional. The conservative variables are evolved via an explicit second order Runge-Kutta scheme. Parabolic terms are also included explicitly; this and the dispersive Hall whistler waves introduce strong constraints on the admissible timesteps satisfying the Courant–Friedrichs–Lewy (CFL) criterion. The CFL constraint due to the Keplerian velocity field is weak in comparison, so we do not activate the FARGO orbital advection scheme \citep{MFSK12}. Inter-cell fluxes are computed with a modified HLL Riemann solver, including the Hall drift in a conservative manner \citep{LKF14}. We use a piecewise linear space reconstruction to estimate the Godunov fluxes at cell interfaces, with the Van Leer slope limiter. The solenoidal condition $\nabla \bm{\cdot B} = 0$ for the magnetic field is maintained to machine precision by the constrained transport method \citep{EH88}. 

\subsection{Computational domain and grid} \label{sec:compgrid}

The computational domain is defined in spherical geometry by $(R,\theta) \in \left[1,10\right] \times \left[\pi/2-\theta_0, \pi/2+\theta_0\right]$, with the extremal polar angle $\theta_0 \equiv 20 \,(h/r) \approx \pi/3$. We found no significant benefit from increasing this polar range toward the poles but additional numerical difficulties. The hydrostatic equilibrium has an excessively low density in this region, and is numerically unstable. The equilibrium velocity field is also very slow; with very little inertia, spurious radial in-/out-flows develop, with high velocities and associated shear $\partial_{\theta} v_R$, disturbing the rest of the corona. 

\begin{figure}[ht]
\centering
\includegraphics[width=\hsize]{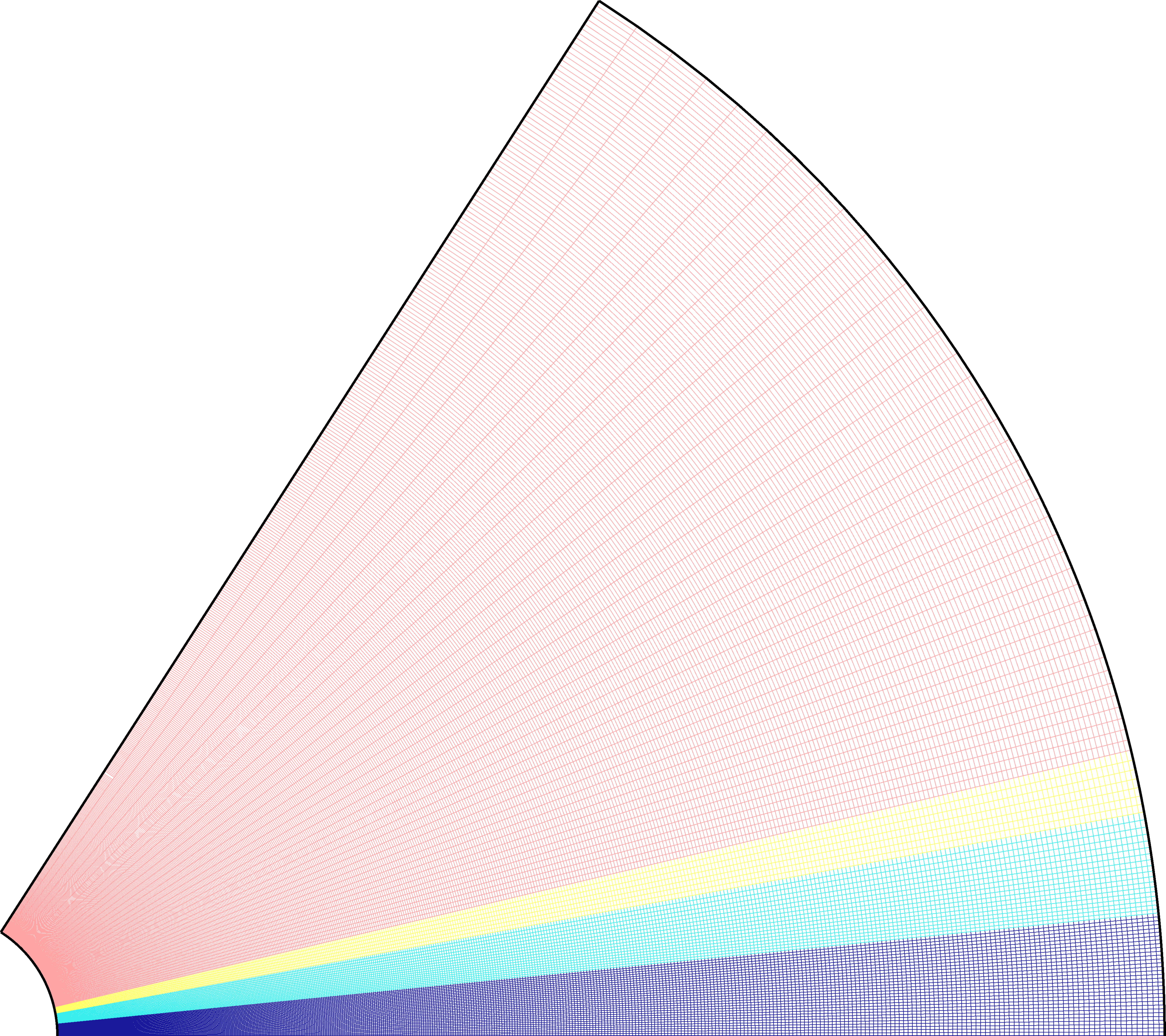}
\caption{Northern half of the computational grid; from the midplane to the pole: deep disk region (dark blue), disk surface region (cyan), disk-corona transition layer (yellow) and corona (red). }
\label{fig:meshmap}
\end{figure}

The polar interval is split into several ranges, as illustrated in Fig. \ref{fig:meshmap}. The deepest $4h$ are meshed with $64$ uniform grid cells (dark blue region). The remaining range, from two to twenty scale heights (cyan + yellow + red), is meshed with $64$ `stretched' cells, increasingly larger as we go to the polar boundary. This separation allows us to properly resolve the small-scale dynamics of the disk while not over-resolving the large volume of the corona. 

The radial interval is meshed with a logarithmic grid of $512$ cells, i.e., the increment $\delta r$ is proportional to the local radius $r$. This grid gets coarser as we go from the inner to the outer radial boundary, keeping a constant aspect ratio of grid cells at a given polar angle, and thus a constant numerical resolution per scale height. With this radial range, the initial mass in the disk region is always $m_d \equiv \int_{\mathcal{D}} \rho \approx 7.10 m_0$ in code units. 

Three-dimensional runs have an azimuthal extension $\varphi \in \left[0,\pi/2\right]$ to reduce computational cost compared to a full $2\pi$ disk. They are meshed with $128$ uniformly distributed azimuthal cells.

\subsection{Initial, boundary and internal conditions} \label{sec:fullcond}

We present here only the qualitative features of our numerical setup, and refer the reader to appendices \ref{app:initcond} to \ref{app:nonideal} for the details. 

\begin{figure}[ht]
\centering
\includegraphics[width=\hsize]{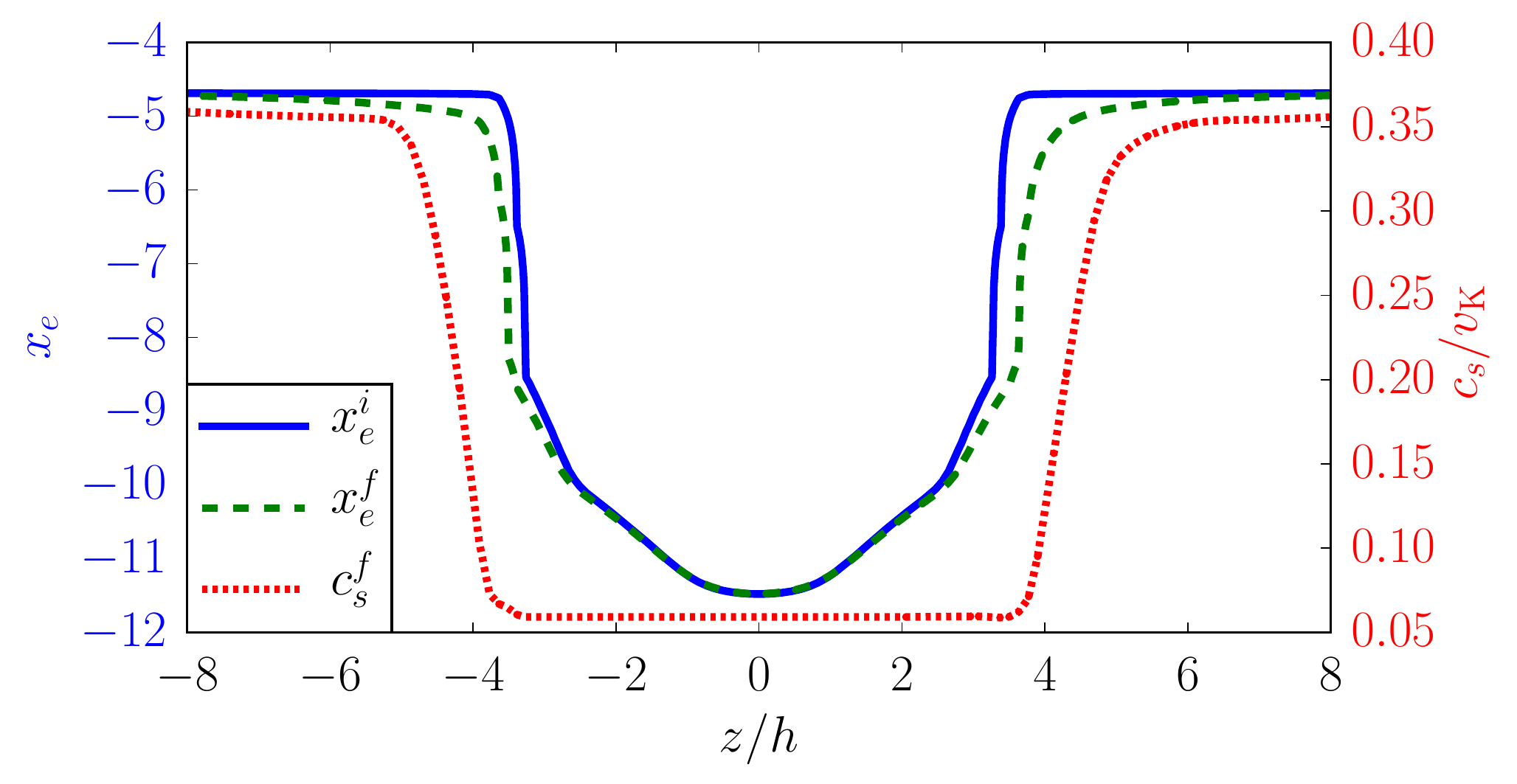}
\caption{Vertical profiles in run R1-P4 (from $1\au$ to $10\au$, with weak $B_z>0$) at $r=5 r_0 = 5 \au$: initial (solid blue) and final (dashed green at $300T_0$) ionization fractions $x^i_e$ and $x^f_e$, and final sound speed $c^f_s$ (red dots)	 in units of the midplane orbital velocity.}
\label{fig:profz_xecs}
\end{figure}

We initialize a cold disk plus warm corona hydrodynamic equilibrium. This thermodynamical configuration is held fixed during the simulation. A proper treatment of the thermodynamics would imply coupling the dynamics to radiative heating due to non-thermal radiation (typically X-rays and extreme UV) which is beyond the scope of this paper. We however motivate our temperature structure from full-blown radiative transfer calculations \citep[e.g.][]{AKM11}. The temperature transition is smoothed at the disk surface (yellow transition layer in Fig. \ref{fig:meshmap}). The vertical profiles of temperature and ionization fraction at $r=5\au$ in a typical run are shown in Fig. \ref{fig:profz_xecs}. The overall ionization fraction does not evolve considerably over the duration of a simulation, but it is generally reduced at the disk surface due to the screening by the outflow. The transition from $x_e<10^{-9}$ to $x_e>10^{-5}$ is steep, but as desired, it is properly matched by our prescribed temperature increase. The initial magnetic field is vertical everywhere; its intensity decreases with cylindrical radius, so that the midplane $\beta$ is constant with radius. The sign of the initial magnetic field is given with respect to the rotation axis; it can be either positive ($\bm{\Omega \cdot B} > 0$) or negative ($\bm{\Omega \cdot B} < 0$). We take $\bm{\Omega}$ along $+\bm{e}_z$. 

We impose a minimal outflow velocity through the polar and outer radial boundaries. Because the global wind tends to flow radially outward, outflow conditions at the inner radial boundary are inappropriate; we thus allow matter to flow into the computational domain from this side. Buffer zones are added in the vicinity of the radial boundaries to avoid spurious wave reflections and reduce the influence of the boundaries on the flow. 

We limit the Alfv\'en velocity to the Keplerian velocity at the inner disk radius, $\id{v}{A} \leq v_0$ by artificially increasing the density in the concerned cells. This cap occasionally affects a few percent of the corona region. We relax the temperature toward its initial distribution over a characteristic time of $1/10\Omega$, so that the flow is locally isothermal to a good approximation. 

\begin{figure}[ht]
\centering
\includegraphics[width=\hsize]{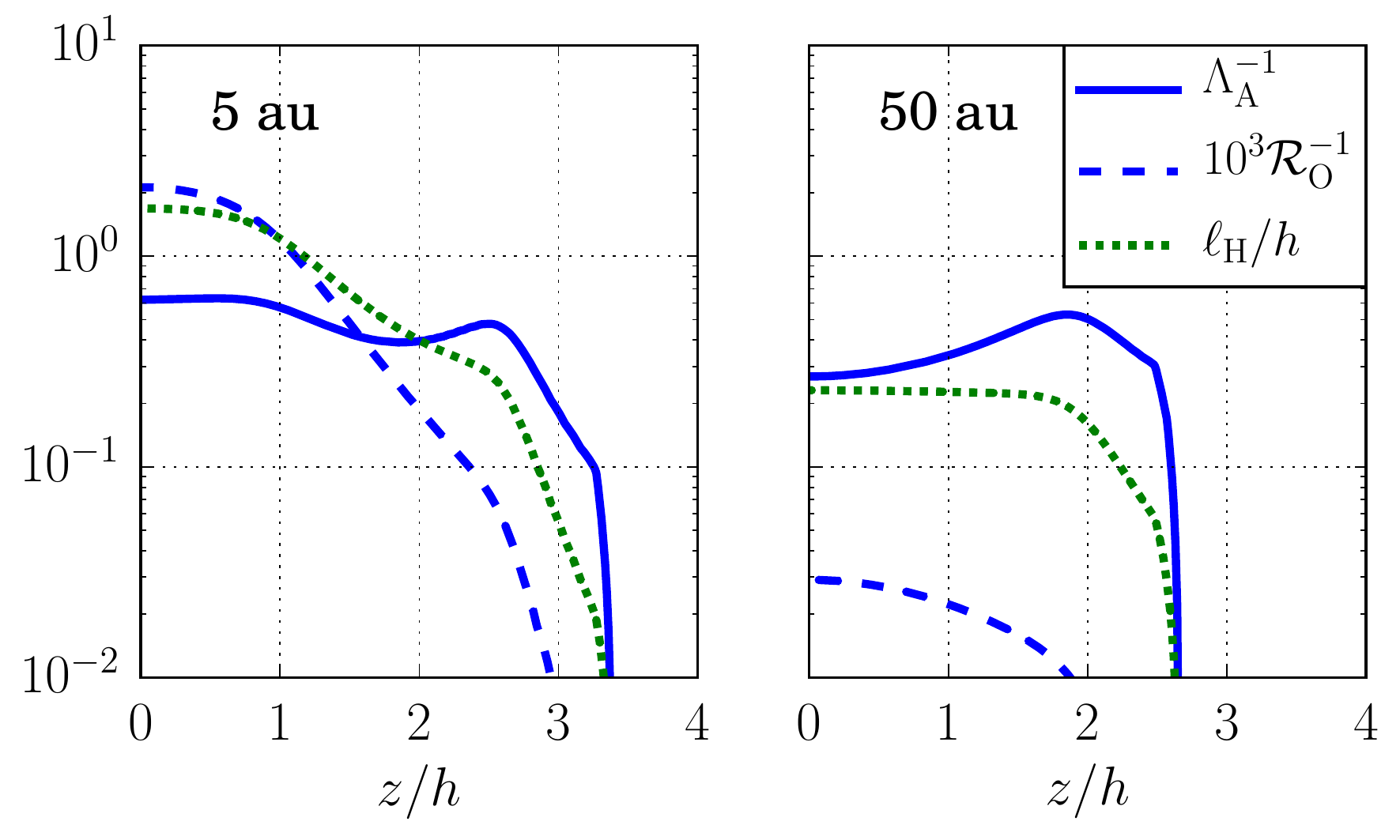}
\caption{Vertical profiles of dimensionless numbers characterizing non-ideal effects in the initial conditions at $r=5 \au$ (left panel) and $r=50 \au$ (right panel): inverse of the ambipolar Elsasser number $\Lambda_{\mathrm{A}}^{-1}$ (solid blue), inverse of the ohmic Reynolds number $\mathcal{R}_{\mathrm{O}}^{-1}$ (dashed blue, multiplied by $10^3$ for visibility, both at $5$ and $50 \au$), and ratio of the Hall length to the local pressure scale height $\lH/h$ (green dots). }
\label{fig:profz_nonideal}
\end{figure}

We draw in Fig. \ref{fig:profz_nonideal} the initial profiles of dimensionless numbers characterizing non-ideal effects. The ambipolar Elsasser number $\id{\Lambda}{A}$ and ohmic Reynolds number $\mathcal{R}_{\mathrm{O}}$ are defined by
\begin{equation}
\id{\Lambda}{A}^{-1} \equiv \frac{\eta_{\mathrm{A}} \Omega}{v_{\mathrm{A}}^2} = \Omega \tau_{in} \quad ; \quad \mathcal{R}_{\mathrm{O}}^{-1} \equiv \frac{\eta_{\mathrm{O}}}{\Omega h^2}. 
\end{equation}
These numbers do not depend on the magnetic field. Their radial distribution are comparable to those derived by \cite{SLKA15} (see their Fig. 1). Two procedures are applied to non-ideal coefficients deduced from the ionization fraction, as described in Appx.~\ref{app:nonideal}:
\begin{enumerate}
\item We reduce diffusivity coefficients in the corona, progressively from $z \gtrsim 3h$ to $z \lesssim 8h$ to account for the fact that the gas should become fully ionized in the corona.
\item We restrict the intensity of ohmic diffusion, the Hall effect and ambipolar diffusion to avoid extremely small time steps. This restriction effectively affects the Hall effect below $2.3 r_0$, and ambipolar diffusion only in runs with the highest magnetization $\beta = 5 \times 10^2$. 
\end{enumerate}

\section{Results} \label{sec:results}

We list in Table \ref{tab:setup} the parameters of the runs presented in this paper. Runs are labeled according to their inner radius (R1 at $r_0=1\,\au$, R10 at $r_0=10\,\au$) and initial magnetic field (P for $\bm{\Omega \cdot B}>0$, M in the opposite case, followed by the magnitude of the midplane beta). Three dimensional runs are listed at the bottom, the rest being two-dimensional axisymmetric simulations. All axisymmetric runs are integrated over $1000 T_0$, corresponding to approximately $32$ orbits at the outer radius. Run 3D-R1-P4 is evolved for $300T_0$, and the two other 3D runs for $200T_0$ only. It should be pointed out that each simulation shows a long-term (secular) evolution, where the large-scale configuration evolves over hundreds of local orbits. 

The main scalar diagnostics are gathered in Table \ref{table:results}, averaged over representative time intervals. We describe the properties of our reference simulation in Sect. \ref{sec:fiducial}. This run is not generic in every aspect, so we characterize a variety of other processes in the following sections. In Sect. \ref{sec:hallpol}, we examine the magnetic polarity dependence introduced by the Hall drift. The properties of the wind in a warm corona are described in Sect. \ref{sec:ejec}. In Sect. \ref{sec:nonejec}, we presents disks in which the vertical flux of angular momentum causes a large-scale meridional circulation, with no net mass accretion. Sect. \ref{sec:selforgathor} describes a vertical symmetry breaking, leading to a one-sided magnetic ejection. Sect. \ref{sec:zf} is dedicated to a self-organization process leading to the formation of zonal flows. Finally, our set of simulations is considered as a whole for discussion in Sect. \ref{sec:discussion}. 

\begin{table}
Global parameters for the runs presented in this paper
\centering                          
\begin{tabular}{l c c c c c c c}        
\hline\hline                 
Label & dim & $r_0$ & $\beta/5$ & $k$ & $\Sigma_0$ &Sect. \\
\hline                        
   R1-P4 & 2 & 1 & $10^{4}$ & $6$ & $500$ & \ref{sec:disccoex}\\
   R1-P4-C4 & 2 & 1 & $10^{4}$ & $4$ & $500$ & \ref{sec:nonejec}\\
   R1-P3 & 2 & 1 & $10^{3}$ & $6$ & $500$ & \ref{sec:hallpol}, \ref{sec:zfoverview}\\
   R1-P2 & 2 & 1 & $10^{2}$ & $6$ & $500$ & \ref{sec:thormech}, \ref{sec:zfsatur}\\
   R1-M3 & 2 & 1 & $-10^{3}$ & $6$ & $500$ & \ref{sec:selforgathor}\\
\hline
   R10-P3 & 2 & 10 & $10^{3}$ & $6$ & $500$ & \ref{sec:hallpol}\\
   R10-P2 & 2 & 10 & $10^{2}$ & $6$ & $500$ & \ref{sec:zf}\\
   R10-M3 & 2 & 10 & $-10^{3}$ & $6$ & $500$ & \ref{sec:ejec}\\
   R10-M3-C2 & 2 & 10 & $-10^{3}$ & $2$ & $500$ & \ref{sec:fiducial} (fiducial) \\
\hline
   3D-R1-P4 & 3 & 1 & $10^{4}$ & $6$ & $500$ & \ref{sec:hallpol}\\
   3D-R1-P3 & 3 & 1 & $10^{3}$ & $6$ & $500$ & \ref{sec:zfoverview}\\
   3D-R1-M4 & 3 & 1 & $-10^{4}$ & $6$ & $500$ & \ref{sec:hallpol}\\
   3D-R10-P3 & 3 & 10 & $10^{3}$ & $6$ & $500$ & \ref{sec:zf_3d}\\
\hline                                   
\end{tabular}
\vspace{2mm}
\caption{Name and dimension of the run, inner radius in astronomical units, initial plasma beta in the midplane (a minus sign corresponds to $\bm{\Omega \cdot B} < 0$), ratio of corona over disk sound speed $k$ (see Eq \eqref{eqn:csdiskoro}), mass surface density at $1\au$ in $\mathrm{g}.\mathrm{cm}^{-2}$, section where the run is discussed. }
\label{tab:setup}      
\end{table}

\begin{table*}
Results for the runs presented in this paper
\sisetup{separate-uncertainty=true,table-align-uncertainty=true}
\centering
\begin{tabular}{l S[table-format=2.2,table-figures-uncertainty=1] S[table-format=-2.2,table-figures-uncertainty=1] S[table-format=-2.1,table-figures-uncertainty=1] S[table-format=2.1,table-figures-uncertainty=1] S[table-format=2.1,table-figures-uncertainty=1] r r r}     
\hline\hline
Label & {$10^3 \times \alpha^{\mathcal{T}}_{r\varphi}$} & ${10^3 \times \alpha^{\mathcal{T}}_{z \varphi}}$ & ${10^7 \times \tau_{z}}$ & ${10^6 \times \dot{m}^-_w/m_d}$ & ${10^6 \times \dot{m}^+_w/m_d}$ & $\iota_{\varphi}$ & $\sigma_{\varphi}$ & $n_{\mathrm{zf}}$  \\
\hline                    
   R1-P4 & 3.94 \pm 0.08 & 0.10 \pm 0.2 & 2.0 \pm 2 & 1.4 \pm 0.2 & 1.5 \pm 0.3 & $6\%$ & $-55\%$ & $0$ \\
   R1-P4-C4 & 3.80 \pm 0.10 & -1.10 \pm 0.4 & 5.0 \pm 2 & 0.7 \pm 0.4 & 0.8 \pm 0.3 & $3\%$ & $-81\%$ & $0$\\
   R1-P3 & 14.30 \pm 0.8 & -8.00 \pm 1 & 23.0 \pm 2 & 2.8 \pm 0.5 & 3.0 \pm 0.6 & $-13\%$ & $-51\%$ & $7$\\
   R1-P2 & 75.00 \pm 0.8 & 27.40 \pm 0.6 & -22.0 \pm 2 & 22.0 \pm 8 & 4.0 \pm 2 & $97\%$ & $0\%$ & $5$ \\
   R1-M3 & 6.50 \pm 0.5 & 2.00 \pm 1 & -4.0 \pm 5 & 10.0 \pm 2 & 3.1 \pm 0.7 & $-99\%$ & $6\%$ & $0$ \\
\hline
   R10-P3 & 24.00 \pm 2 & 8.00 \pm 2 & -18.0 \pm 4 & 40.0 \pm 8 & 21.0 \pm 4 & $96\%$ & $6\%$ & $9$ \\
   R10-P2 & 131.00 \pm 6 & 47.00 \pm 6 & -88.0 \pm 35 & 117.0 \pm 44 & 24.0 \pm 9 & $91\%$ & $5\%$ & $9$ \\
   R10-M3 & 10.50 \pm 0.8 & 8.90 \pm 0.7 & -24.0 \pm 3 & 14.0 \pm 3 & 13.0 \pm 3 & $-7\%$ & $12\%$ & $0$ \\
   R10-M3-C2 & 23.00 \pm 2 & 17.00 \pm 2 & -43.0 \pm 11 & 18.0 \pm 4 & 19.0 \pm 4 & $12\%$ & $73\%$ & $0$ \\
\hline
   3D-R1-P4 & 4.00 \pm 0.05 & 0.74 \pm 0.08 & -2.8 \pm 0.4 & 1.2 \pm 0.2 & 1.2 \pm 0.2 & $4\%$ & $-45\%$ & $0$\\
   3D-R1-P3 & 18.90 \pm 0.3 & -2.40 \pm 0.3 & 18.7 \pm 0.2 & 2.4 \pm 0.8 & 2.4 \pm 0.8 & $0\%$ & $-59\%$ & $2$\\
   3D-R1-M4 & 1.09 \pm 0.04 & 0.32 \pm 0.06 & 0.0 \pm 0.1 & 1.3 \pm 0.2 & 1.3 \pm 0.2 & $-7\%$ & $20\%$ & $0$ \\
   3D-R10-P3 & 16.00 \pm 1 & 8.10 \pm 0.5 & -16.0 \pm 5 & 12.0 \pm 2 & 11.0 \pm 2 & $0\%$ & $-24\%$ & $2$ \\
\hline	
\end{tabular}
\vspace{2mm}
\caption{Label of the run; normalized horizontal and vertical stresses, $\alpha^{\mathcal{T}}_{r\varphi}$ and $\alpha^{\mathcal{T}}_{z\varphi}$ as defined by Eq. \eqref{eqn:alphadef}, volume-averaged from $3r_0$ to $8r_0$; accretion rate due to the vertical stress only $\tau_{z}$, as defined in Eq. \eqref{eqn:accretor}; wind mass loss rates in the southern and northern corona, $\dot{m}^{-}_w$ and $\dot{m}^{+}_w$, defined by Eq. \eqref{eqn:mdotwind}, normalized to the disk mass; symmetry coefficients $\iota_{\varphi}$ and $\sigma_{\varphi}$, as defined by Eqs. \eqref{eqn:symphi} and \eqref{eqn:modphi}; number of zonal flows identified at the end of the simulation $n_{\mathrm{zf}}$. Except for $\id{n}{zf}$, all quantities are time-averaged, and the uncertainties are standard deviations over time. }
\label{table:results}
\end{table*}

\subsection{Fiducial case} \label{sec:fiducial}

\subsubsection{Overview of the disk dynamics} \label{sec:fid_overview}

\begin{figure}[ht]
\centering
\includegraphics[width=\hsize]{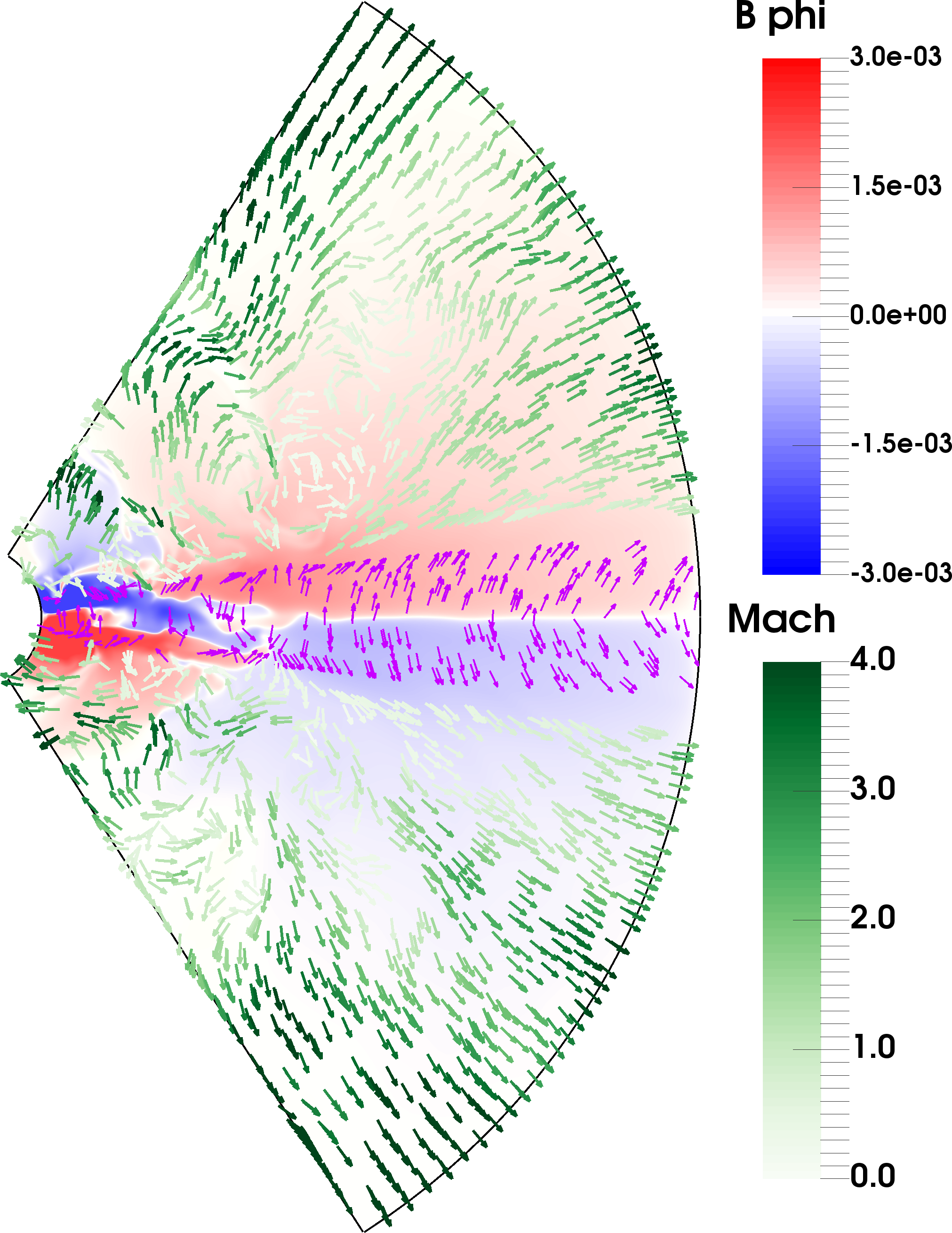}
\caption{Snapshot of run R10-M3-C2 (fiducial) at $t=200 T_0$, showing the toroidal magnetic field in color background (blue to red), the poloidal velocity field in units of local sound speed (green arrows in the corona), and the orientation of the angular momentum flux caused by magnetic stress (purple arrows in the disk); $\brac{B_z}<0$ in this case. }
\label{fig:snap_bp1}
\end{figure}

We show in Fig. \ref{fig:snap_bp1} the instantaneous state of run R10-M3-C2 after $200$ inner orbits of integration time ($6$ outer orbits). The toroidal magnetic field $B_{\varphi}$ changes sign at the midplane, and decreases smoothly with height in the corona. It also changes sign with radius near $r \approx 2.5 r_0$. This change of sign occurs near the inner radial boundary, where the magnetic field follows an inward-pointing wind. This inward orientation of the wind is likely influenced by our outflow polar boundary conditions. 

To this magnetic field is associated a flux of angular momentum $\mathcal{M} \cdot \bm{e}_{\varphi}=-B_\varphi \bm{B}$ (purple arrows in the disk of Fig. \ref{fig:snap_bp1}). In the outer half of the disk, this flux is oriented from the midplane toward the corona. The velocity field is organized, with magnitudes reaching four times the local sound speed. In the inner half of the disk, the flux of angular momentum is directed from the disk surface toward its midplane. In this half, the corona is turbulent, with no signature of organized ejection. 

The length scale of magnetic field variations is comparable to the disk extent, both in the vertical and radial direction. Its characteristic dynamical time is counted in hundreds of local orbits, while the fluid may be streaming at near sonic velocities. The evolution of the magnetic field in the disk may therefore be considered as secular with respect to accretion and ejection. Given a time window, we can decompose the magnetic field in a time and azimuthally averaged component $\brac{\bm{B}}_{\varphi,t}$ plus a fluctuating component $\bm{\tilde{B}}$. Under this decomposition, the average magnetic stress is split into two terms:
\begin{equation} \label{eqn:lamimaxy}
\brac{\mathcal{M}}_{\varphi,t} = -\brac{\bm{B}}\bm{\otimes}\brac{\bm{B}} - \brac{\bm{\tilde{B}\otimes\tilde{B}}} \equiv \mathcal{M}_{\mathrm{laminar}} + \mathcal{M}_{\mathrm{fluctuating}},
\end{equation}
representing a laminar plus a fluctuating contribution to $\mathcal{M}$. We compute the ratio of laminar to total magnetic stress over $30T_0$ ($\approx$ one outer orbit), average it over the poloidal surface of the disk, and find that it exceeds $\mathcal{M}_{\mathrm{laminar}}/\brac{\mathcal{M}} > 90\%$ for every run. Therefore, the magnetic stress in the disk is not due to a turbulent component, but is rather a quasi-steady (laminar) stress. The variability near the inner radius is likely caused by the outflow polar boundary conditions.

\subsubsection{Disk vertical structure} \label{sec:fid_verti}

\begin{figure}[ht]
\centering
\includegraphics[width=\hsize]{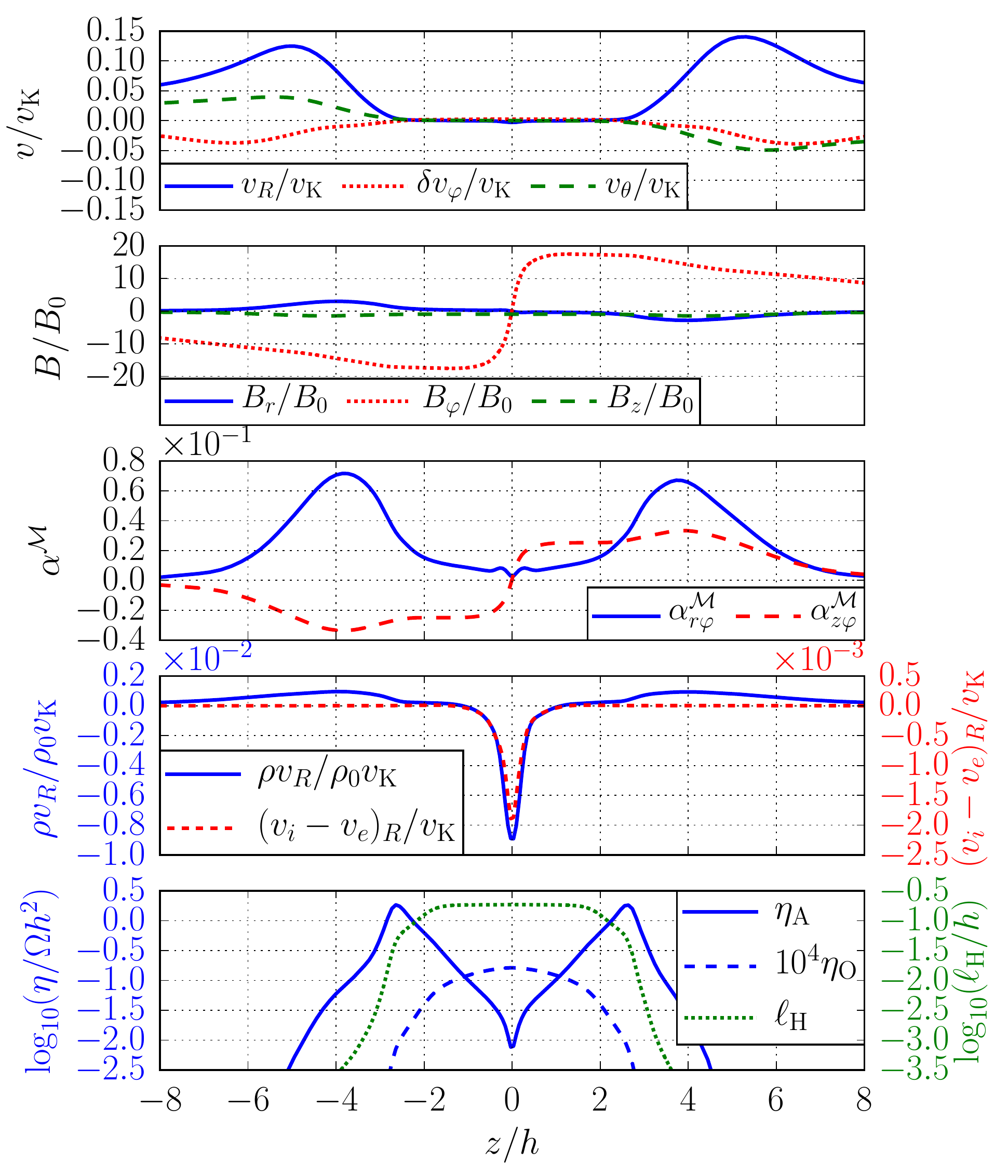}
\caption{Vertical profiles in run R10-M3-C2 (fiducial), averaged in time from $200T_0$ to $600T_0$, and in spherical radius from $5r_0$ to $8r_0$. \emph{First panel}: fluid velocity, normalized by the local Keplerian value $v_{\mathrm{K}}$; the disk mean value has been subtracted from the azimuthal component. \emph{Second panel}: magnetic field, normalized by the vertically averaged value of the vertical field $B_z$. \emph{Third panel}: horizontal (solid blue) and vertical (dashed red) magnetic stresses $\mathcal{M}_{r\varphi}$ and $\mathcal{M}_{z\varphi}$, normalized by the vertically averaged pressure (cf. Eq. \eqref{eqn:alphadef}). \emph{Fourth panel}: radial mass flux (solid blue), and ion minus electron radial velocity, normalized by the local Keplerian velocity and the vertically averaged density. \emph{Fifth panel}: ambipolar diffusivity $\id{\eta}{A}$ (solid blue), ohmic diffusivity $\id{\eta}{O}$ (dashed blue, increased by a factor $10^4$ for visibility), and Hall length $\lH$ (green dots). }
\label{fig:fid_vb}
\end{figure}

We show in Fig. \ref{fig:fid_vb} a series of vertical profiles in the outer region of the fiducial run R10-M3-C2. In the top panel, we see that the deviations from Keplerian velocity become significant above $z \gtrsim 2h$. The radial velocity is the dominant component, with $v_R / v_{\mathrm{K}} \approx 10\%$, corresponding to $v_R / c_s \approx 1$ at $z = 5h$. The polar velocity $v_{\theta}$ is second in magnitude, reaching $v_{\theta}/\id{v}{K} \approx 5\%$. The azimuthal velocity decreases by $5\%$ in the corona, compared to the disk midplane value at the same radius. This last trend could be due to the outflow itself, or to the thermal pressure support against gravity in the corona. %

The second panel shows that the horizontal magnetic field has an odd symmetry about the midplane. The initial vertical component is negative in this case, and it is the weakest component during the evolution of the simulation. The toroidal component is anti-correlated to the radial one, with a magnitude twenty to a hundred times higher in the disk. The location where $B_{\varphi} = B_r = 0$ corresponds to a thin \emph{current sheet}.

The third panel shows the horizontal and vertical magnetic stresses. The horizontal magnetic stress $\mathcal{M}_{r\varphi}$ is positive everywhere. It is maximal at the disk surface, where $\alpha^{\mathcal{M}}_{r\varphi} \approx 0.08$ is ten times higher than in the midplane. The vertical stress $\mathcal{M}_{z\varphi}$ changes sign at the midplane, transporting angular momentum downward in the southern hemisphere and upward in the northern one. The large-scale magnetic field thus extracts angular momentum from the disk, and transports it toward the corona. 

In the fourth panel of Fig. \ref{fig:fid_vb}, the radial mass flux displays a negative peak at the current sheet (i.e., the midplane is accreting), and becomes positive at the base of the wind $z \gtrsim 2h$. The accretion stream is very narrow, and accurately fits the profile of radial electric current. We show instead the relative ion-electron radial velocity, $\bm{v}_i - \bm{v}_e \equiv \bm{J} / n_e$, which drives the Hall drift. Because this disk sees a net vertical magnetic field $B_z<0$, the Lorentz force $F_{\varphi} \simeq -J_r B_z < 0$ is extremal at the current sheet, thereby slowing matter and causing it to fall inward. 

The profile of radial electric current $\brac{J_r}_{\varphi} = -\partial_z \brac{B_{\varphi}}$ can be linked to the ambipolar diffusivity $\id{\eta}{A}$, drawn in the bottom panel of Fig. \ref{fig:fid_vb}. The magnetic energy dissipation rate due to ambipolar diffusion is $\int \id{\eta}{A} J_{\perp}^2$, where $\bm{J_{\perp}}$ is the electric current perpendicular to the local magnetic field. To minimize energy dissipation, this current must flow along the circuit with the smallest resistivity. Low density and strong field regions are poorly conducting due to ambipolar diffusion (cf. Eq. \eqref{eqn:defetaA}). This enforces the current layer to be localized near the midplane and near the current sheet, where $\id{\eta}{A}$ takes its smallest value. Ohmic resistivity is unimportant in this region, making ambipolar diffusion the dominant dissipation process. The presence of dust grains should not alter this ordering in the outer regions of protoplanetary disks \citep[see Fig. 1 of][]{SLKA15} . With the present values of $\lH$ and $B_0$, the region below $z \lesssim 2h$ is stable with respect to the Hall-Shear instability \citep[HSI,][]{BT01,K08}.

\subsubsection{Mass and momentum fluxes} \label{sec:fid_massmom}

\begin{figure}[ht]
\centering
\includegraphics[width=\hsize]{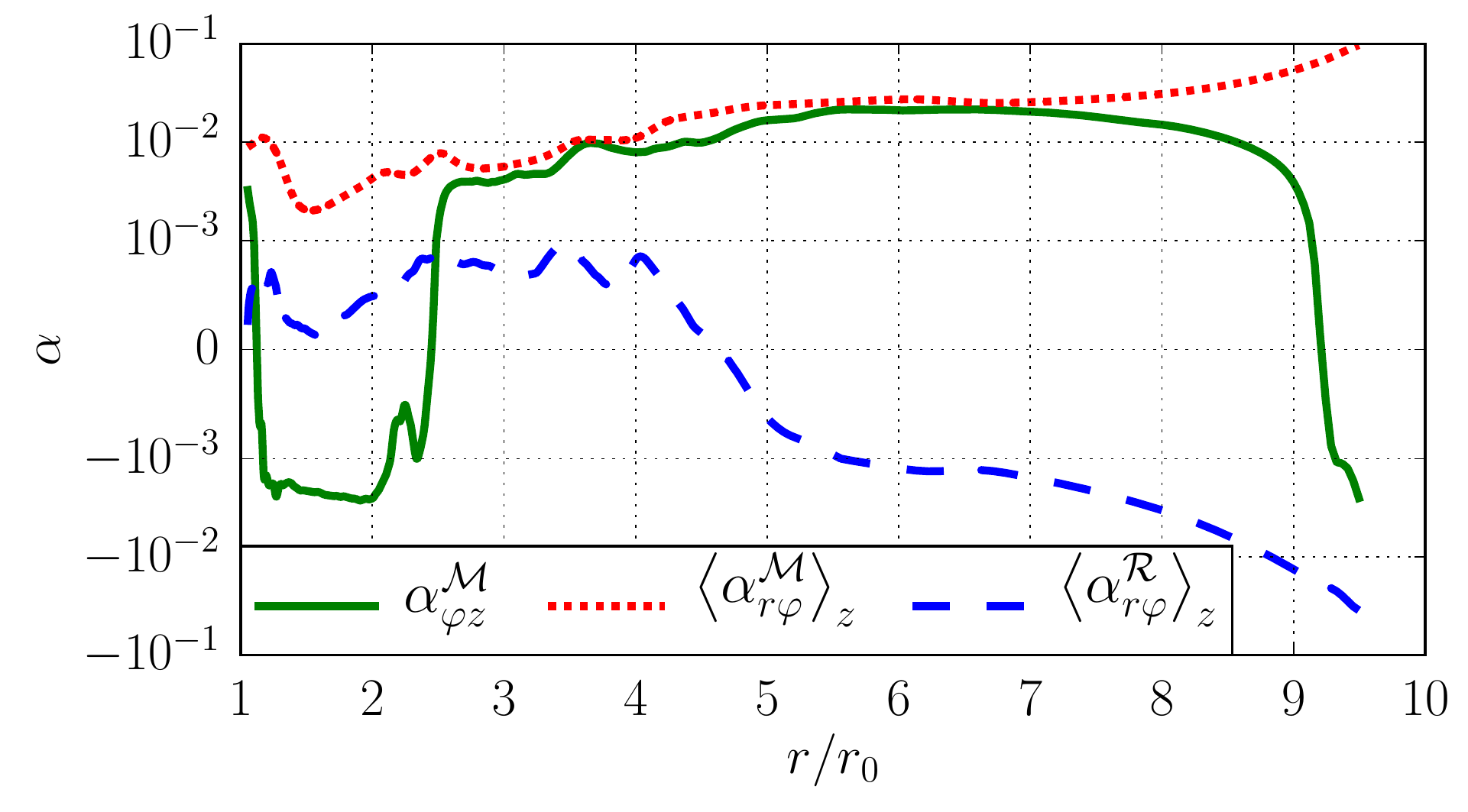}
\caption{Radial profiles of normalized stress in R10-M3-C2 (fiducial), averaged in time between $200T_0$ and $600T_0$. The radial components of Maxwell (red dots) and Reynolds (dashed blue) are averaged over the disk height, whereas the vertical Maxwell stress (solid green) is measured at the disk surface $z=\pm H$. }
\label{fig:fid_alpha}
\end{figure}

We draw in Fig. \ref{fig:fid_alpha} the radial profiles of stress in our fiducial run R10-M3-C2. The horizontal Maxwell stress $\mathcal{M}_{r\varphi}$ is positive at all radii, transporting angular momentum radially outward, with $\alpha^{\mathcal{M}}_{r\varphi}$ increasing from $10^{-2}$ to $10^{-1}$ from $10\,\au$ to $100\,\au$. The vertical component $\mathcal{M}_{z\varphi}$ is positive beyond $2.5r_0$, transporting angular momentum from the disk to the corona. The fact that $\mathcal{M}_{\varphi z}<0$ in the inner region is possibly related to our polar outflow boundary conditions. The Reynolds stress $\alpha^{\mathcal{R}}_{r\varphi}$ is negative for $r \gtrsim 4.6 r_0$. This is because above $z > 1h$, the velocity fluctuations $\bm{\tilde{v}} \equiv \bm{v} - \small\langle \bm{v} \small\rangle_{\rho}$ display $\tilde{v}_{r} > 0$ and $\tilde{v}_{\varphi} < 0$. Because $\vert \mathcal{M}_{r\varphi}\vert > 10 \,\vert\mathcal{R}_{r\varphi}\vert$, we will focus on the Maxwell stress for transport-related processes.

\begin{figure}[ht]
\centering
\includegraphics[width=\hsize]{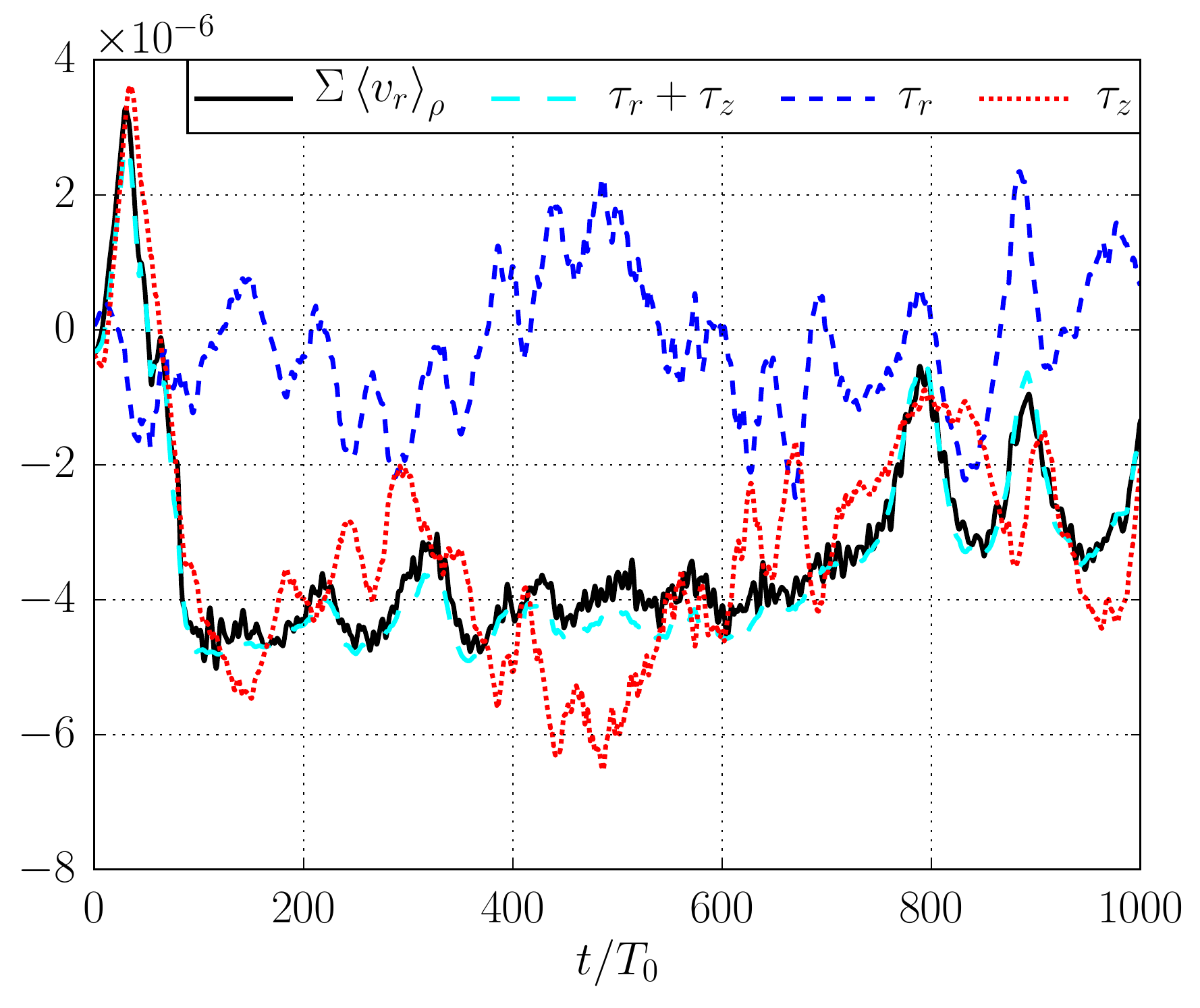}
\caption{Accretion in the disk region of run R10-M3-C2 (fiducial), between $r \in \left[ 3, 8 \right] r_0$; the actual mass flux (solid black) is decomposed into radial (dashed blue) and vertical (red dots) torques as explicited in Eq. \eqref{eqn:accretor}; the sum is drawn (dashed cyan) to validate this decomposition. }
\label{fig:fid_accretor}
\end{figure}

The absence of correlation $\brac{\mathcal{R}} \not\propto \brac{\mathcal{M}}$, so as the fact that $\brac{\mathcal{R}}_{\varphi,z} < 0$, are discrepant with a state of MRI turbulence \citep{PCP06}. It was mentioned in Sect. \ref{sec:fid_overview} that the flow cannot be considered as turbulent, and that there is no clear spatial scale separation between the magnetic structures and the entire disk. For these reasons, the $\alpha$ viscosity prescription should not be applied in this context \citep{BP99}. However, conservation of angular momentum still applies and we can use Eq. \eqref{eqn:accretor} to compute the accretion rate separately caused by radial and vertical angular momentum fluxes; this decomposition is represented in Fig. \ref{fig:fid_accretor}. We find that the vertical transport of angular momentum $\tau_z$ is the main driver of accretion. Conversely, the radial contribution $\tau_r$ oscillates about zero and causes no accretion on average. We recall that it does not contradict the presence of a significant radial stress (cf. Fig. \ref{fig:fid_alpha}) since accretion is controlled by the \emph{stress divergence}. Finally, we find the average mass flux in this box is $\Sigma \brac{v_r}_{\rho} \approx - 4 \times 10^{-6}$ in code units, corresponding to an average accretion rate $\dot{m}_r \equiv \vert 2\pi r \Sigma \brac{v_r}_{\rho} \vert \approx 1.1 \times 10^{-7} \,\mathrm{M}_{\odot}.\,\mathrm{yr}^{-1}$ at $50\,\au$ around a solar-mass star. 

We find that $\tau_r \approx 0$ in all our simulations, meaning that the radial flux of angular momentum is divergence-free. This is a fortuitous consequence of our initial $\Sigma(r)$, $\beta(r)$, etc.. We will therefore focus on the vertical transport of angular momentum. 

\subsubsection{Cold, magnetized wind} \label{sec:fid_ejec}

We have shown that a wind is launched by the outer half of the disk, where the Maxwell stress transports angular momentum from the disk to the corona. We qualify these winds as `cold' since the ratio of sound speeds between the disk and the corona is set to $k=2$ (which corresponds to a factor 4 in temperature). 

To characterize these winds, we first look at the dynamics of a fluid element. We average every quantity in azimuth and time, and compute the characteristic velocities along a streamline in the poloidal plane. We recall the definition of the slow (minus sign) and fast (plus sign) magnetosonic waves velocity:
\begin{equation}
v_{\pm}^2 \equiv \frac{1}{2}\left(v_{\mathrm{A}}^2 + c_s^2 \pm \sqrt{(v_{\mathrm{A}}^2 + c_s^2)^2 - 4 c_s^2 v_{\mathrm{A}p}^2}\right), 
\end{equation}
where the index $p$ stands for the poloidal component of a vector. These velocities are relevant only in the ideal MHD regime. 

\begin{figure}[ht]
\centering
\includegraphics[width=\hsize]{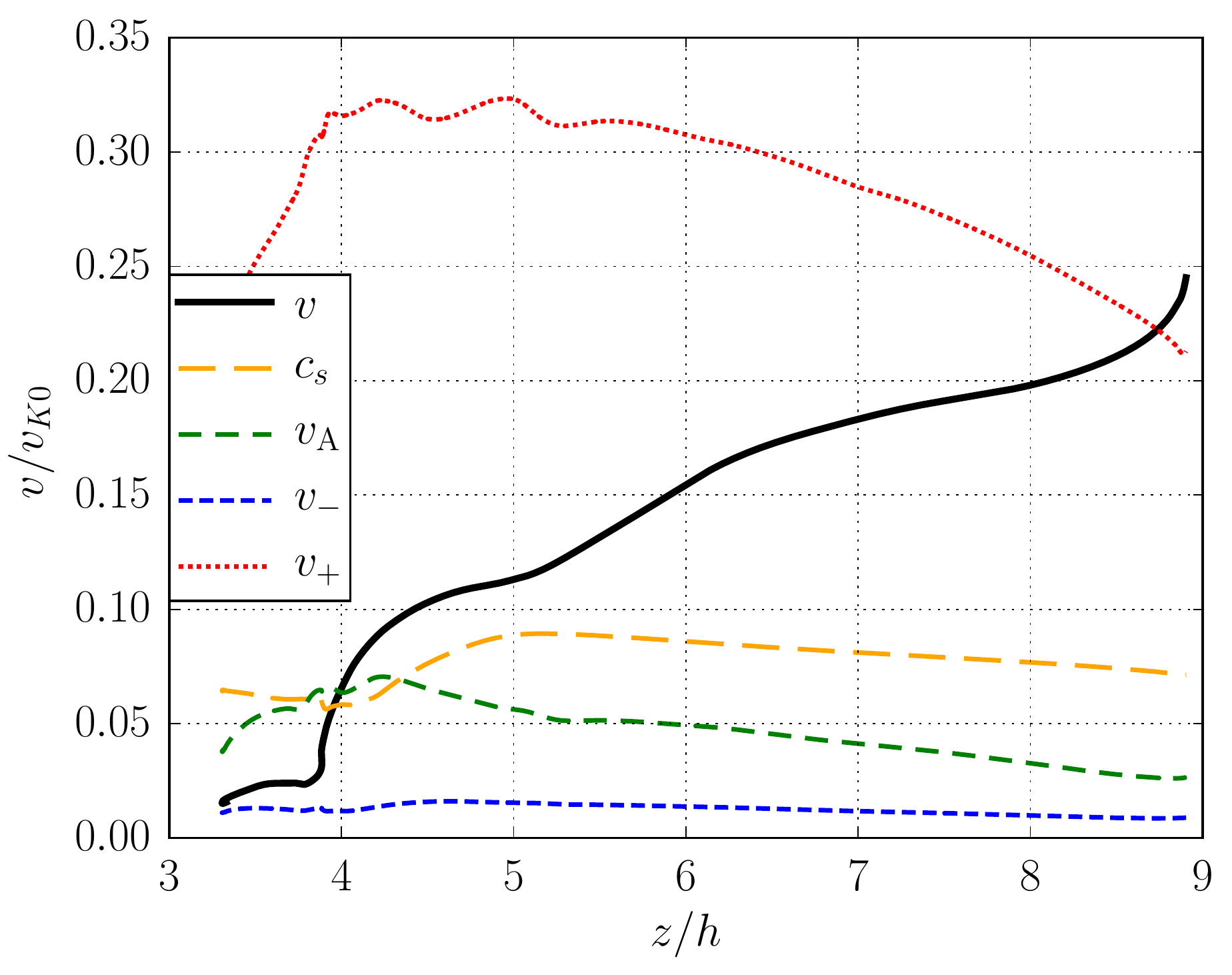}
\caption{Velocities projected on a streamline passing through $(r=6r_0,z=5h)$ in run R10-M3-C2 (fiducial), averaged from $400T_0$ to $500T_0$, and normalized by the Keplerian velocity at the launching radius $v_{\mathrm{K}0}$; flow velocity $v$ (solid black), sound speed $c_s$ (dashed orange), Alfv\'en velocity $\id{v}{A}$ (dashed green), slow magnetosonic speed $v_{-}$ (dashed blue) and fast magnetosonic speed $v_{+}$ (red dots). }
\label{fig:fid_vp}
\end{figure}

As apparent in Fig. \ref{fig:fid_vp}, the fluid poloidal velocity monotonically increases, and crosses all characteristic MHD velocities. 
Surprisingly, the fast-magnetosonic point is located right before the domain boundary. The same is true in several, but not all runs (see for example Fig. \ref{fig:wiwi_vp}). This fact was already observed in stratified, shearing-box simulations \citep{FLLO13}. It indicates that our boundary conditions are still, somehow, constraining the flow structure down to the disk in this run. 

The mass transported by the wind is computed via Eq. \eqref{eqn:mdotwind}. We estimate its average value $\dot{m}_W \approx 2.6 \times 10^{-4}$ in code units, corresponding to approximately $2.3 \times 10^{-7} \,\mathrm{M}_{\odot}.\,\mathrm{yr}^{-1}$ for this run.

\begin{figure}[ht]
\centering
\includegraphics[width=\hsize]{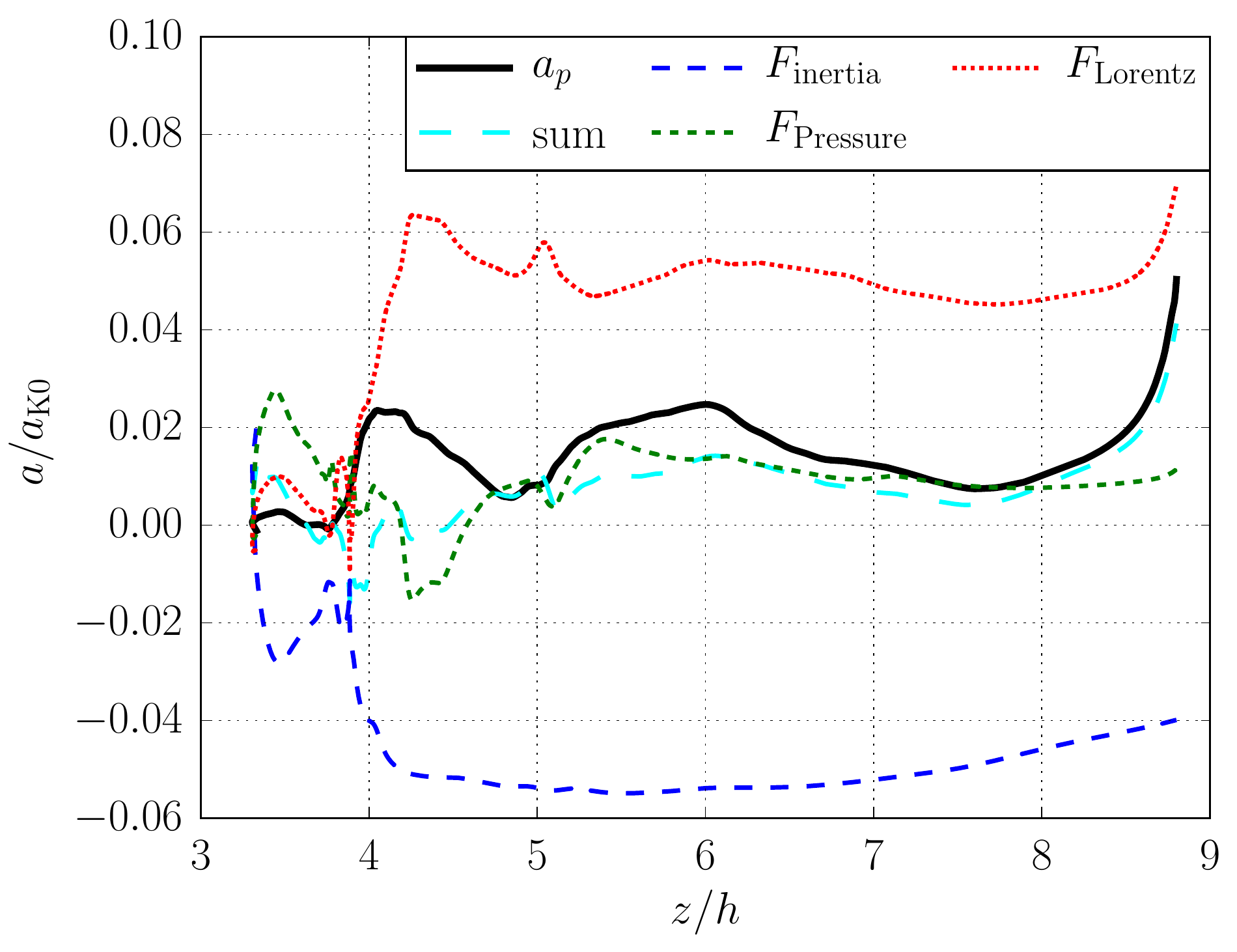}
\caption{Acceleration along a streamline passing through $(r=6r_0,z=5h)$ in run R10-M3-C2 (fiducial), averaged from $400T_0$ to $500T_0$, normalized by the Keplerian acceleration $a_{\mathrm{K}0}$ at the streamline base. The sum of the different forces (dashed cyan) is shown to validate this decomposition. }
\label{fig:fid_windacc}
\end{figure}

We compute separately the acceleration $a_p \equiv v_p \partial_p v_p$, and the acceleration caused by the forces $F$ along the streamline. These include the thermal pressure gradient, the Lorentz force, and the inertial force due to gravitational and centrifugal accelerations:
\begin{equation}
\id{\bm{F}}{inertia} \equiv - \rho \left(\nabla \Phi + \bm{ v \cdot} \nabla \bm{v} - v_p \partial_p \bm{v_p}\right). 
\end{equation}
We present the resulting accelerations $F/\rho$ in Fig. \ref{fig:fid_windacc}, normalized by the Keplerian value $a_{\mathrm{K}} \equiv \id{v}{K}^2/r$ at the streamline base. The sum of the forces decently reproduces the true acceleration in spite of the variability of the flow. The thermal pressure gradient helps accelerating the flow, but it is significantly weak than the Lorentz force. The latter barely compensates the inertial term, resulting in an overall small wind acceleration $a_p / a_{\mathrm{K}0} \lesssim 2\%$. 

\begin{figure}[ht]
\centering
\includegraphics[width=\hsize]{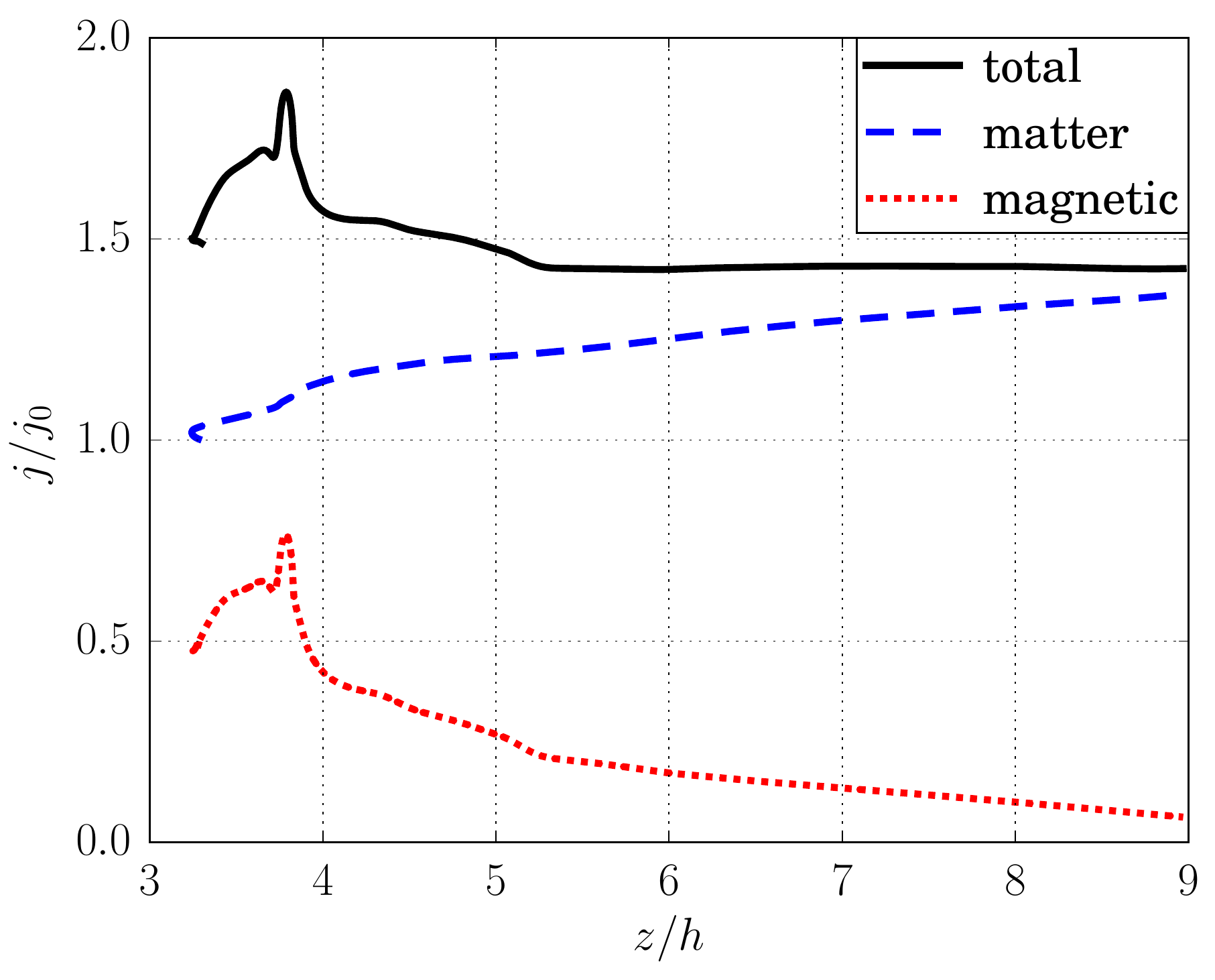}
\caption{Angular momentum along a streamline passing through $(r=6r_0,z=5h)$ in run R10-M3-C2 (fiducial), averaged from $400T_0$ to $500T_0$, normalized to its value at the streamline base; the total angular momentum (solid black) is decomposed into matter (dashed blue) and magnetic (red dots) components as explicited in Eq. \eqref{eqn:split_angm}. }
\label{fig:fid_angm}
\end{figure}

We split the specific angular momentum of a fluid element into its matter and magnetic contributions:
\begin{equation}
j \equiv \underbrace{r v_{\varphi}}_\text{matter} - \underbrace{r B_{\varphi}/\kappa}_\text{magnetic}, \label{eqn:split_angm}
\end{equation}
where $\kappa \equiv \rho v_p / B_p$ is the ratio of mass over magnetic flux along the streamline. Both $j$ and $\kappa$ should be invariant along streamlines for stationary, axisymmetric, ideal MHD flows \citep{Chan56,PP92}. We show in Fig. \ref{fig:fid_angm} that the total specific angular momentum $j$ is approximately constant above $z \gtrsim 5h$. The magnetic torque transfers angular momentum back to the outflowing plasma, leading to an increase of the kinetic content up to the domain boundary. This demonstrates that the wind is magnetocentrifugally accelerated. Below $5h$, the variations of $j$ indicate that the average flow does not behave as an ideally conducting plasma. This is caused by ambipolar diffusion above the disk surface. It should be noted that the outflow is already super-Alfv\'enic when entering the ideal MHD regime $z \gtrsim 5h$ (cf. Fig. \ref{fig:fid_vp}). As a consequence, only the sonic and fast-magnetosonic critical points can impose regularity conditions on the outflow \citep{FP95}. 

The magnetic contribution to $j$ is only $0.4j_0$, corresponding to a magnetic lever arm $\lambda \approx 1.4$, much smaller than expected from strongly magnetized jets \citep[see Fig. 4 of][]{Ferreira97}. For a magnetic lever arm smaller than $3/2$, the Bernoulli invariant of a cold flow should be negative \citep{CF00B}; this is indeed the case in this simulation. It follows that the outflow cannot escape the gravitational field on its own energy content when it reaches the domain boundary. The outflow can have $\mathcal{B} < 0$, and still be both stationary and super fast-magnetosonic. Effectively, $\mathcal{B}<0$ corresponds to an outflow which is gravitationally bound when $z\rightarrow\infty$, but it is not necessarily bound for the gravitational potential restricted to the computation box.

\subsection{Hall and magnetic field polarity} \label{sec:hallpol}

It is known that the Hall drift discriminates between the two polarities of the vertical magnetic field \citep{BT01}. When $\bm{\Omega\cdot B}>0$, the Hall drift acts to enhance angular momentum transport; in the opposite case $\bm{\Omega\cdot B}<0$, the Hall drift tends to stabilize the flow, and greatly reduce angular momentum transport \citep{LKF14}. 

\begin{figure}[ht]
\centering
\includegraphics[width=\hsize]{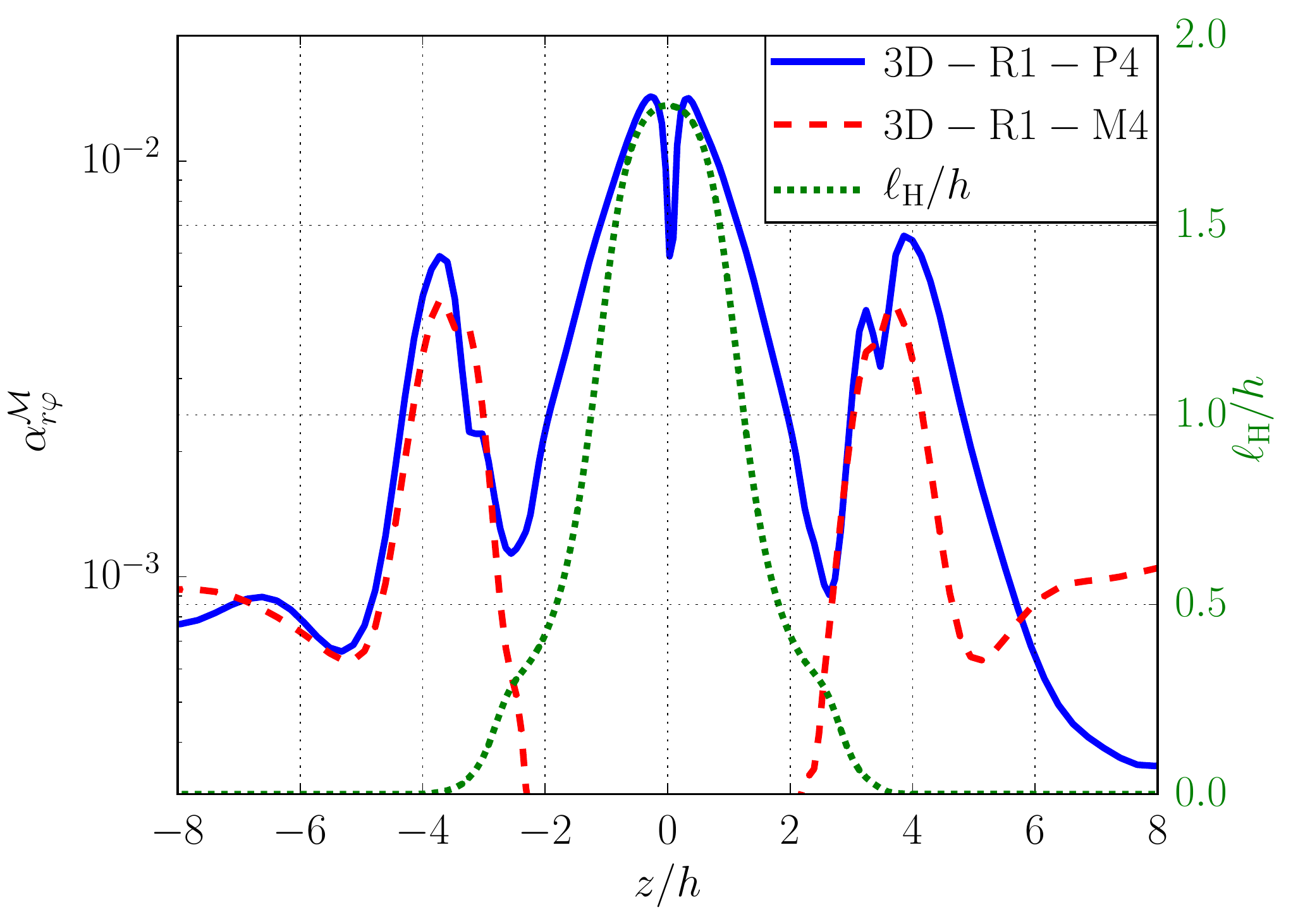}
\caption{Vertical profiles of horizontal magnetic stress, normalized by the vertically averaged pressure in the disk $\mathcal{M}_{r\varphi}/\brac{P}_{z}$ in runs 3D-R1-P4 (solid blue, $\bm{\Omega\cdot B}>0$) and 3D-R1-M4 (dashed red, $\bm{\Omega\cdot B}<0$), averaged in time over $50T_0$ and radially from $3r_0$ to $6r_0$; the green dots show the average ratio $\lH/h$ in this same region. }
\label{fig:hall_maxwel}
\end{figure}

We draw in Fig. \ref{fig:hall_maxwel} the vertical profiles of horizontal stress $\mathcal{M}_{r\varphi}$ for the two orientations of the initial magnetic field. When $B_z>0$ (run 3D-R1-P4), there are three regions of important magnetic stress: the midplane and the two disk-corona interfaces. The midplane stress reaches $\alpha^{\mathcal{M}}_{r\varphi} \approx 10^{-2}$ and decreases exponentially fast with height. Near $z\approx 2.7h$, the surface stress becomes dominant, and increases up to $z \approx H$, where it reaches $\alpha^{\mathcal{M}}_{r\varphi} \approx 5\times 10^{-3}$. This profile is similar to what was found in local shearing-box simulations (see Fig. 13 of \cite{LKF14}). In the opposite case $B_z<0$ (run 3D-R1-M4), the surface stress is essentially the same, but the midplane stress is now $\alpha^{\mathcal{M}}_{r\varphi} \approx 0$. 

The vertical distribution of $\lH$ is overlaid in Fig. \ref{fig:hall_maxwel}. It reaches $\lH/h \sim 1$ deep in the disk, and decreases rapidly with height. In run 3D-R1-M4, below $z \lesssim 2h$, the combination of a strong Hall length and a negative vertical magnetic field stabilizes the disk with respect to the HSI. At $z\gtrsim 3h$, the Hall drift becomes negligible regarding the linear stability of the flow, and we retrieve a polarity-independent system. 

The volume-averaged values of $\alpha^{\mathcal{T}}_{r\varphi}$ differ only by a factor two between run R1-P3 (R10-P3), and the equivalent run with a reversed field R1-M3 (resp. R10-M3, cf. Table \ref{table:results}). This estimated $\alpha^{\mathcal{M}}_{r\varphi}$ includes the disk up to $z \leq H$, so part of it is due to the polarity-independent surface stress. Also, the midplane is not magnetically dead when $B_z < 0$ and the initial magnetization $\beta \leq 5 \times10^{3}$ (see third panel of Fig. \ref{fig:fid_vb}). As the intensity of the net magnetic flux increases, so does the magnetic stress at the disk surface, and the ambipolar diffusivity. The magnetic field can then diffuse, from the surface to the midplane, over a few tens of local orbits (cf. Sect. \ref{sec:recap_magconf}). The surface over midplane stress contrast is smaller at larger radii, where the Hall drift is weaker (see fifth panel of Fig. \ref{fig:fid_vb}). 

We conclude that $\mathcal{M}_{r\varphi} \approx 0$ in the midplane only when the midplane is Hall-shear stable, and the intensity of the background field is sufficiently weak. For stronger magnetic fields, the surface-driven torque can reach the midplane \citep[see for example][]{WK93,SKW11}.

\subsection{Warm winds} \label{sec:ejec}

We described the launching and acceleration of a cold, magnetized wind in our reference simulation with $k=2$. In this section, we investigate the role of thermal pressure with a warmer corona. We consider a corona over disk temperature ratio of $36$, i.e., a ratio of $k=6$ in isothermal sound speed. 

\subsubsection{Overview}

\begin{figure}[ht]
\centering
\includegraphics[width=\hsize]{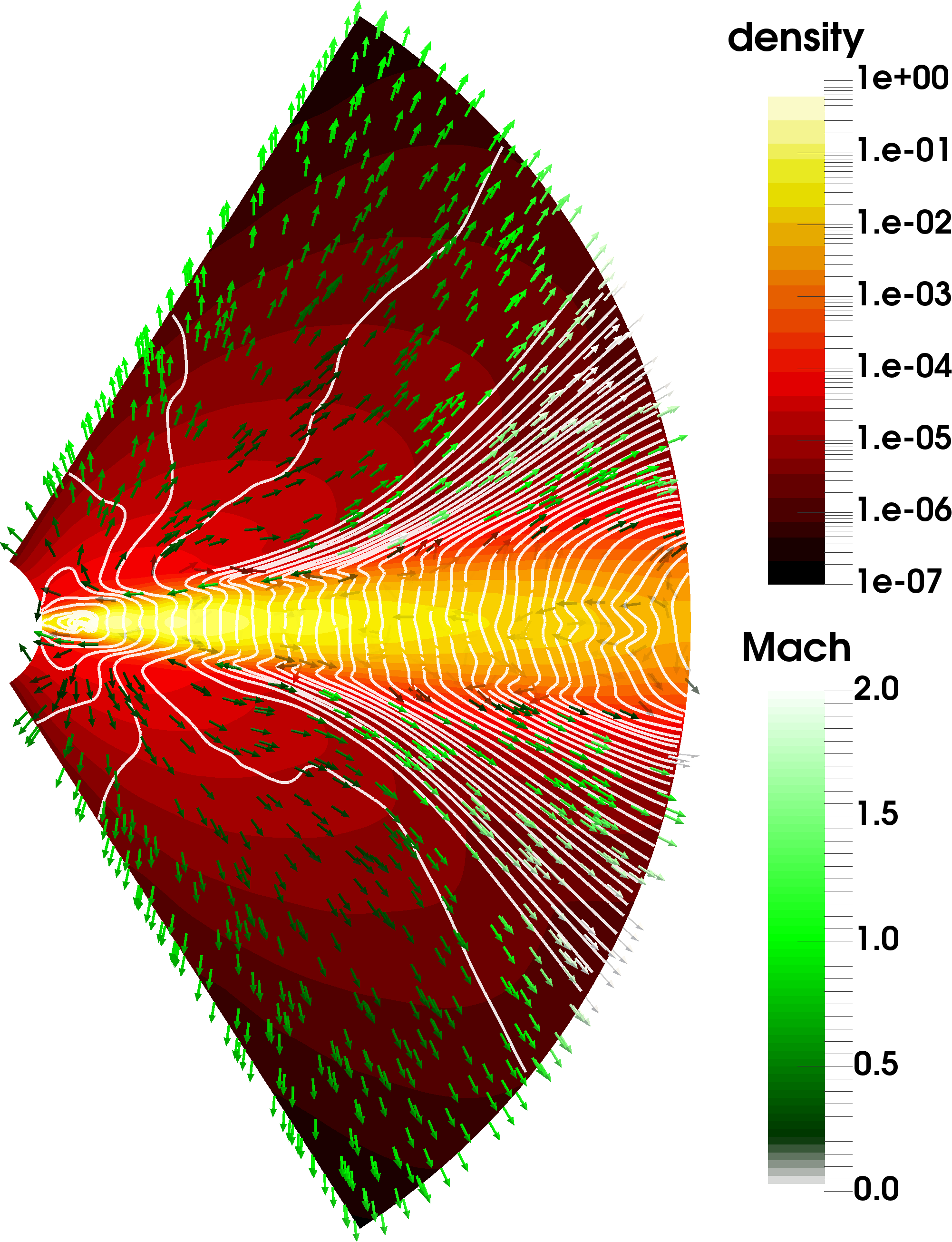}
\caption{Averaged flow poloidal map for run R10-M3 (from $10\au$ to $100\au$, with $B_z < 0$) from $400T_0$ to $700T_0$; magnetic field lines are regularly sampled along the midplane, and the velocity field is indicated with green arrows over the background density field. }
\label{fig:snap_wind}
\end{figure}

We show in Fig. \ref{fig:snap_wind} the average flow morphology in run R10-M3, initially threaded by a negative $B_z < 0$. The density in the disk has homogeneously decreased by $10\%$, maintaining its radial and vertical stratification profiles. Magnetic field lines are straight in the disk, as one can expect in a diffusion-dominated regime. In the inner regions of the corona, the field lines remain vertical or bend inward. Because the magnetic field is the weakest in these regions, it exerts no constraints on the fluid. The velocity field in the innermost region appears to be affected by our outflow polar boundary conditions. 

The initial magnetic field has been substantially reduced for radii as far as $3r_0$, where positive magnetic flux has entered the domain from the inner radial boundary. Although this phenomenon is artificially caused by our boundary conditions, it calls for a dedicated study of the Hall-driven transport of magnetic flux in protoplanetary disks \citep{BS16arxiv}. The outer corona displays a laminar structure. Field lines bend outward, with an inclination greater than $60^{\circ}$ at the disk surface, favorable to magnetocentrifugal acceleration. The poloidal velocity and magnetic field lines are aligned in this region, consistently with a quasi-steady and ideal MHD wind. The average wind mass loss rate is $\dot{m}_W \approx 2.7 \times 10^{-5}$ in code units. This corresponds to approximately $1.7 \times 10^{-7} \,\mathrm{M}_{\odot}.\,\mathrm{yr}^{-1}$ at $100\au$, comparable to our fiducial case despite the temperature difference. 

\begin{figure}[ht]
\centering
\includegraphics[width=\hsize]{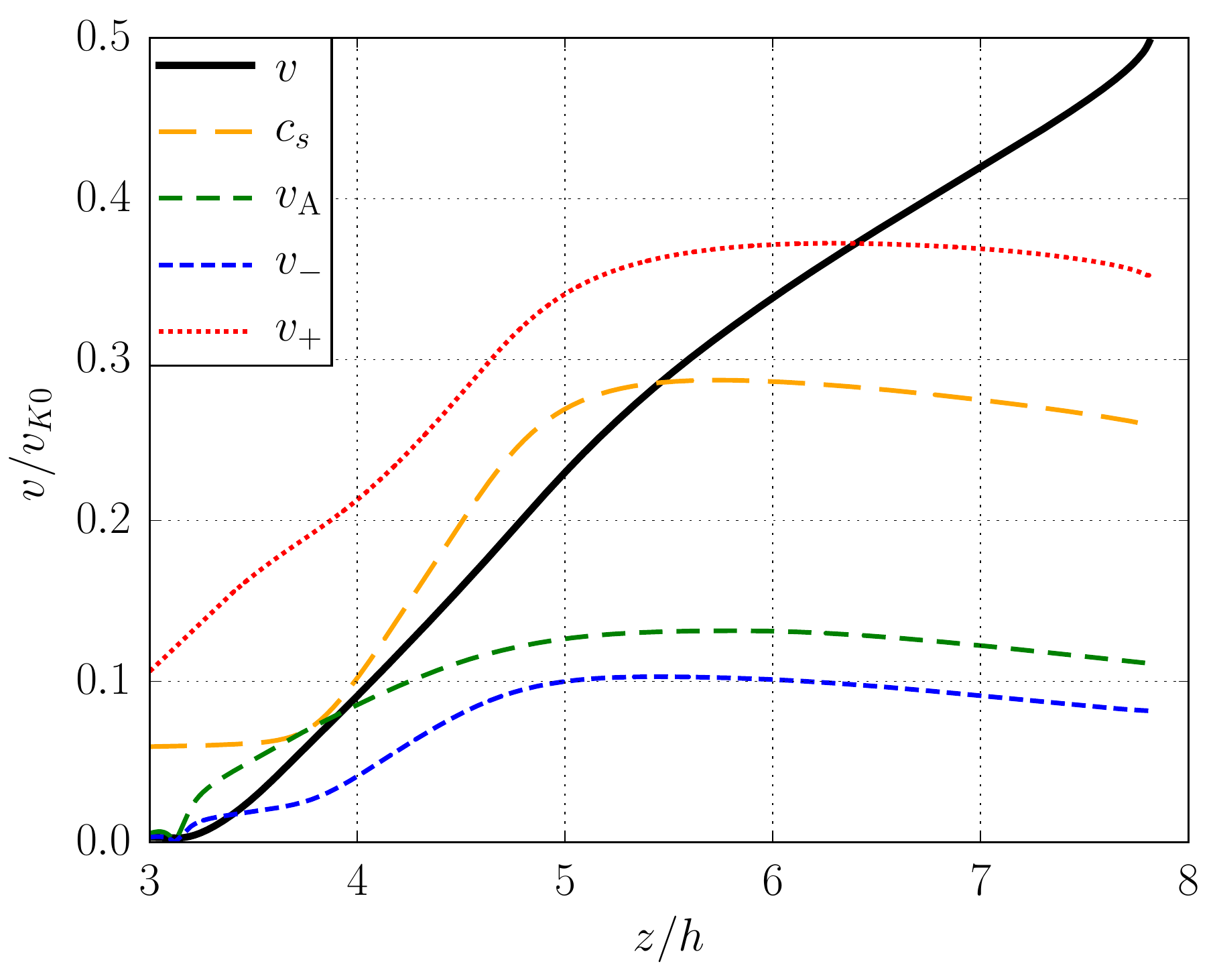}
\caption{Same as Fig. \ref{fig:fid_vp} for a streamline passing through $(r=7r_0,z=5h)$ in run R10-M3, averaged from $400T_0$ to $700T_0$. The fast-magnetosonic point is clearly crossed within the computational domain. }
\label{fig:wiwi_vp}
\end{figure}

We select a streamline in the poloidal plane, passing through $(r=7r_0,z=5h)$. The flow and the characteristic MHD velocities are projected on this streamline in Fig. \ref{fig:wiwi_vp}. The increase in sound speed, from $z=3.7h$ to $4.7h$, marks the disk-corona temperature transition. As for the fiducial, cold corona case (cf. Fig. \ref{fig:fid_vp}), the slow-magnetosonic and Alfv\'en critical points are crossed at $z\approx 4h$. The fast-magnetosonic critical point is now clearly within the computational domain. It is crossed by the outflow way before reaching the outer radial boundary, so the base of the wind should be causally disconnected from the boundary. 

The magnetic field again contributes to $0.4j_0$ in the specific angular momentum along this streamline. According to steady-state MHD jet theory, a small magnetic lever arm $\lambda \approx 1.4$ corresponds to an ejection efficiency $\xi \equiv \partial \log \dot{m}_W / \partial \log r \sim 1$ \citep{FP93,Ferreira97}. This is consistent with the high ratio of mass outflow over mass accretion rates observed in our simulation. In this case, there would be too much matter loaded into the wind to obtain a steady and trans-Alfv\'enic outflow without coronal heating \citep[cf. Eq. (40) of][]{Ferreira97}.

\subsubsection{Energy budget} \label{sec:streamflux}

We examine the energetics of the outflow by means of the Bernoulli invariant. Following \cite{SI09}, we decompose the ideal MHD Poynting flux as:
\begin{align}
\label{eqn:Poynting}
\begin{split}
\bm{\Pi} &\equiv \bm{E \times B} = (-\bm{v\times B}) \bm{\times B} \\
 &= (B_{\perp} \cdot B_{\perp}) \bm{v} - (v_{\perp} \cdot B_{\perp}) \bm{B} \equiv B^2_{\perp} \bm{v} + \bm{w}. 
\end{split}
\end{align}
The first term can be thought of as the advection of magnetic energy by the flow; the second term gives the wave-like transport of energy. Given a streamline, we can decompose the integral
\begin{align} \label{eqn:enthalpy}
\int \frac{\nabla P}{\rho} \bm{\cdot} d\bm{l} = \frac{\gamma}{\gamma - 1} \frac{P}{\rho} - \int T \,\nabla s \bm{\cdot} d\bm{l} \equiv \mathcal{H} - Q,
\end{align}
respectively an enthalpy contribution minus a heating term. The first corresponds to the adiabatic work exerted by the fluid, whereas the second measures its variation of specific entropy $s$. By dotting Eq \eqref{eqn:dyn-v} with $\bm{v}$, we can construct a quantity constant along stationary streamlines, the Bernoulli invariant: 
\begin{align} \label{eqn:bernoulli}
\begin{split}
\mathcal{B}' &\equiv \frac{v^2}{2} + \Phi + \frac{\Pi_p}{\rho v_p} + \int \frac{\nabla P}{\rho} \bm{\cdot} d\bm{l} \\
&= \frac{v^2_{\varphi}}{2} + \frac{v^2_{p}}{2} + \Phi + \frac{B^2_{\perp} v_p + w_p}{\rho v_p} + \mathcal{H} - Q. 
\end{split}
\end{align}
Because $Q$ keeps an integral form, the value of $\mathcal{B}'$ along a streamline depends on the choice of integration bounds. We choose to subtract from $\mathcal{B}'$ the heat input $\mathcal{Q}$ evaluated at the domain boundary: $\mathcal{B} \equiv \mathcal{B}' + Q_{\mathrm{end}}$. With this definition, a fluid element having $\mathcal{B}>0$ is able to escape the gravitational potential on its own energetic content, without additional heating past the domain boundary. 

\begin{figure}[ht]
\centering
\includegraphics[width=\hsize]{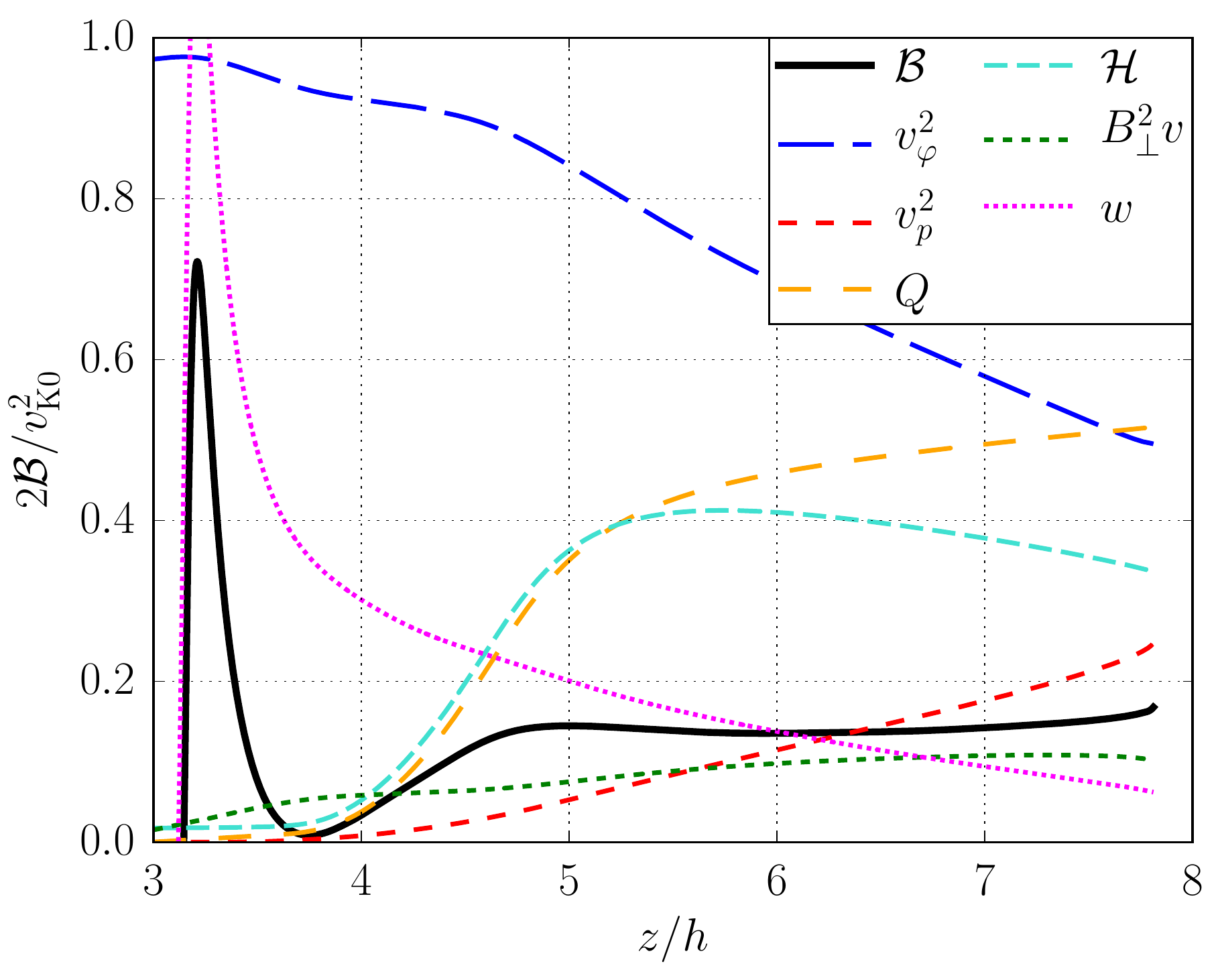}
\caption{Contributions to the Bernoulli invariant $\mathcal{B}$ (solid black)  as explicited in Eq. \eqref{eqn:bernoulli}, along a streamline passing through $(r=7r_0,z=5h)$ in run R10-M3, averaged from $400T_0$ to $700T_0$; the gravitational contribution is not shown. }
\label{fig:wiwi_bernie}
\end{figure}

We draw the various contributions to $\mathcal{B}$ along the streamline in Fig. \ref{fig:wiwi_bernie}. The poloidal component $v^2_p/2$ keeps increasing, as already apparent in Fig. \ref{fig:wiwi_vp}. The toroidal component $v^2_{\varphi}/2$ is initially the main energy reservoir against gravity, and slowly decreases. This is because the fluid specific angular momentum $r v_{\varphi}$ is approximately constant above $z \gtrsim 5h$, so $v_{\varphi}$ decreases as $1/r$ along the streamline. The wave-like Poynting flux $w_p \sim (v_{\mathrm{K}0}^2/2)$ is strong at the disk surface $z \approx 3.2h$, and most of it has been consumed at the end of the streamline. However, it is efficiently converted into kinetic energy only in the ideal-MHD region $z \gtrsim 5h$. The thermal contributions to the Bernoulli invariant, namely $\mathcal{H}$ and $\mathcal{Q}$, both increase at the disk-corona transition. Beyond $z \gtrsim 5.5h$, the enthalpy $\mathcal{H}$ decreases, while the heating $\mathcal{Q}$ keeps increasing. The decrease in enthalpy corresponds to the adiabatic cooling of the fluid, exerting a mechanical work and thus contributing to the outflow acceleration. The increase in $\mathcal{Q}$ means that the fluid keeps receiving heat, because it is always colder than the prescribed temperature at a given location. 

The final value of $\mathcal{B} \approx 1.6\times 10^{-2} \,(v_{\mathrm{K}0}^2/2)$ is positive in this case, so the outflow has the potential to escape from the gravitational field. We note that the heating $\mathcal{Q}$ is necessary to make $\mathcal{B} > 0$. The quantity $\mathcal{B}' + \mathcal{Q}$ increases along the streamline, and it becomes positive at $z \approx 5.1h$. In principle, we could stop heating above this height, and keep a potentially released flow. 

The analysis of the Bernoulli invariant suggests that magnetic acceleration is important only at the wind basis, whereas most of the remaining acceleration along the streamline is due to coronal heating. This association constitutes the key mechanism at the origin of the magnetothermal winds we observe in these simulations. This is an extension of the situation described by \cite{CF00B}, with the important difference that the flow becomes super-Alfv\'enic in the non-ideal MHD zone.

\subsubsection{Ejection mechanism} \label{sec:massload}

\begin{figure}[ht]
\centering
\includegraphics[width=\hsize]{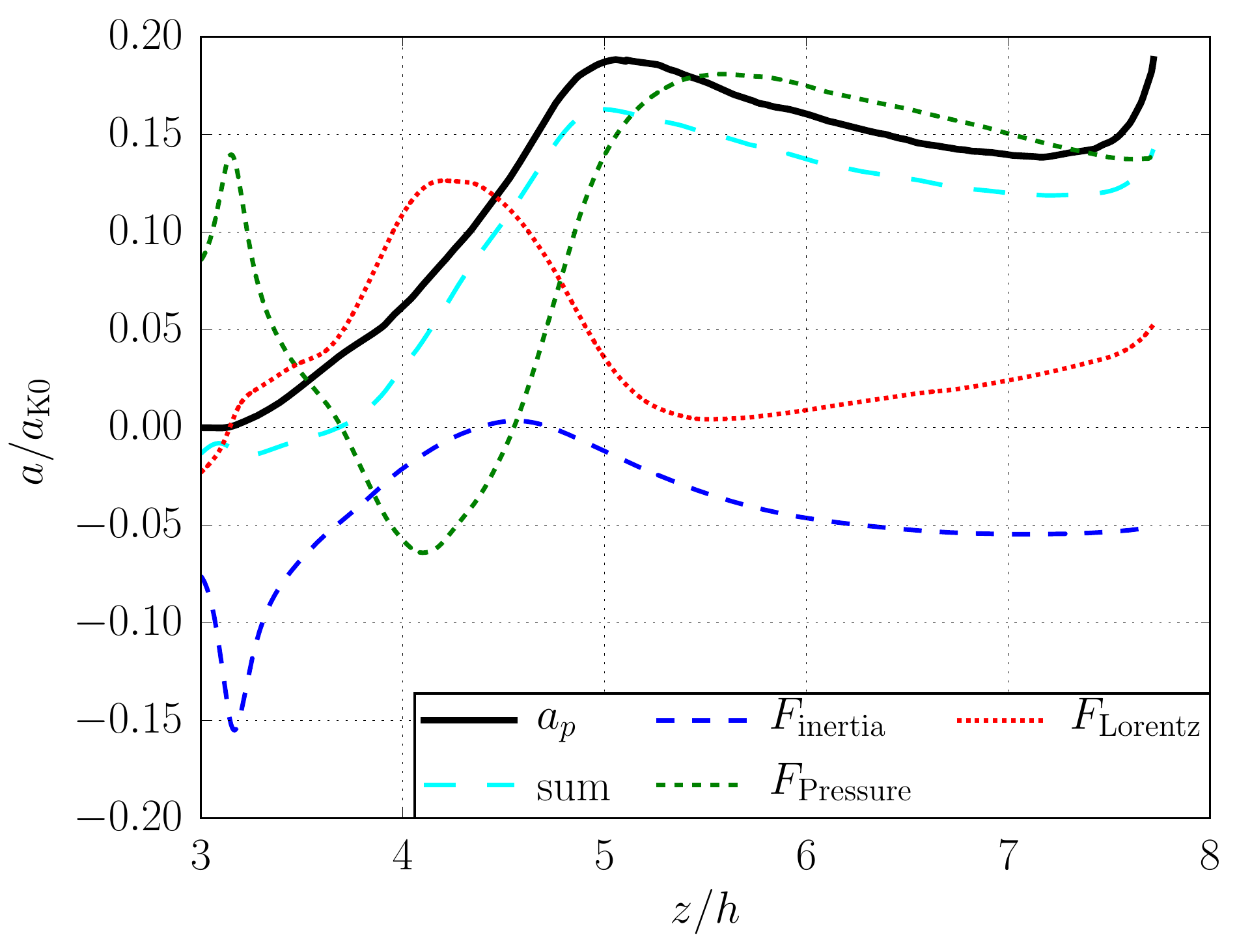}
\caption{Same as Fig. \ref{fig:fid_windacc} in run R10-M3, averaged in time between $400T_0$ and $700T_0$, with a streamline passing through $(r=7r_0,z=5h)$. }
\label{fig:wiwi_force}
\end{figure}

We examine the role of coronal heating in the lifting and acceleration of material along a streamline in run R10-M3. The acceleration of a fluid element and its decomposition into pressure, Lorentz and inertial forces are drawn in Fig. \ref{fig:wiwi_force}. The true acceleration $a_p$ is about $15\%$ stronger than the sum reconstructed from the averaged forces. This bias may be due to the correlated fluctuations of density with the Lorentz force, or between velocity components in the inertial term. We verified that it is entirely removed in runs showing better stationarity properties.  

At the surface of the disk $z \lesssim H$, the streamline is nearly vertical. In this direction, the inertial potential precisely cancels the vertical pressure gradient, as required from the initial, hydrostatic equilibrium. In the transition region $z/h \in \left[3.7, 4.7\right]$, matter from the disk is being pushed down toward the midplane by the hot gas at the base of the corona. The inertial acceleration is negligible in this region, and it is the Lorentz force that provides the kick lifting matter up into the hotter wind region. 

Because the streamlines strongly bend outward at $z > 4.7h$, the radial pressure gradient becomes favorable to the outflow, whereas the Lorentz force is negligible in this region. For $z > 6h$, the acceleration produced by thermal pressure gradients decreases, whereas magnetic acceleration increases again. This is generic to all magnetically-ejecting runs with a warm corona. 

We conclude that in our warm corona simulations, mass is still loaded into the wind by the Lorentz force. The vertical temperature gradient acts against the launching of a wind. The radial temperature gradient bends the magnetic field lines, and causes the outward acceleration of the flow along these lines.

\subsection{Non-accreting disks} \label{sec:nonejec}

Some of our simulations do not exhibit accretion streams in the midplane. In this section, we describe the opposite configurations, where the radial mass flux is directed outward in the midplane and inward at the disk surface, resulting in a large-scale meridional circulation \citep{FLM11}. We focus on run R1-P4-C4, which adopted this configuration over the entire duration of the simulation. Although R1-P4-C4 differs from our fiducial case R10-M3-C2 by four control parameters, only the orientation of the net magnetic field seems to affect this outcome in our simulations (cf. Sect. \ref{sec:recap_magconf}).

\subsubsection{Overview} \label{sec:nonejec_overv}

\begin{figure}[ht]
\centering
\includegraphics[width=\hsize]{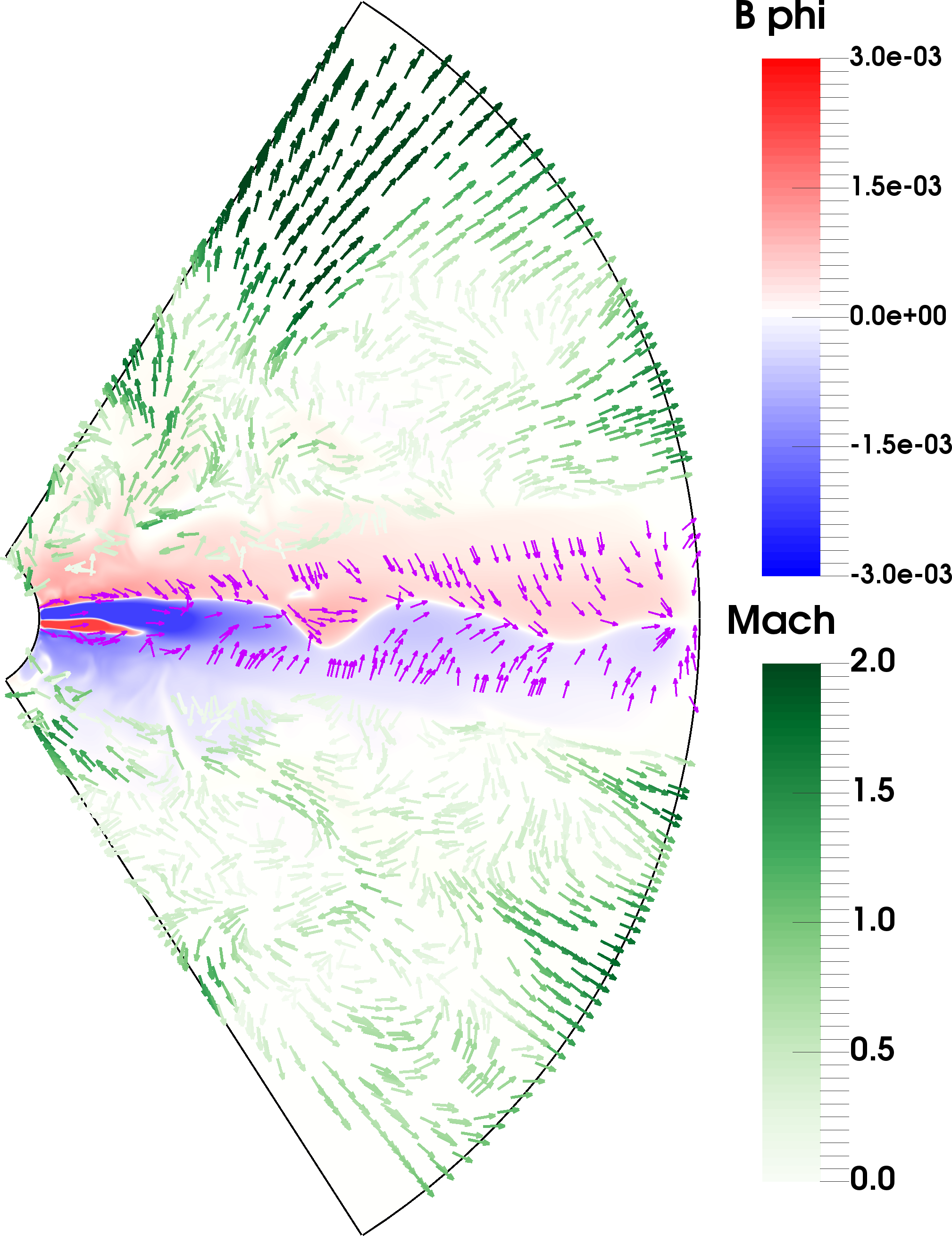}
\caption{Snapshot of run R1-P4-C4 (from $1\au$ to $10\au$, with weak $B_z > 0$) at $t=1000 T_0$, showing the toroidal magnetic field in color background (blue to red), the poloidal velocity field in units of local sound speed (green arrows in the corona), and the orientation of the angular momentum flux caused by magnetic stress (purple arrows in the disk). The orientation of the angular momentum flux is reversed compared to the fiducial, accreting case. }
\label{fig:snap_bp2}
\end{figure}

We show in Fig. \ref{fig:snap_bp2} the morphology of the flow in run R1-P4-C4. We note several differences with our fiducial case (cf. Fig. \ref{fig:snap_bp1}). First, the entire corona exhibits a disorganized velocity structure. There are no signatures of acceleration above the disk, and the flow reaches moderate velocities. Second, the toroidal magnetic field does not expand to the corona, but remains confined within the disk instead. Finally, the large-scale magnetic field transport angular momentum from the disk surface to its midplane. 

As mentioned in Sect. \ref{sec:fid_overview}, studying this configuration isolately makes sense because magnetic structures evolve over long time scales. However, the accreting property can vary in space (cf. Sect. \ref{sec:disccoex}), and in time for a given simulation.

\subsubsection{Vertical structure} \label{sec:nonejec_zstruct}

\begin{figure}[ht]
\centering
\includegraphics[width=\hsize]{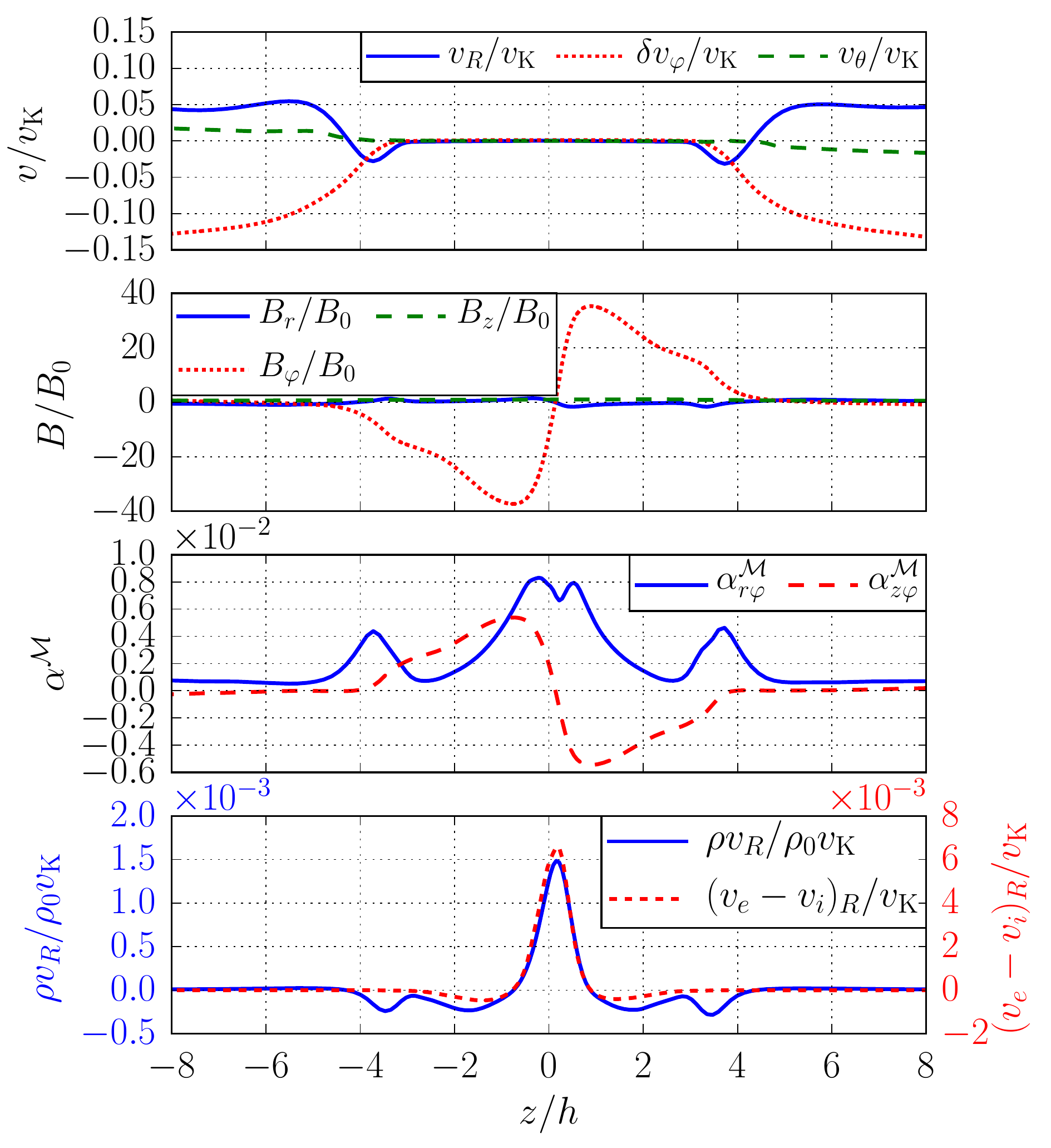}
\caption{Vertical profiles in run R1-P4-C4, averaged in time from $400T_0$ to $1000T_0$, and in spherical radius from $5r_0$ to $8r_0$. \emph{First panel}: fluid velocity, normalized by the local Keplerian value $v_{\mathrm{K}}$; the disk mean value has been subtracted from the azimuthal component. \emph{Second panel}: magnetic field, normalized by the vertically averaged value of the vertical field $B_z$. \emph{Third panel}: horizontal (solid blue) and vertical (dashed red) magnetic stresses, normalized by the vertically averaged pressure (cf. Eq. \eqref{eqn:alphadef}). \emph{Fourth panel}: radial mass flux (solid blue), and electron minus ion radial velocity, normalized by the local Keplerian velocity and the vertically averaged density. }
\label{fig:nowi_vb}
\end{figure}

We show in Fig. \ref{fig:nowi_vb} the vertical structure of the flow in run R1-P4-C4. In the first panel, the radial velocity is negative at the disk surface, where it reaches $v_R/c_s \approx -0.4$. It becomes positive above $z \approx 4.3h$, with $v_R/\id{v}{K} \approx 5\%$, two times smaller than the fiducial case. The azimuthal velocity drops at $z \approx H$, where the gas is heated. The polar velocity $v_{\theta}$ is negligible up to $z \approx 4.3h$, beyond which it increases to $2\%$ of $\id{v}{K}$, again small compared to the fiducial case (cf. Fig. \ref{fig:fid_vb}). This outflow is launched twice higher than in the fiducial case, near the temperature transition at $z \approx H$, suggesting a predominantly thermal wind launching. 

The second panel shows that the toroidal magnetic field is dominant in the disk, twenty to eighty times greater than the radial one. Unlike the fiducial accreting case, the toroidal field $B_{\varphi}$ vanishes at $z \approx 4.7h$, and does not expand to the corona. 

The resulting vertical stress $\mathcal{M}_{z\varphi}$ is negligible in the corona. The disk thus receives no angular momentum from the corona, and a fortiori from our polar boundary conditions. Within the disk, the Maxwell stress $\mathcal{M}_{z\varphi}$ is positive in the southern hemisphere, and negative in the northern one. This induces a flux of angular momentum from the disk surface toward its midplane. 

The fourth panel shows that mass is now streaming outward in the midplane, and inward at the surface. This is consistent with the idea that angular momentum is extracted from the surface, and provided to the midplane by magnetic stress. We emphasize that the net radial mass flux is approximately zero through this meridional circulation. The disk does not receive angular momentum from the corona, and mass accretion is solely caused by the radial flux of angular momentum, $\tau_r \approx 0$, as shown in Fig. \ref{fig:fid_accretor}. 

The property of being accreting or not essentially depends on the sign of $\mathcal{M}_{z\varphi} \equiv - B_z B_{\varphi}$. Because the disk is threaded by a net vertical magnetic field, and because the flow is laminar, the orientation of the angular momentum flux is given by the sign of $B_{\varphi}$ in both hemispheres. We will refer to this as the vertical \emph{phase of the toroidal field}. When $B_{\varphi}$ has a non-accreting phase, the radial velocity is oriented outward in the midplane, and inward at the disk surface. The radial pressure gradient is negative in the corona, so it opposes the launching of a wind directed inward. This imposes a meridional circulation within the disk. 

The pressure gradient forces the radial velocity to change sign above $z \gtrsim H$. The resulting outflow is thermally driven. It reaches moderate velocities, and it is not organized on large scales. We were not able to reconstruct streamlines leaving the disk surface, because of the time-variability of the flow in the corona. We presume that the outflow mass loading results from turbulence above the accretion layers. Forcing an outflow velocity at the domain boundaries greatly improved the numerical stability of our setup in such configurations. 

We measured the phase of $B_{\varphi}$ in every run via $\sigma_{\varphi}$, as defined by Eq. \eqref{eqn:modphi}. Runs having $\sigma_{\varphi} > 0$ should be accreting, whereas $\sigma_{\varphi} < 0$ corresponds to predominantly non-accreting configurations. Runs with $\sigma_{\varphi} \approx 0$ are the object of Sect. \ref{sec:selforgathor}. Time-averaged values of $\sigma_{\varphi}$ for every runs are given in Table \ref{table:results}.

\subsubsection{Magnetic equilibrium} \label{sec:mageq}

We raise two questions regarding the magnetic equilibrium of the non-accreting configuration. First, if there is no outflow to remove toroidal magnetic flux, another process must be responsible for the saturation of the MRI. Second, because the corona rotates slower than the disk, the net vertical field should always exert a global torque on the disk, favoring the accreting configuration. To address these issues, we split the induction equation \eqref{eqn:dyn-b} into several terms, labeled by the associated physical process:
\begin{align}
\begin{split}
\brac{\frac{\partial B_R}{\partial t}}_{\varphi} &= \frac{1}{R} \left( \cot(\theta) + \frac{\partial}{\partial_{\theta}} \right) \cdot \left[ \underbrace{\brac{v_R B_{\theta}}}_\text{stretch} - \underbrace{\brac{v_{\theta} B_{R}}}_\text{outflow} \right.\\
&\left. -\underbrace{\brac{\id{\eta}{O} J_{\varphi}}}_\text{Ohm} -\underbrace{\brac{\id{\eta}{H} \left( J \times e_b \right)_{\varphi}}}_\text{Hall} - \underbrace{\id{\eta}{A} \left(\left( J \times e_b \right) \times e_b \right)_{\varphi}}_\text{ambipolar} \right] \label{eqn:split_br}
\end{split} \\
\begin{split}
\brac{\frac{\partial B_{\varphi}}{\partial t}}_{\varphi} &= \left( \frac{1}{R} + \frac{\partial}{\partial R} \right) \cdot \left[ \underbrace{ \brac{ v_{\varphi} B_R}}_\text{shear} -\underbrace{\brac{v_R B_{\varphi}}}_\text{advection} \right]\\
& - \frac{1}{R} \frac{\partial}{\partial \theta} \left[ \underbrace{\brac{v_{\theta} B_{\varphi}}}_\text{outflow} - \underbrace{\brac{v_{\varphi} B_{\theta}}}_\text{stretch} \right] - \underbrace{\left( \nabla \times \brac{ \mathcal{E}_{\mathrm{O}, \mathrm{H}, \mathrm{A}}} \right)_{\varphi}}_\text{Ohm, Hall, ambipolar}, \label{eqn:split_bp}
\end{split} 
\end{align}
where the electromotive field $\mathcal{E}_{\mathrm{O}, \mathrm{H}, \mathrm{A}}$ contains the three non-ideal effects, as developped in Eq \eqref{eqn:split_br}.  

\begin{figure}[ht]
\centering
\includegraphics[width=\hsize]{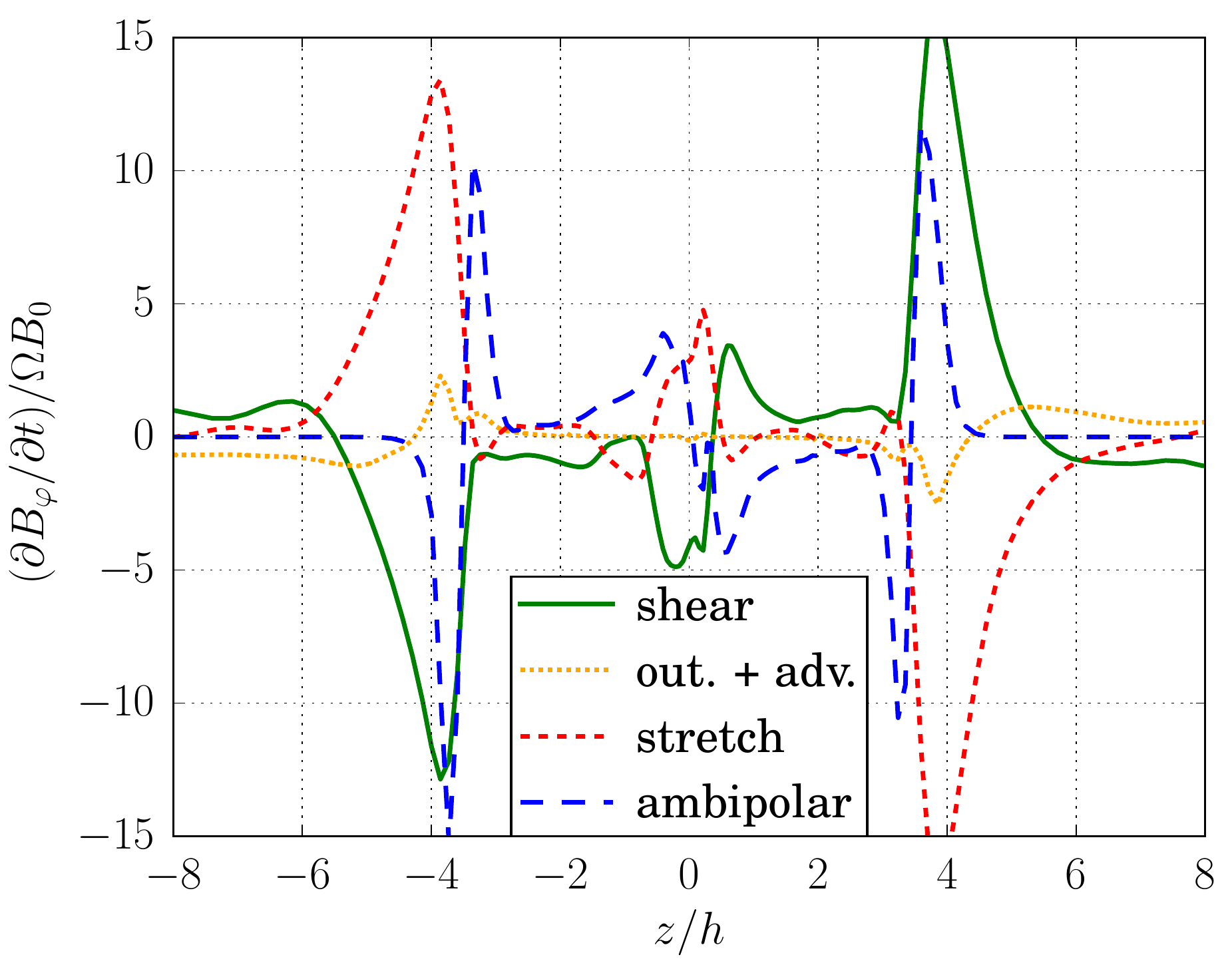}
\caption{Contributions to the induction of toroidal magnetic field in run R1-P4-C4, as explicited in Eq. \eqref{eqn:split_bp}, averaged in time from $400T_0$ to $1000T_0$ and in spherical radius from $5r_0$ to $8r_0$, normalized by the local Keplerian frequency and the local initial magnetic field. }
\label{fig:nowi_dbpdt}
\end{figure}

We represent in Fig. \ref{fig:nowi_dbpdt} these contributions to the induction of toroidal magnetic field $\partial_t B_{\varphi}$. The ohmic and Hall induction terms are not shown for they are negligible in this case. 

The sum of the outflow plus advection terms has a minor effect, located at $z \approx H$. Deep in the disk, the shearing of $B_R$ is only balanced by ambipolar diffusion. At $z\approx 3.4h$, where the toroidal Alfv\'en velocity is maximal, ambipolar diffusion transports the toroidal field toward the surface, and not toward the midplane. In the surface layers $z \approx H$, we find a competition between the stretching of $B_z$ and the shearing of $B_R$. Since the stretching term favors an accreting phase for $B_{\varphi}$, the non-accreting configuration sustains a counteracting $B_R$ at the surface. 

\begin{figure}[ht]
\centering
\includegraphics[width=\hsize]{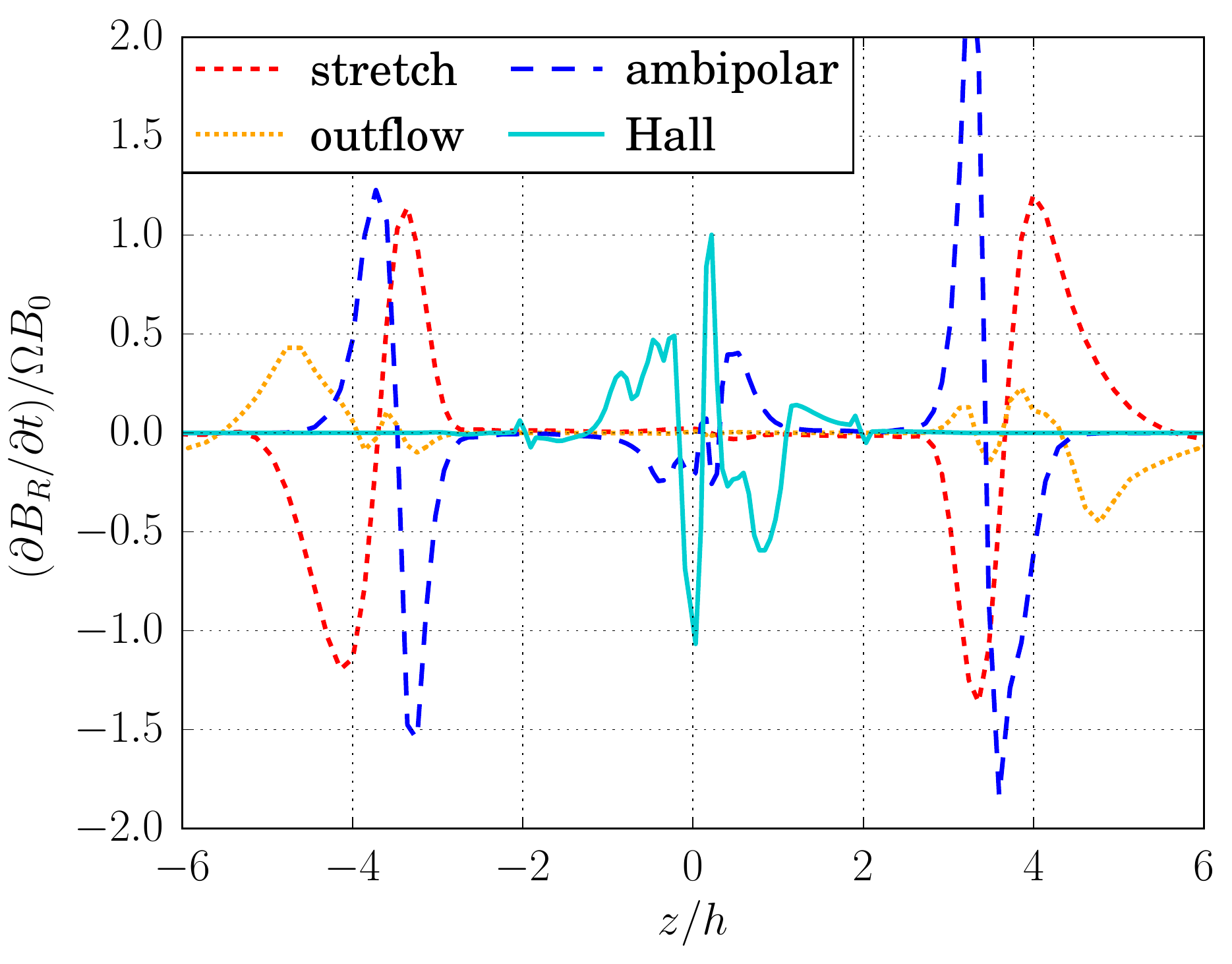}
\caption{Same as Fig. \ref{fig:nowi_dbpdt} for the radial magnetic field $B_R$; see Eq. \eqref{eqn:split_br}. }
\label{fig:nowi_dbrdt}
\end{figure}

The same decomposition is performed for the radial component $B_R$ in Fig. \ref{fig:nowi_dbrdt}. The ohmic contribution is negligible again, so we do not plot it. The Hall term is effective below $z \lesssim 2h$, balanced by ambipolar diffusion. At the surface $z \gtrsim 3h$, the outflow term has a moderate amplitude, and we find a competition between ambipolar diffusion and the stretching of $B_{\theta}$. 

The region $z < 2h$ is separated from the disk surface. Near the midplane, it is the HSI that causes the growth of horizontal magnetic field. Near the surface, the feedback loop corresponds to the MRI: the accretion layers stretch $B_z$ into $B_r$, which is sheared into $B_{\varphi}$; the resulting stress $\mathcal{M}_{z\varphi}$ removes angular momentum from the surface, thereby enhancing the accretion layers. Ambipolar diffusion is responsible for the saturation of the HSI in the midplane, and of the MRI at the disk surface.

\subsubsection{Coexistence of accreting and non-accreting regions} \label{sec:disccoex}

\begin{figure}[ht]
\centering
\includegraphics[width=\hsize]{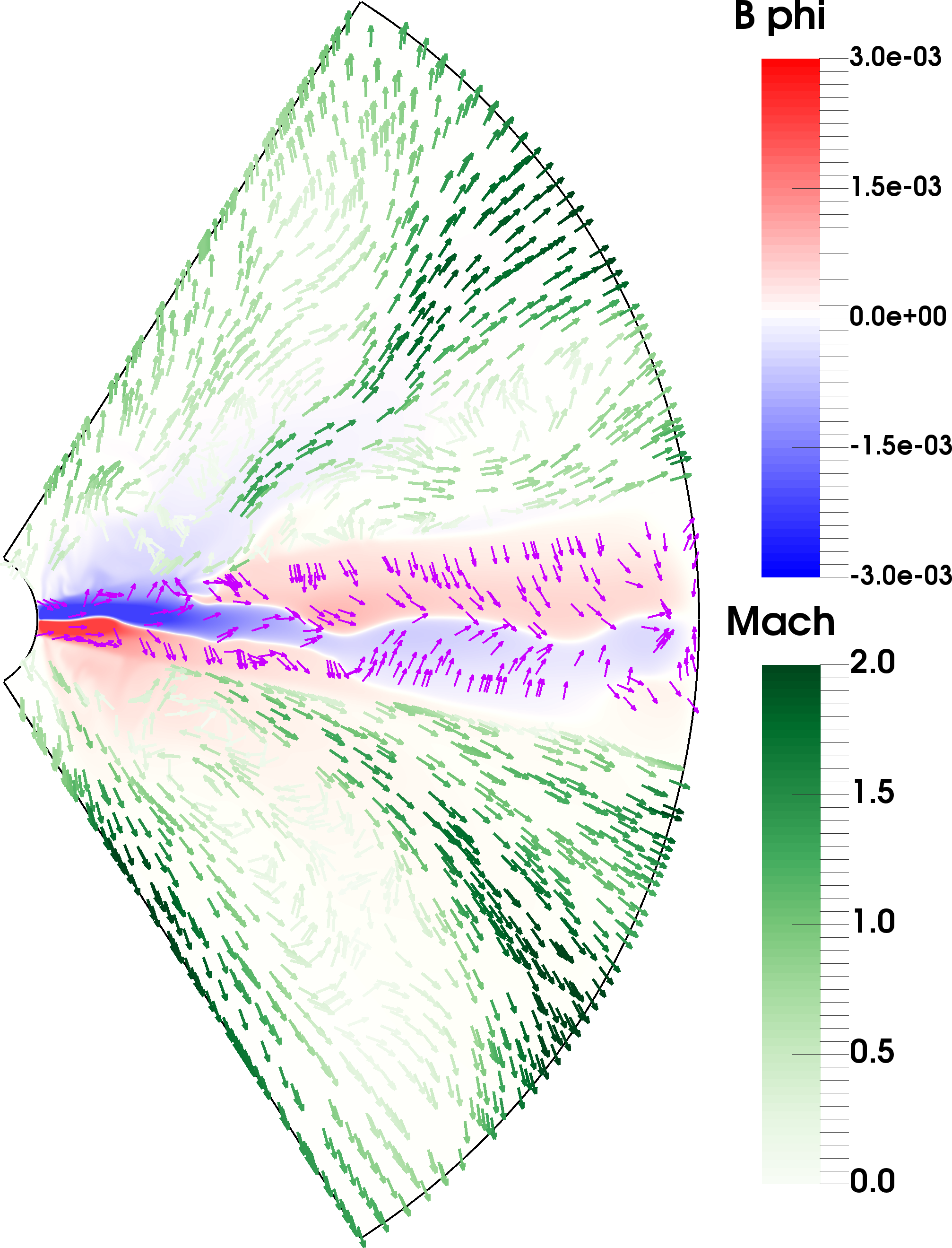}
\caption{Same as Fig. \ref{fig:snap_bp2} for run R1-P4 at $t=1000 T_0$. The disk exhibits both an accreting (inner) and a non-accreting (outer) region. }
\label{fig:snap_bp3}
\end{figure}

All the cases presented so far were relatively symmetric with respect to the disk midplane. In particular, the inward or outward mass streams were located near the midplane. Yet we have disks exhibiting both accreting and non-accreting behaviors at the same time. We show the radial transition between two different portions of the same disk in Fig. \ref{fig:snap_bp3}. 

As mentionned in Sect. \ref{sec:fid_verti}, the mass flux actually follows the $B_{\varphi} = 0$ layer, where the electric current is extremal. The transition from an accreting to a non-accreting disk region necessarily comes with the current sheet reaching the surface of the disk. In a region where $B_{\varphi}$ has a non-accreting phase, there is no magnetized outflow, and $B_{\varphi} \approx 0$ above the disk. In this case, the mass flux follows closed circulation loops along the current sheet, inward at the surface and outward in the midplane. At the intersection with an accreting region, part of this mass flux is reoriented into the wind, and part of it goes to the midplane.

\subsection{Vertical symmetry breaking} \label{sec:selforgathor}

This section describes a spontaneous breaking of the up/down symmetry identified in our simulations. It is related to the emergence of a favored magnetic polarity over long time scales. 

\subsubsection{Overview} \label{sec:thorover}

Within our stratified setup, the horizontal magnetic flux is free to leave the disk in the vertical direction. One polarity can be removed or amplified faster than the other, leaving the disk with only one sign of $B_{\varphi}$. This was observed in stratified shearing-box simulations \citep{LKF14,Bai15}, but considered unlikely to be directly connected to a global flow geometry. Focusing on the horizontal magnetic field, we will refer to these as \emph{even configurations} with respect to the disk midplane in opposition to states showing an odd symmetry. 

\begin{figure}[ht]
\centering
\includegraphics[width=\hsize]{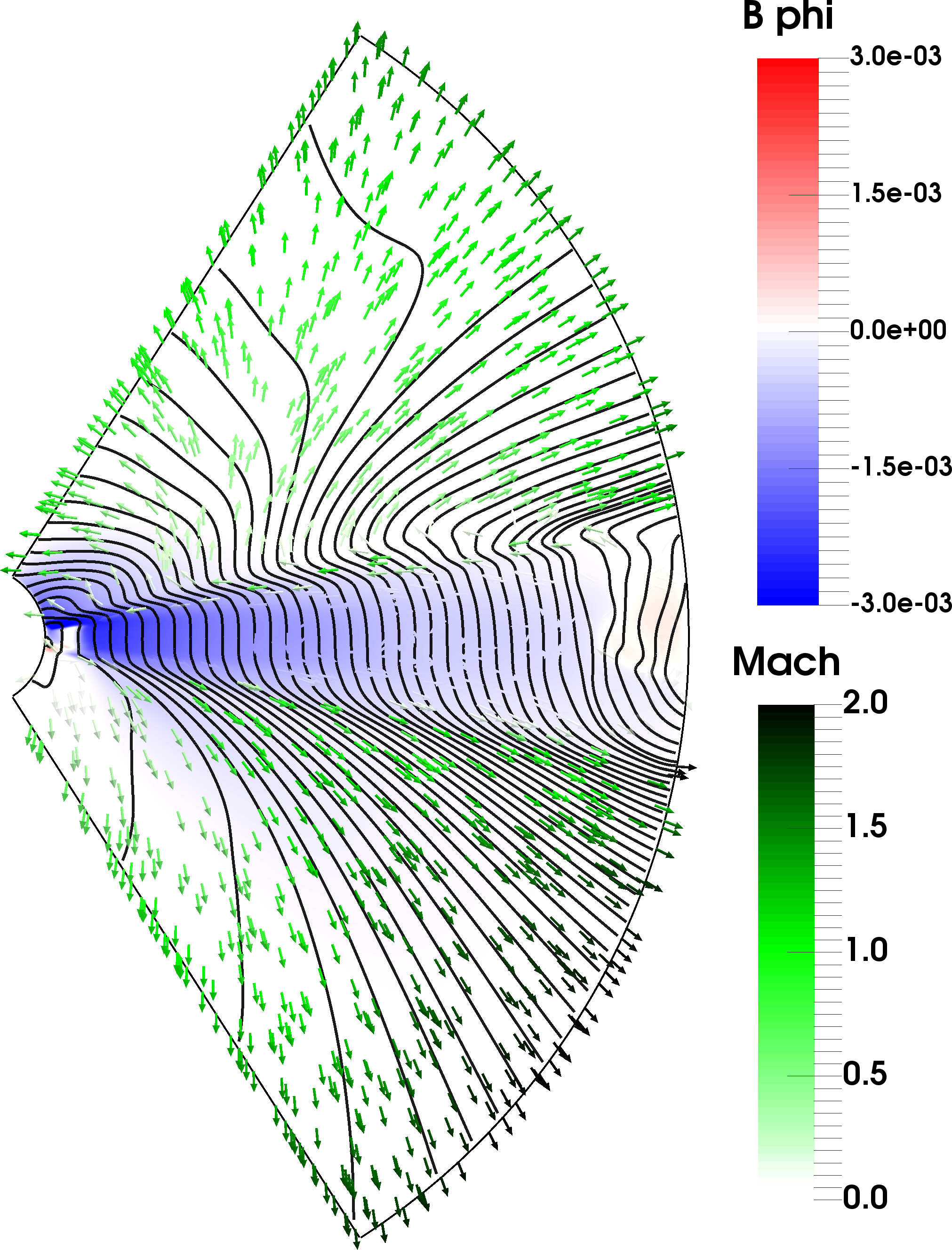}
\caption{Averaged flow poloidal map for run R1-M3 (from $1\au$ to $10\au$, with moderate $B_z < 0$) from $800T_0$ to $1000T_0$; magnetic field lines are sampled along the midplane, and the velocity field is indicated with green arrows over the background toroidal field. The polar asymmetry and the absence of $B_{\varphi}$ sign reversal within the disk are obvious. }
\label{fig:thor_snap}
\end{figure}

Such a configuration is illustrated in Fig. \ref{fig:thor_snap} for run R1-M3. The entire disk sees $B_{\varphi} < 0$; the Keplerian and Hall shears ensure that $B_r > 0$ is also even about the midplane. In the inner half of the northern corona, the magnetic field lines do not guide the velocity field. Because the corona obeys ideal MHD, this implies that this part of the flow is turbulent, resulting in an effective ``turbulent diffusion'' for the time-averaged flow. On the contrary, the outflow in the southern corona is very laminar and stationary. A magnetic collimation effect is observed in this hemisphere. 

\subsubsection{Vertical structure} \label{sec:thorstruct}

\begin{figure}[ht]
\centering
\includegraphics[width=\hsize]{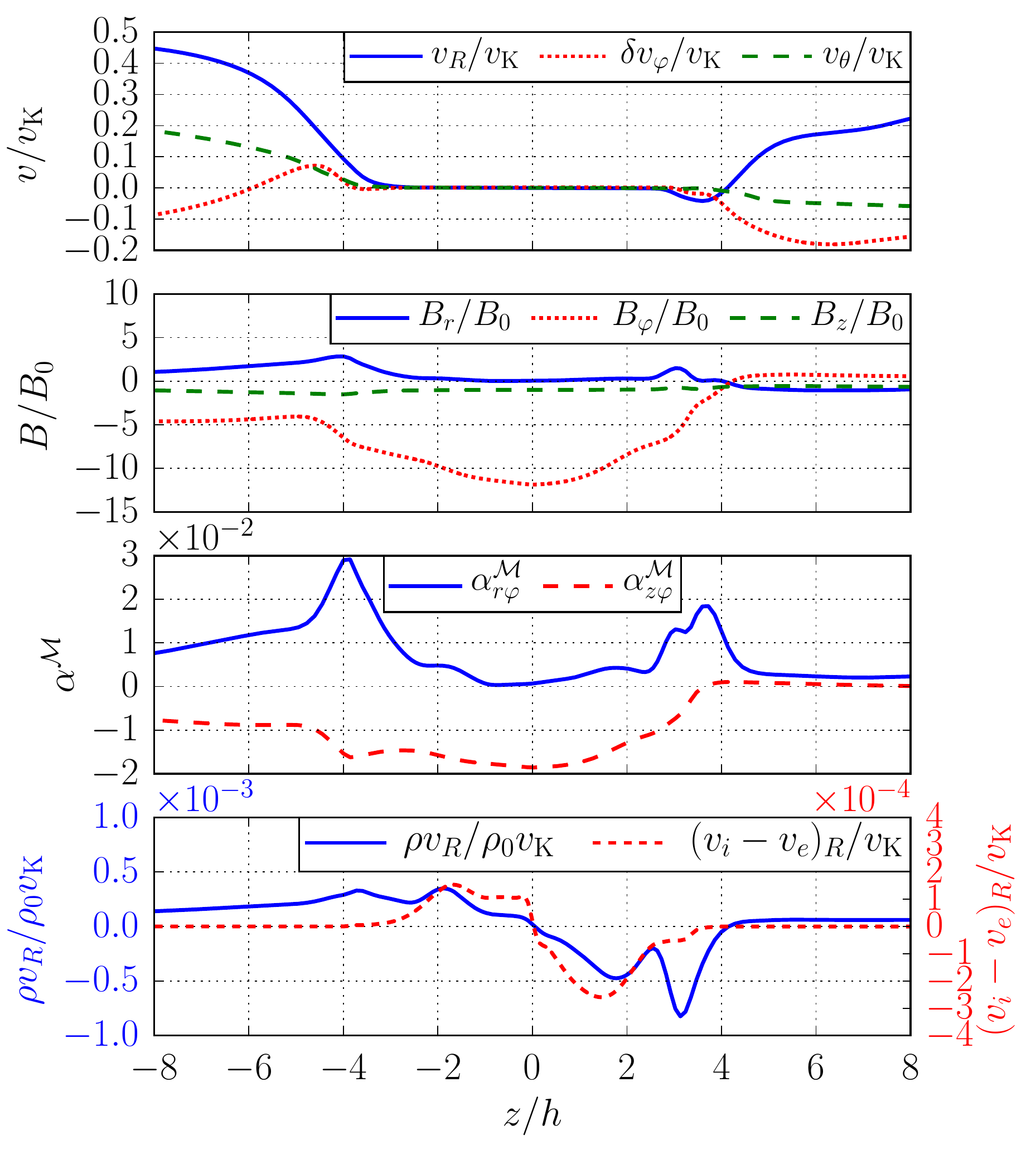}
\caption{Vertical profiles in run R1-M3, averaged in time from $800T_0$ to $1000T_0$, and in spherical radius from $6r_0$ to $8r_0$. \emph{First panel}: fluid velocity, normalized by the local Keplerian value $v_{\mathrm{K}}$; the disk mean value has been subtracted from the azimuthal component. \emph{Second panel}: magnetic field, normalized by the vertically averaged value of $B_z$. \emph{Third panel}: horizontal (solid blue) and vertical (dashed red) magnetic stresses, normalized by the vertically averaged pressure (cf. Eq. \eqref{eqn:alphadef}). \emph{Fourth panel}: radial mass flux (solid blue), and ion minus electron radial velocity, normalized by the local Keplerian velocity and average density. }
\label{fig:thor_vb}
\end{figure}

The vertical structure of run R1-M3 is represented in Fig. \ref{fig:thor_vb}. A strong outflow is launched in the southern hemisphere, with $v_{\theta} / \id{v}{K} \approx 20\%$ at $z \approx -8h$. The toroidal velocity is higher than the local Keplerian value, down to $z \approx -6h$. The outflow receives angular momentum via the Maxwell stress $\mathcal{M}_{z\varphi}$, and transports mass at a rate $\dot{m}^{-}_w \approx 2.6 \times 10^{-7} \,\mathrm{M}_{\odot}.\,\mathrm{yr}^{-1}$. Picking a streamline passing through $(r=6r_0,z=-5h)$, its total angular momentum is $1.5$ times larger than the Keplerian value at the wind base, it crosses the fast-magnetosonic velocity before reaching the domain boundary, and its Bernoulli invariant $\mathcal{B} = 0.28 \,(v_{\mathrm{K}0}^2/2) > 0$. 

On the other side, the northern hemisphere displays the properties of a non-accreting disk (cf. Fig. \ref{fig:nowi_vb}). The radial velocity is oriented inward at the disk surface, and the outflow has a moderate velocity. The toroidal field $B_{\varphi} = 0$ at $z \approx 4h$, beyond which the vertical flux of angular momentum is essentially zero. 

Because $B_z$ and $B_{\varphi}$ keep the same sign within the disk, the vertical flux of angular momentum $\mathcal{M}_{z\varphi}$ is unidirectional. We find $\mathcal{M}_{z\varphi} \approx 0$ for $z \gtrsim +4h$, and $\mathcal{M}_{z\varphi} < 0$ below. The gradient $\partial_z \mathcal{M}_{z\varphi}$ thus takes angular momentum from the disk northern surface, and transports it toward the southern hemisphere. This causes both the radial flows at the disk surface, and the magnetic wind launching in the southern hemisphere. Since $\mathcal{M}_{z\varphi} \approx 0$ in the northern corona, the disk is not pumping angular momentum from the northern domain boundaries. 

The fourth panel of Fig. \ref{fig:thor_vb} shows an accretion layer at the northern surface, and a decreting layer at the southern one. On average, the net mass flux is slowly accreting, with $\tau_z = \left( -4 \pm 5 \right) \times 10^{-7}$. We have separated the southern and northern wind mass loss rates $\dot{m}^{-}_w$ and $\dot{m}^{+}_w$ (cf. Eq. \ref{eqn:mdotwind}). Table \ref{table:results} confirms that $\dot{m}^{-}_w > \dot{m}^{+}_w$ when $\iota_{\varphi} \times \mathrm{sign} \left( B_z \right)  > 0$, i.e., the magnetically ejecting side transports more mass.

\subsubsection{Secular evolution of vertical symmetries} \label{sec:thormech}

\begin{figure}[ht]
\centering
\includegraphics[width=\hsize]{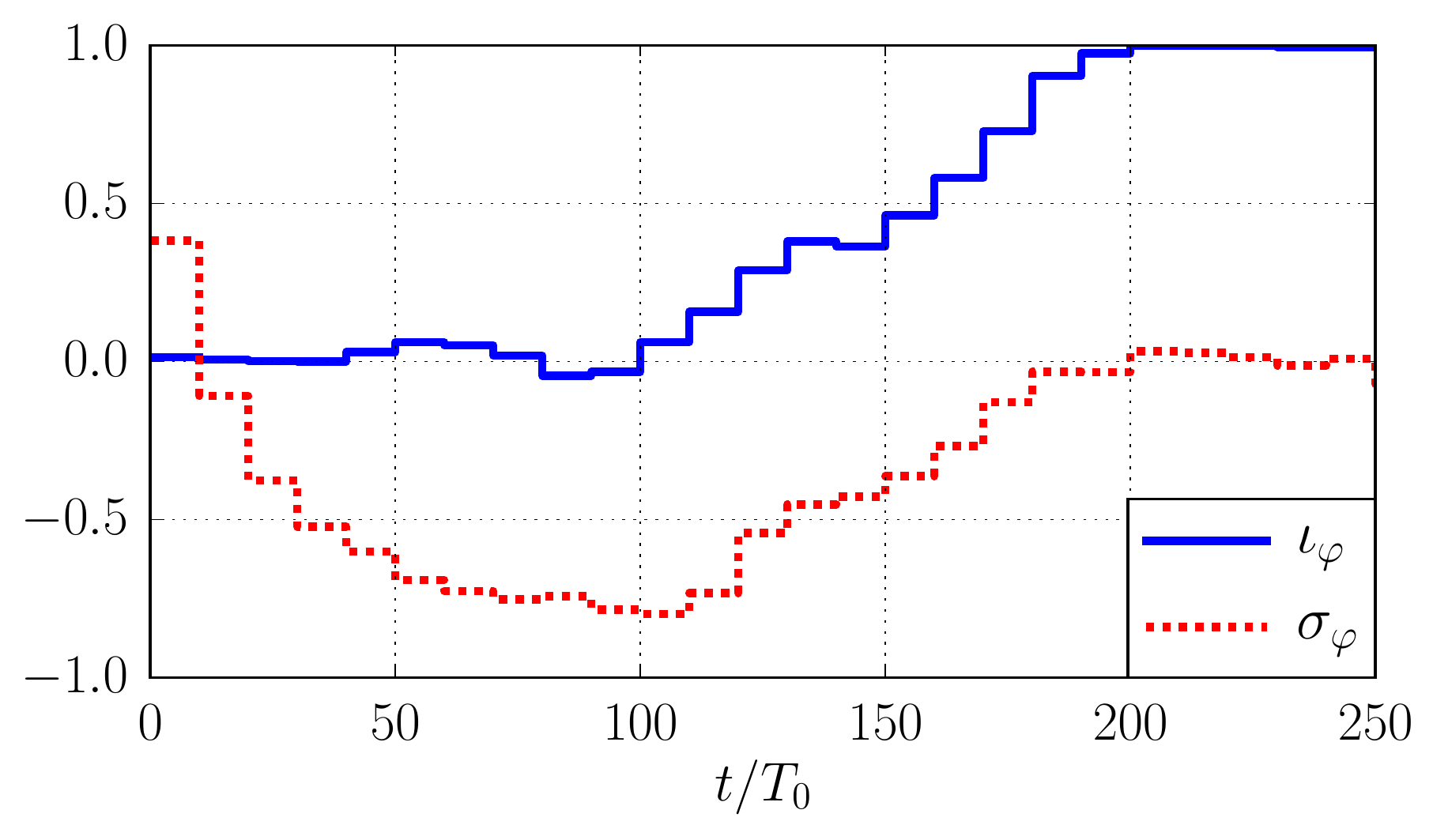}
\caption{Evolution of the symmetry coefficients $\iota_{\varphi}$ and $\sigma_{\varphi}$ in run R1-P2; data have been averaged over bins of $10T_0$. }
\label{fig:thor_symm}
\end{figure}

We examine the onset of an even configuration in run R1-P2, which did not immediately choose a favored polarity. We draw in Fig. \ref{fig:thor_symm} the symmetry coefficients $\iota_{\varphi}$ and $\sigma_{\varphi}$ over time, respectively defined by Eqs. \eqref{eqn:symphi} and \eqref{eqn:modphi}. $\iota_{\varphi}$ defines the average polarity of $B_\varphi$ in the disk, whereas $\sigma_\varphi$ characterizes $B_\varphi$ symmetry across the midplane, the ejecting configurations with accretion in the disk midplane corresponding to $\sigma_\varphi>0$.

During the first $100\,T_0$,  $\iota_{\varphi} \approx 0$ indicates that both polarities of $B_{\varphi}$ are equally present in the disk. The average phase of $B_{\varphi}$ evolves from an accreting configuration $\sigma_{\varphi} > 0$ to a non-accreting one $\sigma_{\varphi} < 0$ in this interval. From $100T_0$ to $200T_0$, the increase in $\iota_{\varphi}$ means that the positive $B_{\varphi} > 0$ polarity is filling the disk. The fact that $\sigma_{\varphi}$ goes to zero confirms that the disk is adopting an even $B_{\varphi}$ symmetry. 

\begin{figure}[ht]
\centering
\includegraphics[width=\hsize]{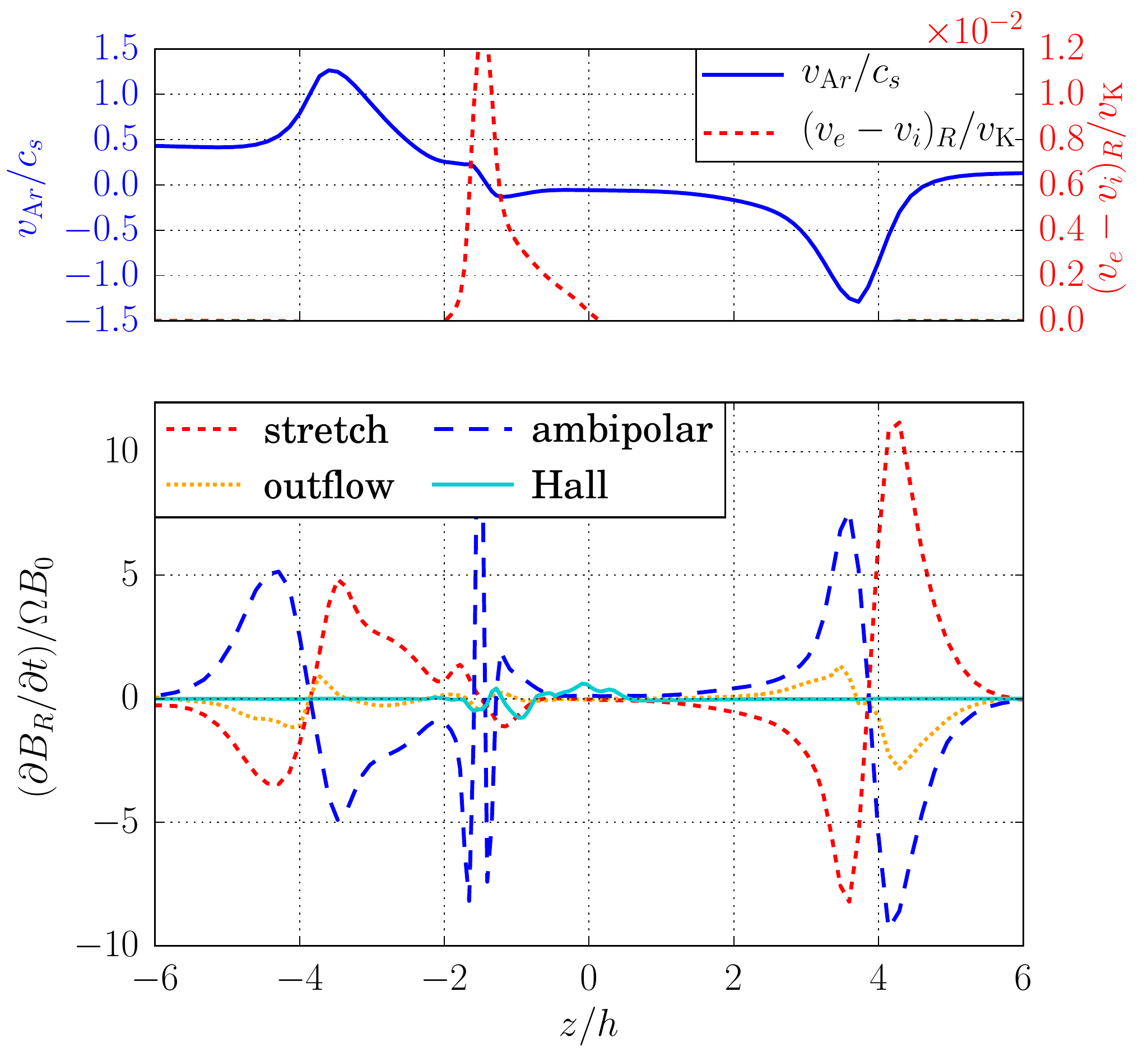}
\caption{Vertical profiles in run R1-P2, averaged in time between $120T_0$ and $130T_0$, and in spherical radius between $5.5r_0$ and $6 r_0$; \emph{upper panel}: sonic Mach number for the radial Alfv\'en velocity (solid blue) and electron minus ion velocity, normalized by the Keplerian velocity (dashed red); \emph{lower panel}: induction of radial magnetic field. }
\label{fig:thor_dbrdt}
\end{figure}

We consider a space-time window in which the transition from an odd to an even $B_{\varphi}$ symmetry is taking place (between $t=120\,T_0$ and $t=130\,T_0$), and look at vertical transport of the horizontal magnetic field. We decompose the induction of radial field $B_R$ in Fig. \ref{fig:thor_dbrdt}. In the upper panel, we see that the electric current layer is displaced of $z \approx -1.5h$ from the midplane, coinciding with the current sheet where $B_r = 0$. This current layer keeps drifting downward in time, and eventually leaves the disk (fourth panel of Fig. \ref{fig:thor_vb}). 

In the lower panel of Fig. \ref{fig:thor_dbrdt}, we show that the transport of negative $B_R < 0$, from the midplane toward the southern surface, is mainly driven by ambipolar diffusion. The outflow and Hall contributions have a negligible impact on this process. Similar symmetry-breaking solutions have been observed in shearing box simulations \citep{LKF14, Bai15}. The cause of this phenomenon is yet unknown, but it appears to be robust.

\subsection{Self-organization: zonal flows} \label{sec:zf}

This section describes the emergence of \emph{zonal flows} in our simulations, characterized by axisymmetric density rings, together with magnetic field accumulations.

\subsubsection{Overview} \label{sec:zfoverview}

\begin{figure}[ht]
\centering
\includegraphics[width=\hsize]{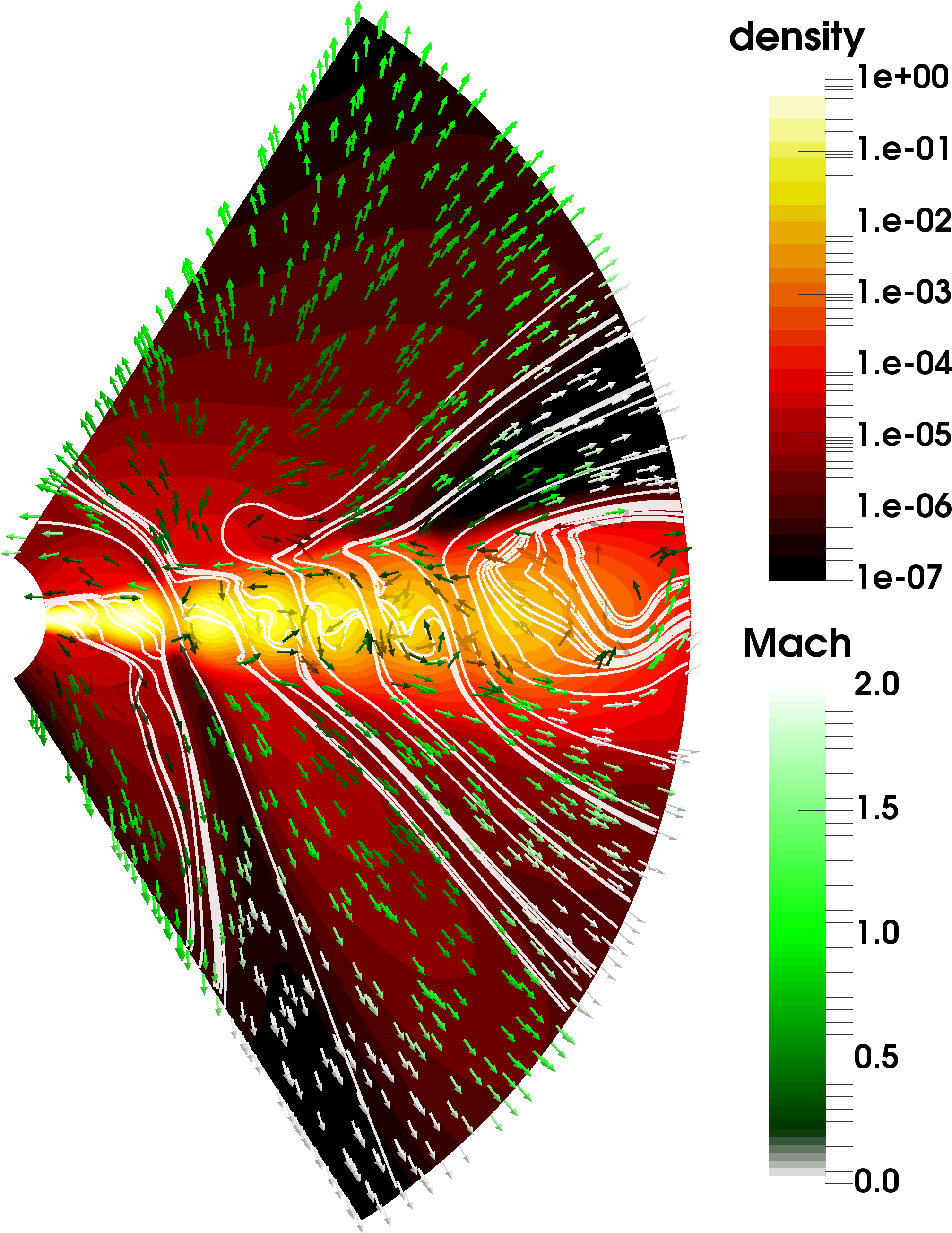}
\caption{Averaged flow poloidal map for run R1-P2 (from $1\au$ to $10\au$, with strong $B_z > 0$), from $500T_0$ to $700T_0$; magnetic field lines are sampled along the midplane, and the velocity field is indicated with green arrows over the background density field. Magnetic field lines are accumulated in low density rings. }
\label{fig:zf_snap}
\end{figure}

We illustrate in Fig. \ref{fig:zf_snap} the morphology of the flow in run R1-P2. We see a series of density bumps in the disk, radially separated by approximately the local disk thickness $2H$. The magnetic flux is concentrated in low density regions, and this foliation is maintained in both coronae. Some magnetic flux accumulations can momentarily be as narrow as eight grid cells in the disk. However, with a magnetic Reynolds number $\mathcal{R}_{\mathrm{A}} \approx 3$, we believe that this width is primarily attributed to physical diffusivity and not numerical diffusion. This run also displays an even $B_{\varphi}$ symmetry, causing the tilting of magnetic field lines through the disk. As in run R1-M3 (cf. Fig. \ref{fig:thor_snap}), ejection is thermally driven in the northern corona, and magnetically enhanced in the southern one. 

Density and magnetic field fluctuations are anti-correlated. Table \ref{table:results} shows that runs with $B_z < 0$ or $\beta > 5 \times 10^{3}$ do not exhibit such structures. These are the runs in which the disk midplane is Hall-shear stable, or has linear growth rates smaller than $0.01 \Omega$. Because our 3D simulations were integrated over shorter time intervals, the indicated number of zonal flows cannot directly be compared between equivalent 2D and 3D runs.

\subsubsection{Self-organization mechanism} \label{sec:zfmech}

The induction of $B_z$ is governed by the toroidal electromotive force (EMF), the same as in Eq. \eqref{eqn:split_br}:
\begin{equation}
\brac{\frac{\partial B_z}{\partial t}}_{\varphi,z} = - \frac{1}{r} \frac{\partial}{\partial r} \brac{ r \left( \mathcal{E}_{\mathrm{I}} + \mathcal{E}_{\mathrm{O}} + \mathcal{E}_{\mathrm{H}} + \mathcal{E}_{\mathrm{A}} \right)_{\varphi} }_{\varphi,z}, \label{eqn:split_bz}
\end{equation}
so the average $B_z$ increases in time where $\mathcal{E}_{\varphi}$ decreases with radius. The ideal, Hall and ambipolar EMFs can be expressed as:
\begin{align}
&\bm{\mathcal{E}}_{\mathrm{I}} \equiv - \bm{v \times B}, \qquad \bm{\mathcal{E}}_{\mathrm{H}} \equiv \lH \bm{J \times B}, \qquad \bm{\mathcal{E}}_{\mathrm{A}} \equiv \id{\eta}{A} \bm{J_{\perp}},\label{eqn:emf}\\
&\bm{J_{\perp}} \equiv \bm{J} - \left(\frac{\bm{J \cdot B}}{\bm{B \cdot B}}\right) \bm{B}. \label{eqn:jperp} 
\end{align}

\begin{figure}[ht]
\centering
\includegraphics[width=\hsize]{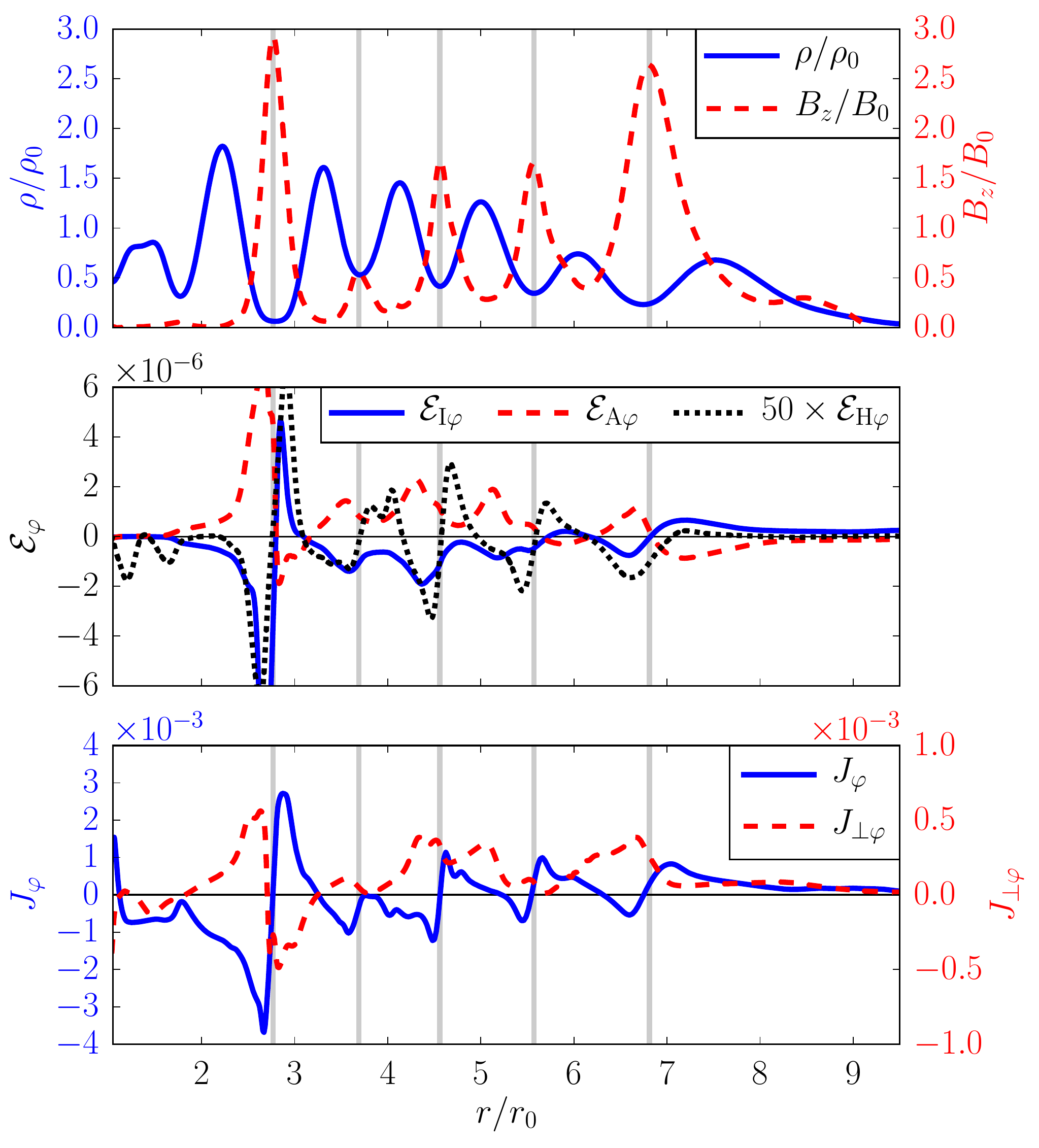}
\caption{Radial profiles in run R1-P2, vertically averaged through the disk from $500T_0$ to $700T_0$; \emph{first panel}: density (solid blue) and vertical magnetic field (dashed red), normalized by their initial profiles; \emph{second panel}: ideal (solid blue) ambipolar (dashed red) and Hall (black dots, multiplied by 50 for visibility) electromotive forces, given by Eq. \eqref{eqn:emf}; \emph{third panel}: toroidal component of the electric current $J_{\varphi}$ (solid blue), and of its projection $J_{\perp \varphi}$ normal to the magnetic field; the vertical lines mark the location of magnetic field maxima. }
\label{fig:zf_mech}
\end{figure}

The first panel of Fig. \ref{fig:zf_mech} shows the anti-correlated fluctuations of density and magnetic field. In the second panel, we plot the averaged EMFs. The ohmic contribution has been omitted for it is negligible and cannot confine magnetic flux. The ideal term $\mathcal{E}_{\mathrm{I}\varphi}$ increases with radius at the location of each magnetic field concentration. The velocity field thus acts as a turbulent diffusion. The ambipolar term precisely balances the ideal one, so it decreases with radius at the location of each band. This was already noted by \cite{BS14} (see their Fig. 8). Ambipolar diffusion is therefore responsible for the accumulation of $B_z$. Apart from being negligible by a factor $50$, the Hall term acts against the accumulation of magnetic flux. Hall-driven self-organization requires the magnetic stress and flux to be anti-correlated \citep{KL13}. This is possible in non-stratified simulations, when the net magnetic flux becomes strong enough to stabilize the HSI. In stratified simulations, the wind-driven stress $-B_{\varphi} \bm{B}_p$ is known to correlate with the net magnetic flux for $\beta \gg 1$ \citep{LKF14}. Self-organization can thus be inhibited if the wind drives the magnetic stress in Hall-shear stable (i.e., strong field) regions.

\begin{figure}[ht]
\centering
\includegraphics[width=0.8\hsize]{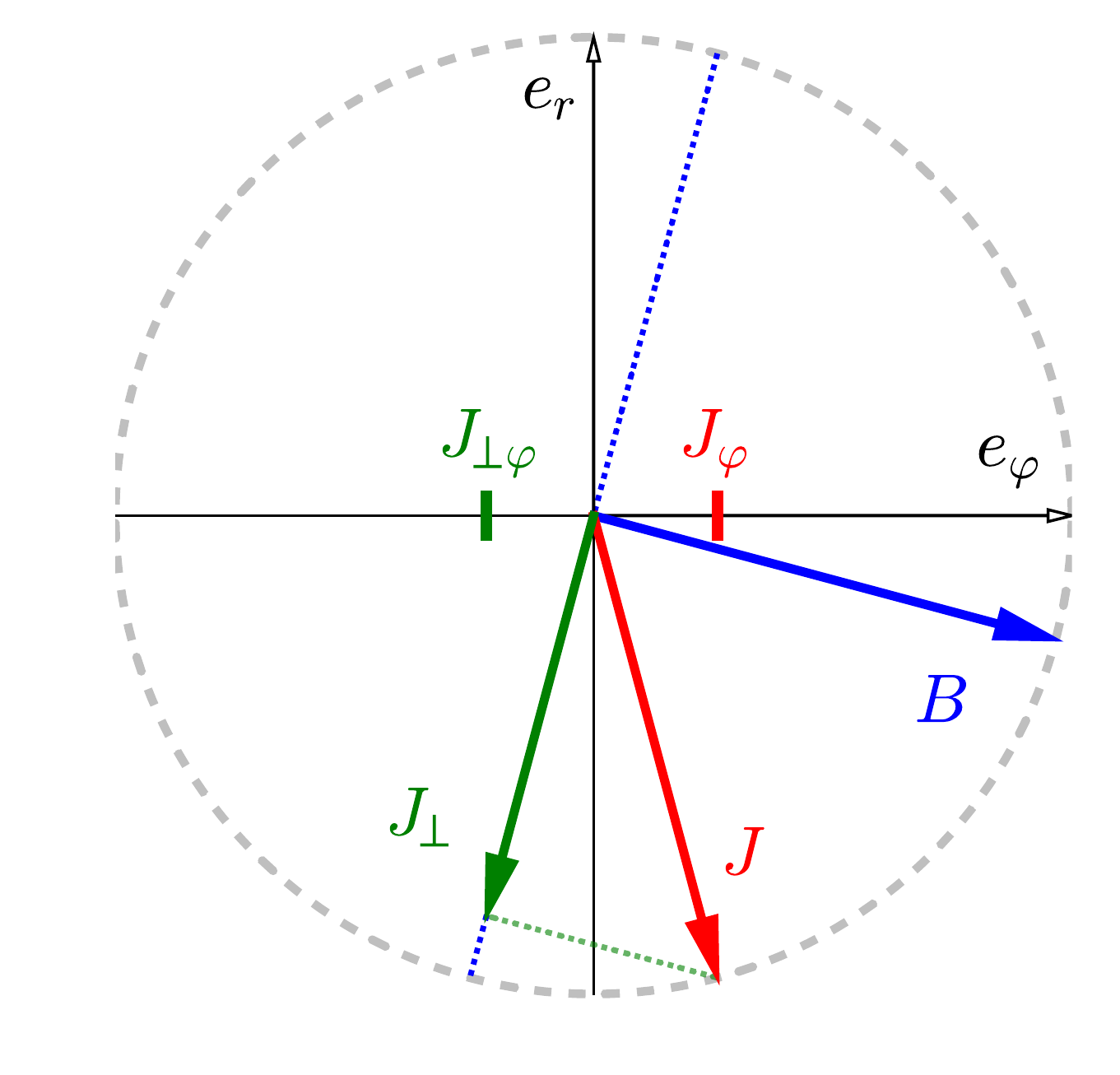}
\caption{Ambipolar-driven self-organization mechanism: the electric current $\bm{J}$ is mainly radial, and the magnetic field $\bm{B}$ is mainly toroidal; as a result, the signs of $J_{\varphi}$ and $J_{\perp\varphi}$ are opposed; this is equivalent to a negative diffusivity for $B_z$ (cf. Eq. \eqref{eqn:emf}). }
\label{fig:zf_proj}
\end{figure}

The direction of the ambipolar EMF is given by the electric current projected perpendicularly to the local magnetic field. Upon projection, the sign of the toroidal component $J_{\perp\varphi}$ can become opposed to the sign of $J_{\varphi}$. This is what happens in our simulations featuring zonal flows, as shown in the bottom panel of Fig. \ref{fig:zf_mech}. In this case, the toroidal component of $\mathcal{E}_{\mathrm{A}}$ yields a negative effective resistivity (cf. Eq. \eqref{eqn:emf}). 

\begin{figure}[ht]
\centering
\includegraphics[width=\hsize]{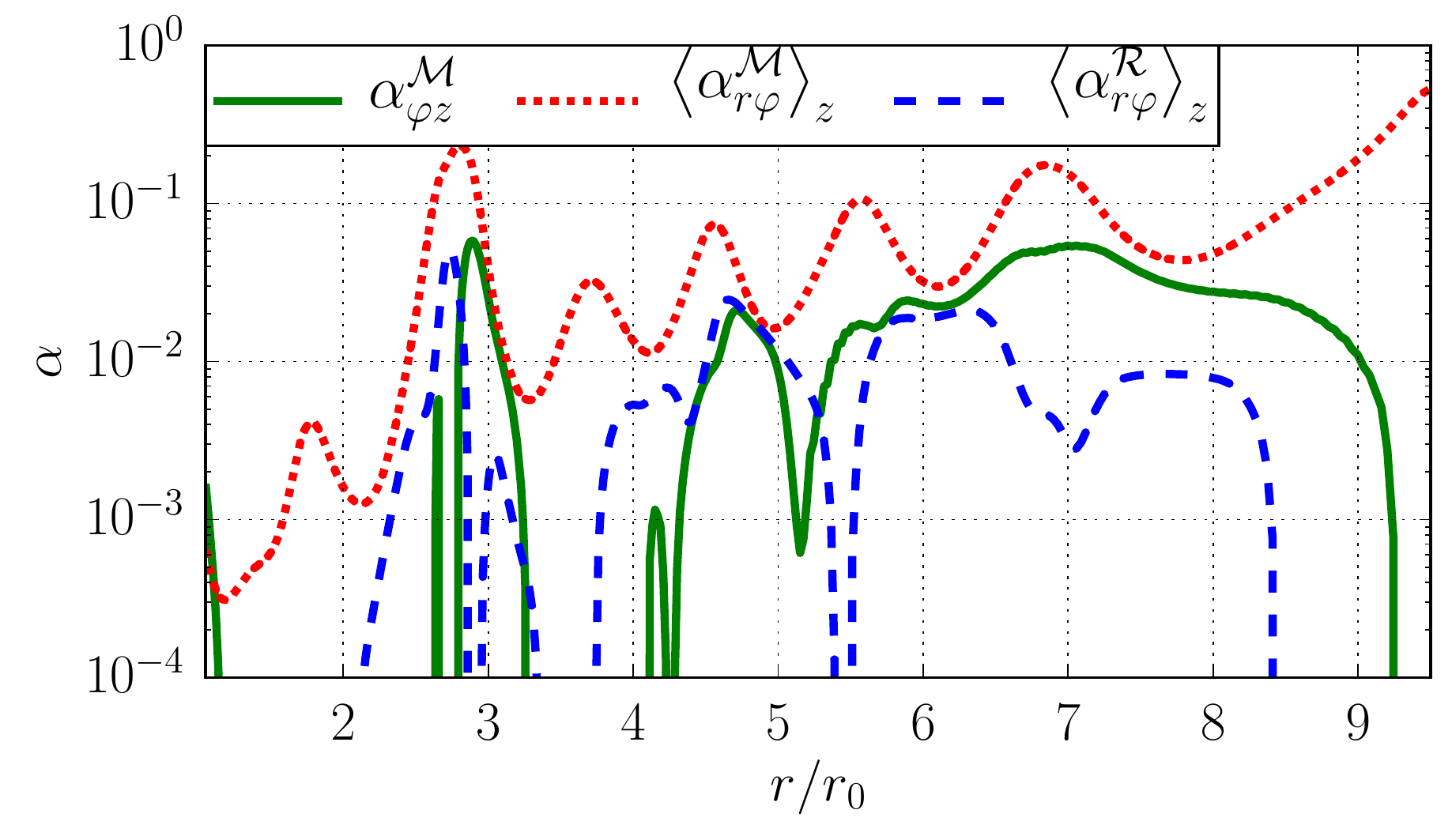}
\caption{Radial profiles of normalized stress in R1-P2, averaged in time between $500T_0$ and $700T_0$. }
\label{fig:zf_alpha}
\end{figure}

The typical configuration occurring in such simulations is sketched in Fig. \ref{fig:zf_proj}. The magnetic field is mainly toroidal in the disk, and the dominant component of $\bm{J}$ is the radial one. The signs of $J_{\varphi}$ and $J_{\perp\varphi}$ are opposed, and they remain opposed when flipping the orientation of $\bm{J}$ and/or $\bm{B}$. Ambipolar diffusion remains a dissipative process when considering all three spatial directions. With $J_{\perp} \simeq J_r$ and $\partial_{\varphi} \simeq 0$, the diffusion occurs primarily on $B_{\varphi}$ in the vertical direction. We verified that the same sign inversion occurred in the ambipolar run ABS presented by \cite{BLF16}, which also featured magnetic flux concentrations without Hall drift. 

We show in Fig. \ref{fig:zf_alpha} the radial profiles of magnetic stress in the saturated state containing zonal flows. The radial stress $\mathcal{M}_{r\varphi}$ is maximal in regions of magnetic field accumulation. The resulting stress divergence pushes mass away from stress maxima, causing the observed anti-correlation between $B_z$ and $\rho$.

\subsubsection{Hydrodynamic properties and saturation mechanism} \label{sec:zfsatur}

\begin{figure}[ht]
\centering
\includegraphics[width=\hsize]{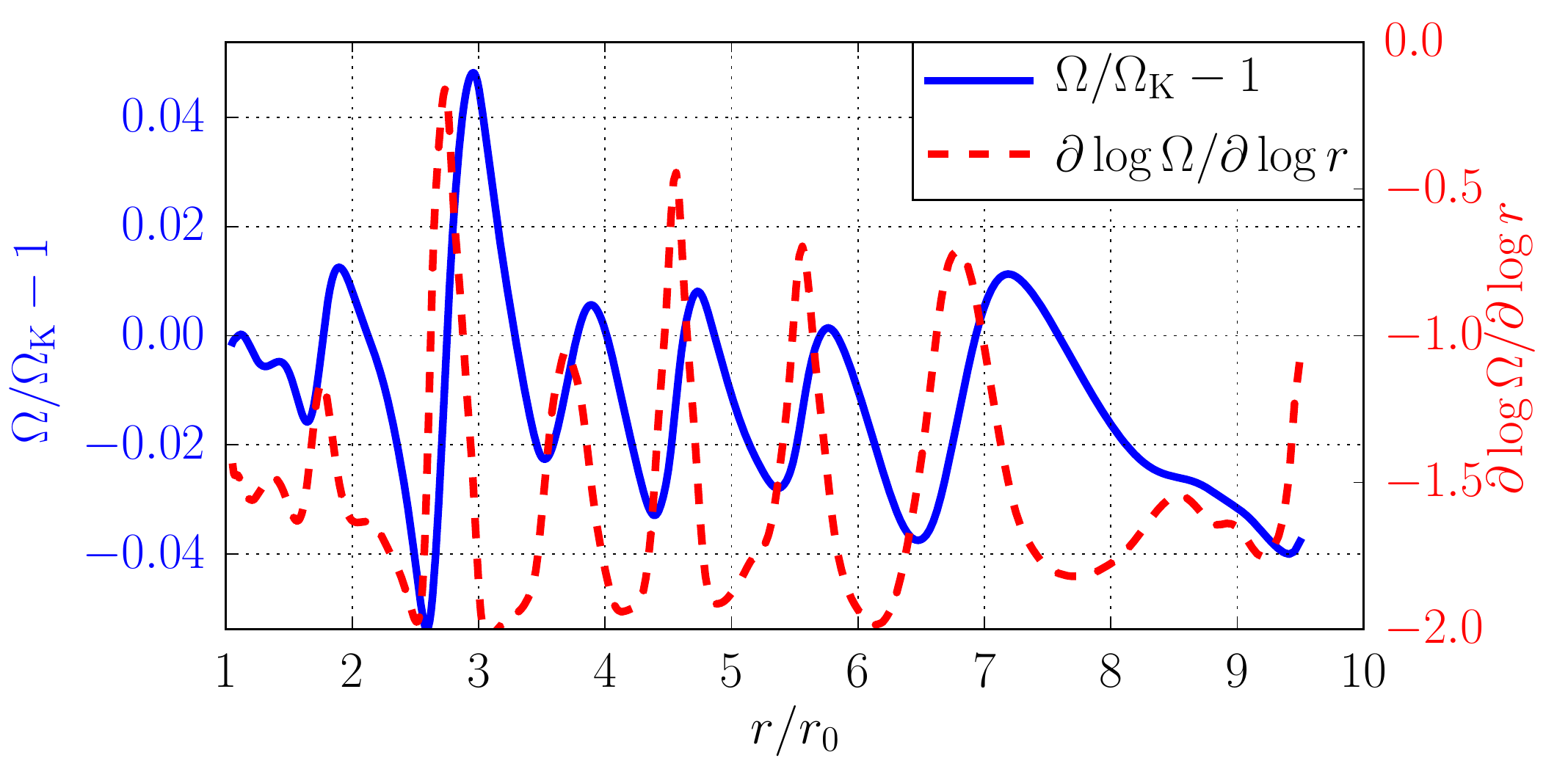}
\caption{Radial profiles in run R1-P2, vertically averaged through the disk from $500T_0$ to $700T_0$, of the variations of angular velocity $\Omega$ relative to Keplerian $\id{\Omega}{K}$ (solid blue), and dimensionless radial shear $\partial \log \rho / \partial \log r$ (dashed red). }
\label{fig:zf_hydro}
\end{figure}

Zonal flows refer to quasi-steady deviations from the Keplerian rotation profile of the disk. We characterize these by the fluctuations of angular velocity $\Omega$ relative to the Keplerian one $\id{\Omega}{K}$, and by the dimensionless shear $q \equiv \partial \log \Omega / \partial \log r$. The radial profiles of these two diagnostics are drawn in Fig. \ref{fig:zf_hydro} for run R1-P2, showing five distinct oscillations. Since $\beta \gg 1$, magnetic pressure gradients do not provide a significant support against gravity. Given the density fluctuations, the deviations from a Keplerian profile follow the condition of geostrophic equilibrium:
\begin{equation} \label{eqn:geostrophic}
v'_{\varphi} \equiv v_{\varphi} - \id{v}{K} \simeq \frac{c_s^2}{2 \id{\Omega}{K}} \frac{1}{\rho_0} \frac{\partial}{\partial r} \left[ \rho - \rho_0\right]
\end{equation}

One important issue regarding zonal flows is their ability to trap dust particles \citep{W77}. Several objects have now revealed axisymmetric structures in their dust distribution, such as rings and gaps \citep{ALMA15, TWHYA16}. One possible scenario involves zonal flows. Dust particles undergo a drag force from the gas as they orbit the star at the local Keplerian velocity, whereas the gas can rotate slower or faster depending on its radial pressure gradient. If the gas rotates faster, it will transfer angular momentum to the dust and make it migrate outward, or inward if it rotates slower. Dust grains are thus accumulated in pressure maxima. We prove in Fig. \ref{fig:zf_hydro} that the zonal flows produced in our simulations are actually able to cause transitions from sub- to super-Keplerian rotation. These zonal flows would therefore act as dust traps.  

The profile of dimensionless shear is superimposed to the rotation profile in Fig. \ref{fig:zf_hydro}, and tells us about the saturation of this process. In the limit $q \rightarrow -2$, the distribution of specific angular momentum is flat and the flow is marginally stable with respect to Rayleigh's criterion. This criterion appears to set the lower bound for the shear, as if any deviation $q < -2$ would instantly reorganize the flow to prevent hydrodynamic instability. However, we do not observe any evidence of purely hydrodynamic turbulence in these simulations. The disk therefore never crosses the Rayleigh line.

\subsubsection{Three-dimensional simulations} \label{sec:zf_3d}

\begin{figure}[ht]
\centering
\includegraphics[width=\hsize]{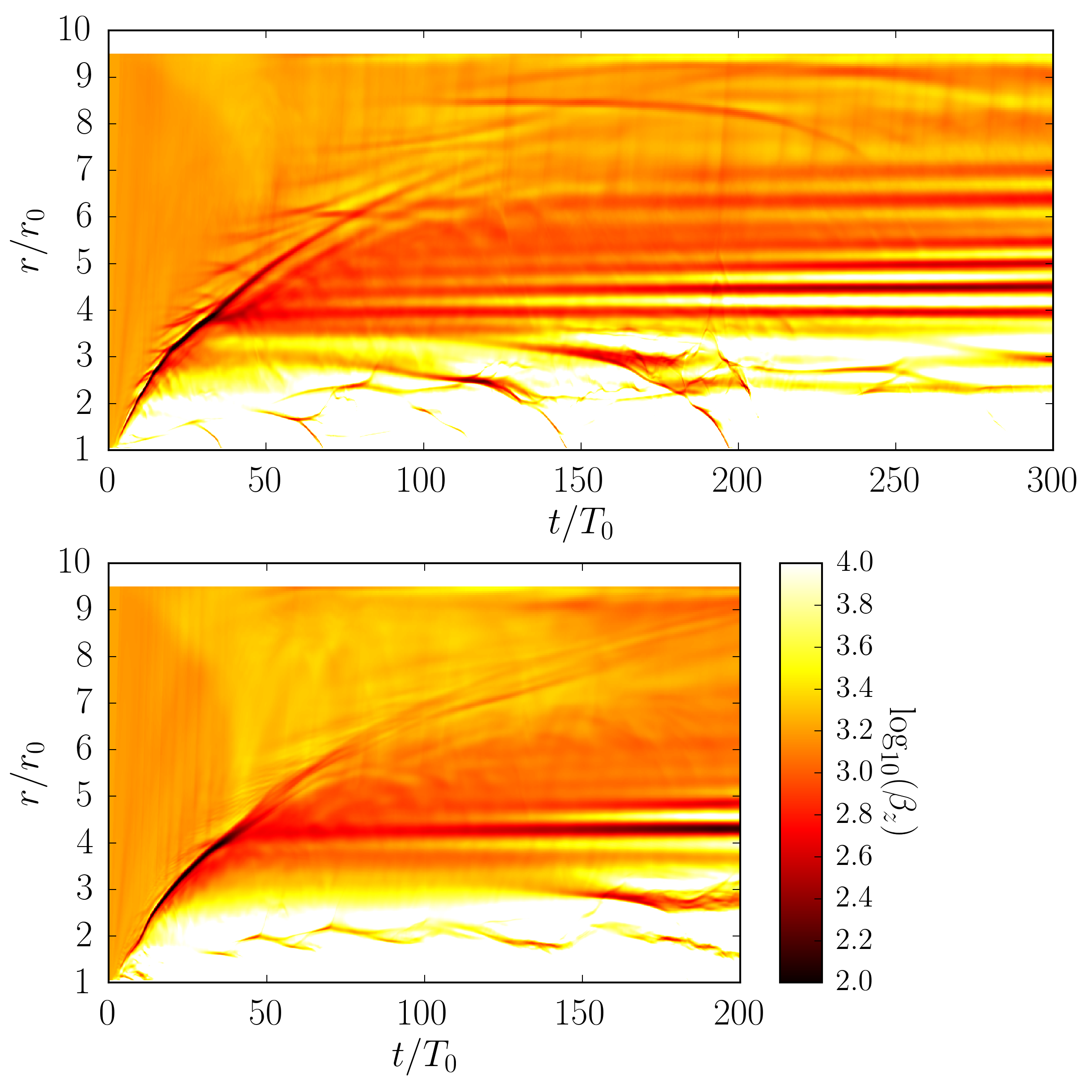}
\caption{Space-time diagram of vertically and azimuthally averaged $\beta_z$ in runs R1-P3 (\emph{upper panel}) and 3D-R1-P3 (\emph{lower panel}); only the first $300T_0$ of the two-dimensional run are shown.  }
\label{fig:zf_st23d}
\end{figure}

We now ascertain that these structures are not restricted to two-dimensional simulations. The formation of axisymmetric rings in run R1-P3 and in its three-dimensional equivalent run 3D-R1-P3 are shown in Fig. \ref{fig:zf_st23d}. We observe the emergence of zonal flows on the same time-scale at about the same location, and with the same radial separation. The structures form very early in the simulation ($50 T_0$ at $4r_0$, i.e., five local orbits), the contrast in $\beta_z \equiv 2P/B_z^2$ increases in time up to two orders of magnitude. They evolve over hundreds of local orbits, but we still count seven bands in run R1-P3 after $1000 T_0$. 

The present self-organization phenomenon is thus not inhibited in three-dimensional simulations. Regarding the non-axisymmetric stability of these structures, they could be Rossby wave unstable for the criterion given by \cite{LLCN99}. Nevertheless, run 3D-R1-P3 proves that they are stable over $20$ local orbits to all modes fitting in a quarter disk. Three-dimensional simulations of a full disk with an increased resolution and/or a less diffusive numerical scheme will be required to confirm their stability over long time scales.

\section{Discussion} \label{sec:discussion}

\subsection{Large-scale configuration of the magnetic field} \label{sec:recap_magconf}

\begin{figure}[ht]
\centering
\includegraphics[width=\hsize]{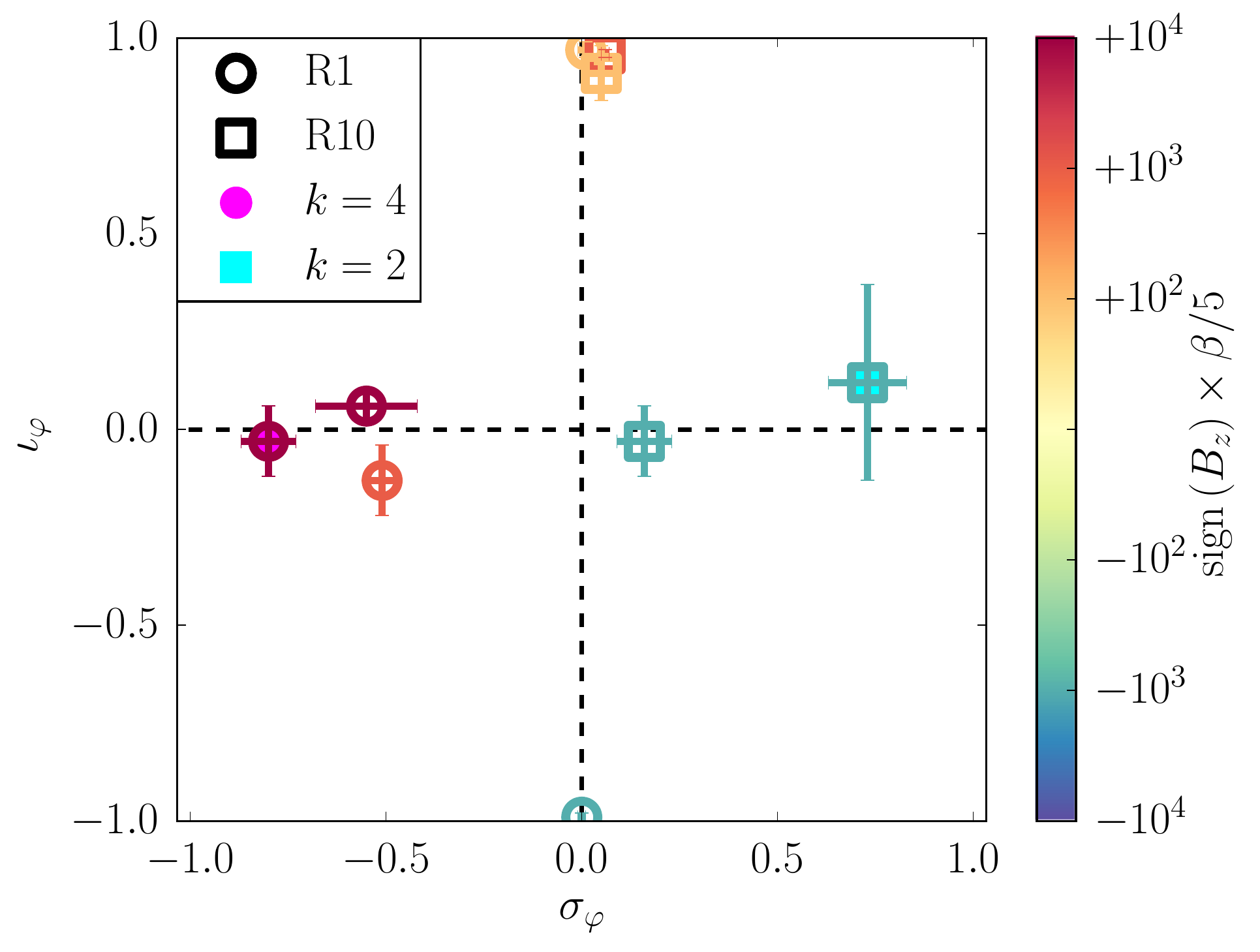}
\caption{Symmetry coefficients $\sigma_{\varphi}$ and $\iota_{\varphi}$ in our set of 2D simulations, averaged over representative time intervals; colors indicate the initial midplane magnetization $\beta$, blue corresponding to $B_z < 0$; runs in $[1,10] \au$ are plotted with circles, runs in $[10,100] \au$ are plotted with squares; different background colors correspond to different disk-corona temperature contrast $k$; the error bars correspond to standard deviations over time. }
\label{fig:recap_iotasigma}
\end{figure}

We gather the symmetry coefficients $\sigma_{\varphi}$ and $\iota_{\varphi}$ from Table \ref{table:results}, and plot them in Fig. \ref{fig:recap_iotasigma}. We distinguish three populations in this area. Runs with $B_z<0$ are the only ones on the right half of the figure ($\sigma_{\varphi} > 0$), i.e., overall loosing angular momentum in the wind and therefore accreting. The blue square associated with run R10-M3 is close to the origin, because its outer half is accreting while the inner half is non-accreting. On the left side $\sigma_{\varphi} < 0$, we find runs with a positive and weak magnetic field. The third population is located near $(\sigma_{\varphi},\iota_{\varphi}) \approx (0,\pm 1)$, i.e., only one sign of $B_{\varphi}$ in the disk, accreting in one hemisphere and non-accreting on the other. These runs mostly have $B_z>0$ and a strong magnetization $\beta \leq 5\times10^{3}$. 

\begin{figure}[ht]
\centering
\includegraphics[width=\hsize]{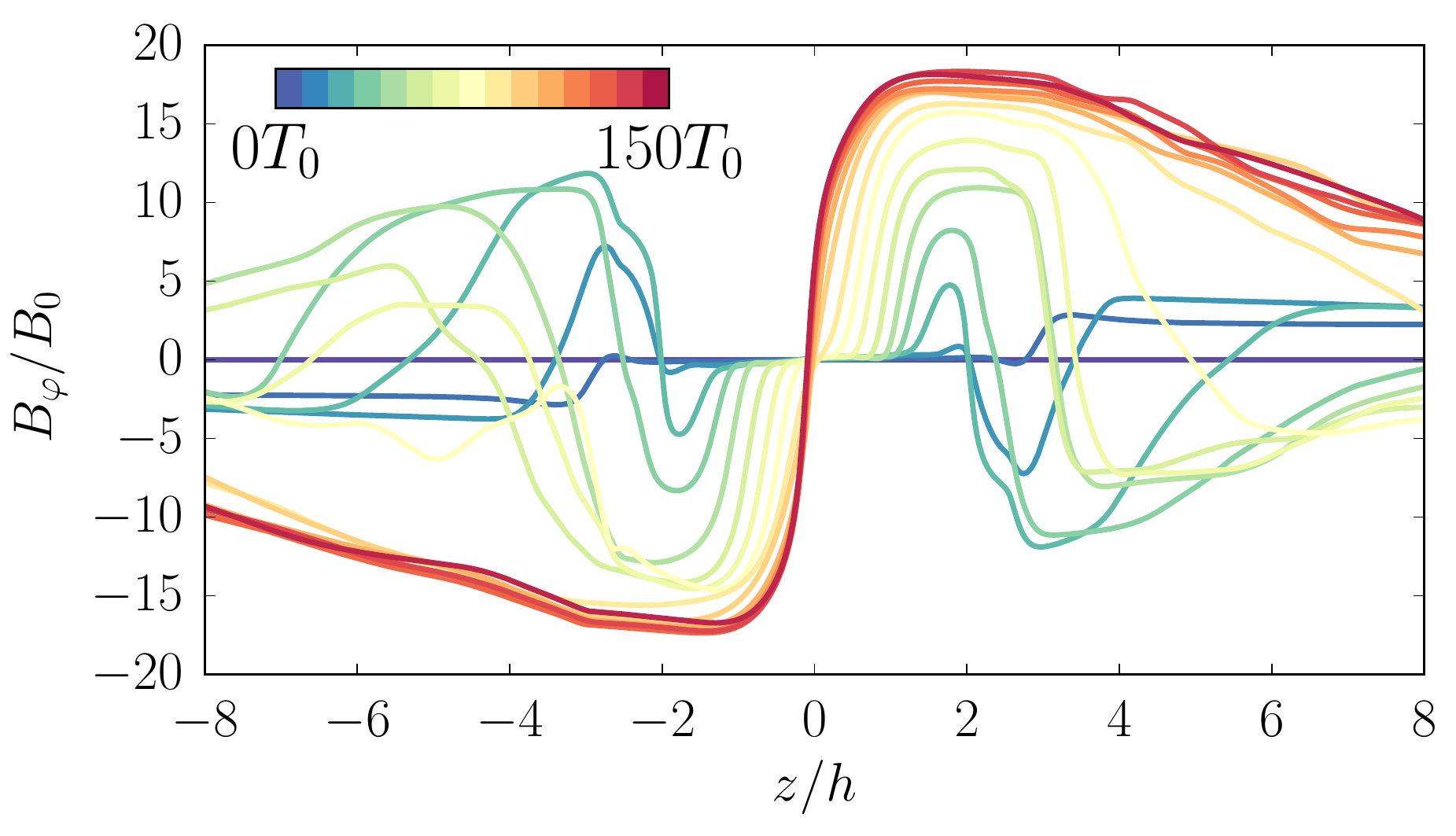}
\caption{Evolution of the toroidal magnetic field $B_{\varphi}$ along a vertical cut at $r=5r_0$ and over the first $150T_0$ (color scale) of run R10-M3-C2 (accreting); values have been normalized by the initial, \emph{unsigned} vertical field at this radius. }
\label{fig:recap_byzcut}
\end{figure}

It is tempting to see a correlation between the sign of $B_z$ and the presence of magnetized outflows. However, we believe that this correlation is fortuitous. Indeed, the vertical phase of $B_\varphi$ seems to be set at least partially by the noise injected in our initial conditions. This is illustrated in Fig. \ref{fig:recap_byzcut} showing the evolution of $B_\varphi$ as a function of time. From $10T_0$ to $40T_0$, the MRI amplifies $B_{\varphi}$ at $z\approx 3h$, causing $\sigma_{\varphi}$ to alternate from positive to negative as the instability saturates and eventually settles with $\sigma_{\varphi}>0$. In this example, $\sigma_{\varphi}>0$ is clearly set by the initial phase and saturation mechanism of the most unstable MRI mode, which itself depends on how the MRI was initially seeded.

Several hundred inner orbits later, one polarity of the toroidal magnetic field can be vertically removed out of the disk, leading to a one-sided outflow. Among these runs with $\iota_{\varphi} = \pm 1$, most have $B_z>0$ and $\beta < 5 \times 10^{4}$, the exception being run R1-M3. These configurations are stable attractors in the sense that these disks never return to an odd symmetry of $B_{\varphi}$ about the midplane.

\subsection{Magnetothermal winds}

We have shown that all runs display a vertical outflow, although it is disorganized for non-accreting disks (cf. Sect. \ref{sec:nonejec}). We observed  no difference between two and three-dimensional winds. Our winds exhibit several properties of steady-state, magnetically-driven jets. However, they become super-Alfv\'enic before reaching the ideal MHD regime, contrarily to all published models of magnetically-driven jets from accretion disks. This could explain why, although they also heated the base of the wind, \cite{CF00B} never achieved an ejection efficiency as high as the one we obtain. It should be possible to design models going continuously from thermally-driven to magnetically-driven winds. Our simulations provide such a link. 

One important control parameter for magnetically-driven winds is the disk magnetization ($1/\beta$). At a comparably low disk magnetization, \cite{MFZ10} also obtained super-fast-magnetosonic winds, but with a larger magnetic lever arm \citep[see their Figs. 8 \& 14, and also][]{SF16}. However, they used constant $\alpha$ viscosity and resistivity coefficients, with prescribed vertical profiles to mimic turbulent transport in the disk. This suggests that different transport regimes within disk may lead to different wind properties. 

\begin{figure}[ht]
\centering
\includegraphics[width=\hsize]{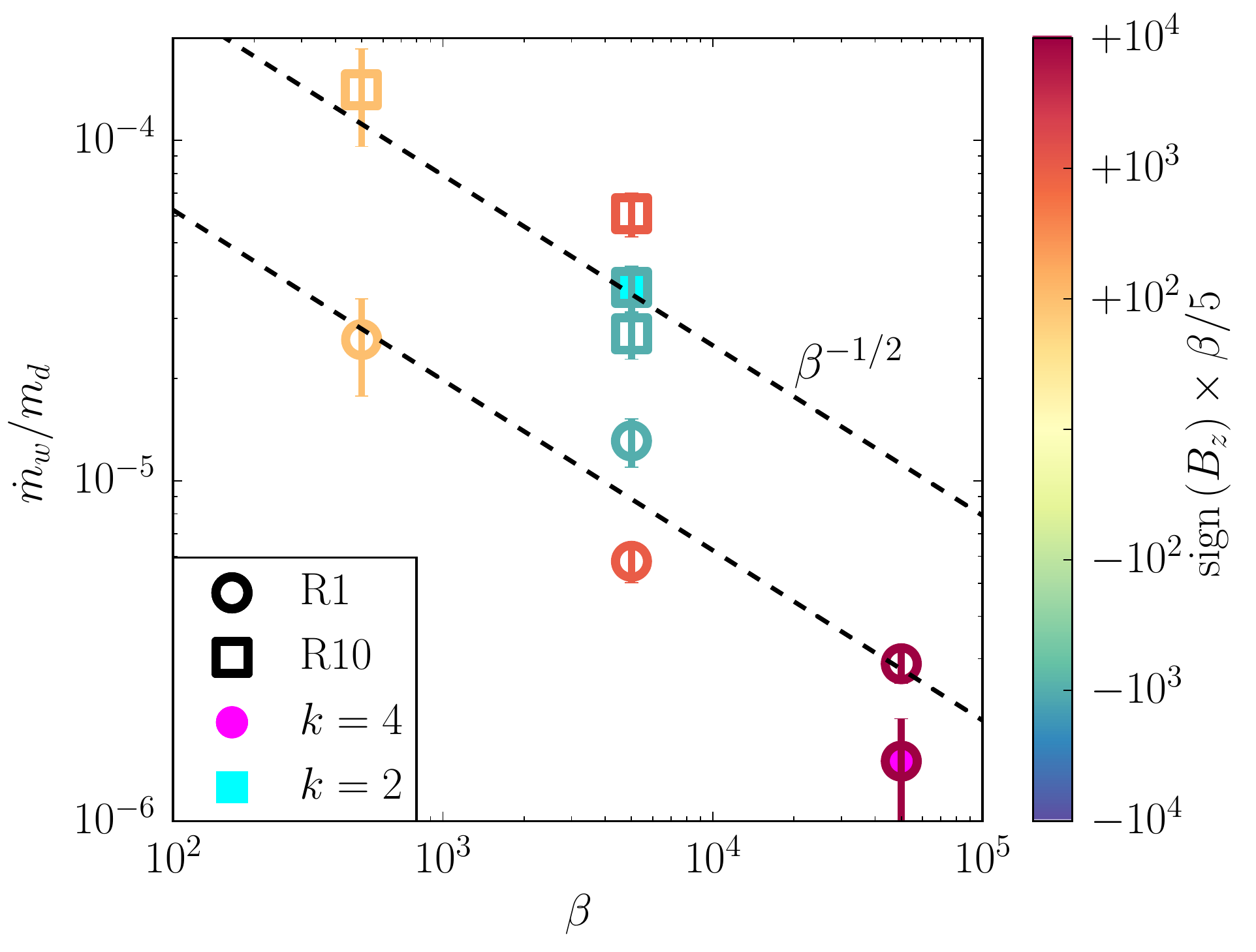}
\caption{Wind mass loss rate relative to the disk mass $\dot{m}_W/m_d$ as a function of the initial midplane magnetization $\beta$; the symbols and color coding are the same as in Fig. \ref{fig:recap_iotasigma}; the two dashed lines indicate the $\dot{m}_W \sim \beta^{-1/2}$ scaling for R1 and R10 runs, bottom and top, respectively. }
\label{fig:recap_mdot}
\end{figure}

We plot the wind mass loss rate $\dot{m}_W$ of our two-dimensional runs in Fig. \ref{fig:recap_mdot} as a function of the midplane magnetization $\beta$. Runs R1-M3 (blue circle) and R10-P3 (red square), both of which have $\iota_{\varphi} = \pm 1$ (cf. Fig. \ref{fig:recap_iotasigma}), also have higher mass loss rates than the other simulations of the same radial range. Given this degree of freedom, we find no clear dependence of $\dot{m}_W$ with the polarity of the initial magnetic field alone. There is no apparent correlation of $\dot{m}_W$ with the temperature contrast $k$ either. 

The wind mass loss rate typically scales as $\dot{m}_W \sim \beta^{-1/2}$ in runs R1 and R10 taken separately. This is consistent with the shearing-box measurements of \cite{BS13}, regardless of the fast-magnetosonic point being crossed in the computational domain \citep[see Sect. \ref{sec:fid_ejec} of this paper, and Sect. 3.4 of][]{LFO13}. In physical units, the wind mass loss rate is
\begin{equation}
\dot{M}_W \approx 2 \times 10^{-2} \left( \frac{r_0}{1 \au} \right)^{-1/2} \left( \frac{\dot{m}_W}{m_d} \right)
\end{equation}
The fact that $\dot{m}_W / m_d$ increases by a factor $4 \approx \sqrt{10}$, from R1 runs to R10 runs, means that $\dot{M}_W$ is roughly independent of the inner radius. We can sum the contributions over $[1,100]\au$ to get
\begin{equation}
\dot{M}_W \approx 3 \times 10^{-5} \beta^{-1/2} ~ M_{\odot}.\,\mathrm{yr}^{-1}. 
\end{equation}

\section{Summary} \label{sec:summary}

We have performed global simulations of protoplanetary disks in the non-ideal MHD framework. Our model includes both radial and vertical density stratification, spanning one decade in radius to separately cover the ranges from $1\au$ to $10\au$ and from $10\au$ to $100\au$, with 20 disk scale heights vertically. The ionization fraction is computed dynamically, and all three non-ideal MHD effects are accounted for. The disk is embedded in a warm corona, with several temperature contrasts. 

In the disk, the flow is essentially laminar and evolves over hundreds of local orbital periods. Angular momentum is transported by a large-scale magnetic stress, unsuited for an $\alpha$ viscosity prescription. The radial transport of angular momentum causes no net mass accretion in our disk model. The vertical flux of angular momentum can take both orientations through the disk. If angular momentum is transported from the disk midplane to the surface, then a magnetized wind is launched and the disk accretes at rates on the order of $10^{-7}\,\mathrm{M}_\odot/\mathrm{yr}$. Conversely, angular momentum can be transported from the disk surface to the midplane, resulting in a large-scale meridional circulation. This circulation is characterized by the gas flowing radially inward at the surface, and outward in the midplane. In this case, a turbulent and low density wind is observed, causing no net mass accretion. Accreting and non-accreting regions can coexist in one single disk, the boundary between the two domains evolving on secular timescales. The Hall drift can reduce the angular momentum flux in the midplane, depending on the polarity of the vertical magnetic field threading the disk, but the overall contrast in stress is only within a factor of three. 

Because of the presence of a warm corona, the winds we obtain are magnetothermal. They constitute a new class of solutions because the plasma can reach super-Alfv\'enic speeds before entering the ideal MHD regime. The mass loading of the wind is driven by the Lorentz force, but the subsequent acceleration is predominantly thermal. We have linked the wind mass loss rate to the disk magnetization, and recovered the $\dot{m}_W \propto \beta^{-1/2}$ relation. Several of our simulations also exhibit asymmetric winds where magnetic ejection occurs on one side of the disk only. We have associated this state with the expulsion of the electric current sheet, usually located near the disk midplane. We speculate that this configuration is a stable attractor for the disk, allowing it to evacuate angular momentum at higher rates than the symmetric situation. One-sided molecular outflows \citep{PGGD06} could be a consequence of such magnetic configurations. 

Finally, we have identified a self-organization process leading to the formation of axisymmetric density rings. To these density rings are associated zonal flows, able to trap dust particles. The vertical magnetic flux becomes concentrated in density minima. We have argued that ambipolar diffusion alone is responsible for this process. \cite{BS14} also observed enhanced magnetic flux concentrations in MRI turbulence simulations. They proposed a mechanism driven by ideal MHD electromotive forces. In our case, we have shown that the concentration mechanism relies solely on ambipolar diffusion. We are therefore looking at a different mechanism from the one they proposed. We presume that \cite{GTNN15} did not observe this phenomenon because of the lower magnetization and ionization fraction in their simulations. We have shown that the Hall drift works against the confinement of magnetic flux in our case. This demonstrates that Hall-driven self-organization \citep{KL13,BLF16} is inefficient in stratified simulations, as already suggested by stratified shearing boxes \citep{LKF14,Bai15}. 

We note that we have not explored systematically the global transport of poloidal magnetic flux in our simulations. The poloidal field being a key ingredient for MHD winds, any long-term prediction for the evolution of these disks implies a parametrization for the evolution of this field \citep{SOMCG16}. This transport depends strongly on the vertical disk structure \citep[e.g.][]{GO12,GO13}, and non-ideal effects add more complications to this picture \citep{BS16arxiv}. In our simulations, we found that transport of flux was highly sensitive to the presence of self-organized structures, which act as physical boundaries for the radial transport of poloidal field. This is to be expected since self-organization is after all a consequence of poloidal flux transport. For this reason, we believe that any description of poloidal flux transport needs to incorporate the possibility of self-organization, without which flux transport could be largely overestimated.

The main caveats of our study are related to the thermodynamics and chemistry of protoplanetary disks. The validity of our prescription for the temperature distribution should be compared with self-consistent, radiative transfer simulations. By omitting small dust grains, we may also have severely underestimated the magnitude of non-ideal plasma effects. Our main justification is practical in nature since both these treatments come with large computational overhead. These limitations will need to be addressed in future work.

\begin{acknowledgements}
We thank Matthew Kunz and Mario Flock, as well as our anonymous referee for providing thoughtful comments on the manuscript. 
This work was granted access to the HPC resources of IDRIS under the allocation 2016-042231 made by GENCI. Some of the computations presented in this paper were performed using the Froggy platform of the CIMENT infrastructure (https://ciment.ujf-grenoble.fr), which is supported by the Rhône-Alpes region (GRANT CPER07\_13 CIRA), the OSUG@2020 labex (reference ANR10 LABX56), and the Equip@Meso project (reference ANR-10-EQPX-29-01) of the programme Investissements d'Avenir supervized by the Agence Nationale pour la Recherche.
\end{acknowledgements}


\bibliographystyle{aa}
\bibliography{biblio}

\begin{appendix}

\section{Initial conditions} \label{app:initcond}

We construct steady, locally isothermal, hydrodynamic disk solutions to the mass and momentum equations \eqref{eqn:dyn-rho} and \eqref{eqn:dyn-v}. We first prescribe an isothermal sound speed distribution $c_s \equiv \sqrt{\partial P / \partial \rho}$. The use of $c_s^2 = c_0^2 (r/r_0)^{\,\alpha}$ helps building simple analytical solutions. The distance unit $r_0 = 1$ in this section. With the gravitational potential $\Phi = -1/R$, the momentum equations in spherical coordinates can be written
\begin{align}
\frac{v_{\varphi}^2}{R} &= c_s^2 \partial_R \log (\rho) + \frac{\alpha c_s^2}{R} + \frac{1}{R^2}, \label{eqn:mom_r}\\
\cot(\theta) v_{\varphi}^2 &= c_s^2 \partial_{\theta} \log (\rho) + \alpha \cot(\theta) c_s^2. \label{eqn:mom_theta}
\end{align}
We can isolate $v_{\varphi}$ and equate the two lines; defining $f \equiv \log(\rho)$, $x \equiv \log(R)$ and $y \equiv - \log(\sin(\theta))$, we obtain
\begin{equation}
\left( \partial_x + \partial_y \right) f = -\frac{1}{R c_s^2} = \frac{1}{c_0^2} \exp\left(- \left(\alpha+1\right) x + \alpha y\right). 
\end{equation}
This can be solved by the method of characteristics, once a midplane density profile $\rho(r,z=0) = \rho_0 (r/r_0)^{\,n}$ has been prescribed:
\begin{equation}
\rho (R,r) = \rho_0 \left( \frac{r}{r_0} \right)^n \exp\left[ \frac{1}{c^2_s(r)} \left( \frac{1}{R} - \frac{1}{r} \right) \right], \label{eqn:rho}\\
\end{equation}
and it is straightforward to obtain for the velocity field 
\begin{equation}
v^2_{\varphi} (R,\theta) = \left( 1 - \left(\alpha+1\right)\left(1- \frac{1}{\sin(\theta)}\right) \right) \frac{1}{R} + \left( n + \alpha \right) c^2_s(R,\theta). \label{eqn:vy2}
\end{equation}

To have a constant opening angle of the disk, we set a constant sound to Keplerian speed ratio: $c_{s, \mathrm{disk}} / \id{v}{Kepler} = 5 \%$, i.e., $\alpha= -1$. In this case, the disk midplane density must decrease with $n=-2$ in order to produce the desired surface density profile $\Sigma(r) \equiv \int \rho dz \sim r^{\,-1}$. 

We define the disk-corona transition as the location where the disk equilibrium density is one thousandth of the local midplane density: $\rho(r,z_i) = 10^{-3} \rho(r,z=0)$. The transition angle is then given by $\sin(\theta_i) = 1 + \log(10^{-3}) \,(h/r)^2$, corresponding to a constant $z_i \approx 3.72h \equiv H$. 

At a given cylindrical radius, the isothermal sound speed is increased up to a factor $k$ higher than the midplane value. This ratio of corona to disk temperature is constant with cylindrical radius, and fixed to $k = 6$ for most runs. The transition from the disk to the warm corona is smoothed in the interval $\sin(\theta) \in \left[\sin(\theta_i) - 0.2(h/r), \sin(\theta_i)\right]$, such that
\begin{align} \label{eqn:csdiskoro}
c_{s,\mathrm{corona}}(r,\theta) &= c_{s,\mathrm{disk}}(r,\theta_i) \times \left( 1 + (k - 1) \frac{\sin(\theta_i) - \sin(\theta)}{0.2 (h/r)}\right),
\end{align}
in order to avoid strong gradients and, if possible, resolve the physics at the transition. The gas in the corona is thus fully heated at a height $z_c / h \approx 4.68$. 

The coronal density distribution takes the same form as in Eq. \eqref{eqn:rho}, except this time the solution is extended from the disk surface and not from the midplane. In order to continuously match the isothermal pressure fields $\id{P}{iso} \equiv \rho c^2_s$ at the interface, the factor $\rho_0$ in Eq. \eqref{eqn:rho} must be replaced by $\rho_0 \mapsto \rho(R,\theta_i) \left[c_{s, \mathrm{disk}}(R,\theta_i) / c_{s, \mathrm{corona}}(R,\theta)\right]^2$. 

The velocity field in the corona takes exactly the same form as \eqref{eqn:vy2}. Due to the vertical temperature gradient, there is a vertical shear in azimuthal velocity in the transition layer $\partial_z v_{\varphi} \neq 0$; we verified via hydrodynamic simulations that this layer quickly relaxes to a steady and stable state. We add a white noise to the three components of the initial velocity field, with amplitude $1\%$ of the local sound speed, so as to quickly trigger the MRI. 

The initial magnetic field has only $B_z \neq 0$. It is initialized via its vector potential $\bm{B} = \nabla \times \bm{A}$ to ensure the solenoidal condition $\nabla \bm{\cdot B} = 0$ from the beginning. To set a constant average / midplane $\beta$, the magnetic field intensity must decrease as $r^{-3/2}$, whence $A_{\varphi} = 2 B_0 \sqrt{r_0/r}$, with $B_0$ the intensity of the magnetic field at the inner radius $r_0$ (see the appendix of \cite{SI14}). The initial conditions deviates from a true MHD equilibrium, but only at order $1/\beta$.

\section{Boundary conditions} \label{app:boundcond}

Three dimensional simulations are periodic in the azimuthal direction. The remaining boundary conditions are listed below. These boundary conditions are designed to prevent boundary-driven flows while remaining relatively simple. To isolate the corners at the inner radial boundary, we define the critical radius $r_c/r_0 \equiv (1 + \sin(\pi/2-\theta_0)) / 2$, with $\theta_0 = 20 (h/r)$ the polar extent of the computational domain in one hemisphere.  
\begin{itemize}
\item $\rho$: zero gradient through all boundaries, with a lower bound $\rho_c=10^{-6}\rho_0$ at the inner radial boundary;
\item $P$: isothermal distribution $\rho c_{s}(r,\theta)^2$ at all boundaries;
\item $v_R$: zero gradient through all boundaries; outward $c_{s}(R,\theta)$ as lower bound at the outer radial boundary in the corona;
\item $v_{\theta}$: zero gradient through all boundaries; outward $c_{s}(R,\theta)$ as lower bound at both polar boundaries;
\item $v_{\varphi}$: zero gradient through polar and outer radial boundaries; at the inner radial boundary: initial Keplerian velocity in the disk, constant angular velocity $v_{\varphi}/R$ in the corona;
\item $B_R$: zero gradient through polar boundaries if $r>r_c$, zero value otherwise;
\item $B_{\theta}$: zero value at inner radial boundary, zero gradient through outer radial boundary;
\item $B_{\varphi}$: zero gradient through radial and polar boundaries if $r>r_c$, zero value otherwise. 
\end{itemize}

We complete these boundary conditions with buffer zones, inside the computational domain to limit the influence of the boundaries on the disk. The inner buffer includes both the disk and the corona in $\left\lbrace (R,\theta) \mid R < 1.05 r_0\right\rbrace$. The outer buffer includes only the disk in $\left\lbrace (R,\theta) \mid R > 9.5 r_0\right\rbrace \cap \mathcal{D}$. Let $w_b$ be the radial width of the considered buffer. Within the buffer, we explicitly relax a given field $X$ to some prescribed distribution $X_0$:
\begin{equation}
\frac{\partial X}{\partial t}(t) = -f \times m \times \left(X(t) - X_0\right),
\end{equation}
with a characteristic frequency scale 
\begin{equation}
f(R,\theta) \equiv 5 \times \left(\id{\Omega}{Kepler}(R,\theta) + c_s(R,\theta)/w_b\right),
\end{equation}
and a linear modulation $m(R)$, equal to $1$ at the domain boundary and $0$ at the other edge of the buffer (in the active domain). Only the three components of the velocity field are modified inside the buffers in continuity with the boundary conditions:
\begin{itemize}
\item $v_R$: relaxed to zero at both boundaries;
\item $v_{\theta}$: relaxed to maintain a constant angular velocity $v_{\theta}/R$, sampled at the edge of the buffer on the active domain side;
\item $v_{\varphi}$: relaxed to the initial velocity in the disk region of both buffers, and to a constant angular velocity $v_{\varphi}/R$ in the corona region of the inner buffer, also sampled at the buffer edge. 
\end{itemize}

\section{Internal conditions} \label{app:inbound}

In addition to the buffer zones, we control the values of each computational grid cell and at each time step. The first reason is to avoid excessively high Alfv\'en velocities, imposing small time steps to satisfy the CFL stability criterion. This is one of the main difficulties with global, stratified MHD simulations including a net magnetic flux. A common remedy is to impose a lower bound, i.e., a floor value on the density. However, we checked that arbitrarily low densities are well handled in the absence of magnetic field, and we found that a density floor could produce significant pressure gradients in the outer corona. We therefore increase the density only when the Alfv\'en velocity becomes too large. With our prescribed temperature distribution, the Alfv\'en velocity seldom reaches $\id{v}{A} \lesssim v_0$, which we set as its upper limit. 

The second control performed on the whole computational domain is related to the gas temperature distribution. Since we do not solve for the radiation field, we enforce a locally isothermal equilibrium, i.e., a constant sound speed distribution $c_0(r,z)$. For a steady toroidal flow, the vertical shear of azimuthal velocity is given by 
\begin{equation} \label{eqn:baroclin}
\frac{\partial}{\partial z} \frac{v^2_{\varphi}}{r} = \left[\frac{\nabla P \bm{\times} \nabla \rho}{\rho^2} \right] \bm{\cdot e}_{\varphi},
\end{equation}
which depends on the thermodynamics of the flow. Our locally isothermal equilibrium has a non-zero vertical shear, and is therefore subject to the vertical shear instability \citep{GS67,Fricke68,UB98}. It was shown by \cite{NGU13} that this instability could be strongly attenuated by letting the entropy distribution reorganize, and then relaxing the temperature distribution over a fraction of orbital period. For this reason, we evolve the total energy density assuming an ideal gas:
\begin{align}
&E \equiv \frac{P}{\gamma - 1} + \frac{1}{2} \left( \rho v^2 + B^2\right) + \rho \Phi, \\
&\partial_t E = - \bm{\nabla \cdot} \left[ \left(E + P + 	B^2/2\right) \bm{v} - \left( \bm{v \cdot B}\right) \bm{B} \right] \label{eqn:dyn-E}
\end{align}
with $\gamma = 7/5$ corresponding to cold diatomic molecules. We then bring the isothermal sound speed $c_s \equiv \sqrt{P/\rho}$ back to its initial distribution, over a fraction of local orbital period:
\begin{equation}
\frac{\partial p}{\partial t}(r,z,t) = - \frac{\Omega(r)}{\epsilon} \left( p(r,z,t) - \rho(r,z,t) \:c_0^2(r,z) \right). 
\end{equation}
Radiative transfer simulations predict short cooling rates in the bulk of the disk \citep{FFGC13,SK14}. We verified via hydrodynamic simulations that $\epsilon = 1/10$ would, indeed, force this instability to saturate at an imperceptibly small level. This value is comparable to the threshold derived by \cite{LY15} with our set of parameters, $\epsilon_0 \sim 1/8$.

\section{Caps on non-ideal MHD effects} \label{app:nonideal}

We apply two different limiters on non-ideal effects. In the first place, our simplified chemical model cannot properly represent the strong degree of ionization high in the corona. We therefore multiply the diffusivities $\id{\eta}{O,H,A}$ in Eq \eqref{eqn:dyn-b} by a factor $\exp\left( - x_e / 10^{-8}\right)$. This coefficient ensures a smooth decrease of the ambipolar diffusivity until up to eight scale heights. 

Because we use an explicit integration scheme, all three non-ideal effects impose severe constraints on the admissible timesteps satisfying the Courant-Friedrich-Lewy (CFL) stability criterion. We add a second cap to prevent excessively small timesteps. The strongest constraint come from the Hall drift, so we choose to limit the ratio $\lH / h(r) \leq 4$. This limit affects only the inner region $r \lesssim 2.3r_0$ for runs at $r_0 = 1\au$, and does not concern runs at $r_0 = 10\au$. The ohmic and ambipolar diffusivities are limited via their associated Reynolds numbers $\id{R}{O,A} \equiv \Omega h^2 / \id{\eta}{O,A} \geq 1/4$. This limiter only affects ambipolar diffusion in the surface layers $z \gtrsim 3h$ of the disk. Once we have applied the first cap on the ionization fraction, this second limiter forces the saturation values of $\id{\eta}{A}$ only in runs R1-P2 and R10-P2, i.e., where the initial magnetization is the strongest. In these two runs, the Reynolds number is saturated at $\id{R}{A} \approx 1/4$ in the surface layers from $z\gtrsim 2.6h$ to $3h$.

\end{appendix}

\end{document}